\documentclass[aps,pra,twocolumn,10pt,superscriptaddress,showpacs,showkeys]{revtex4-1}
\usepackage{color}
\usepackage{graphicx}
\usepackage{amsfonts}
\usepackage{amsmath}
\usepackage{amssymb}
\usepackage{braket}
\usepackage{mathtools}
\usepackage{siunitx}
\usepackage{tablefootnote}
\usepackage[]{combelow}
\usepackage{calc}
\usepackage{verbatim}
\usepackage{textcomp}
\usepackage{scrextend} 
\usepackage[pdftex,
pdfauthor={Matteo Mariantoni},
pdftitle={The Quantum Socket: Three-Dimensional Wiring for Extensible Quantum
Computing},
pdfkeywords={Quantum Computing; Scalable Qubit Architectures; Quantum Socket;
Three-Dimensional Wires; Microwave Interconnects; Superconducting Quantum
Circuits; Quantum Error Correction}]{hyperref}

\DeclarePairedDelimiterX{\abs}[1]{\lvert}{\rvert}{\ifblank{#1}{{}\cdot{}}{#1}}
\DeclareSIUnit\gauss{G}

\begin{document}

\title{The Quantum Socket: Three-Dimensional Wiring for Extensible Quantum
Computing}

\author{J. H.~B\'{e}janin}
\thanks{These two authors contributed equally to this work.}
\affiliation{Institute for Quantum Computing, University of Waterloo, 200
University Avenue West, Waterloo, Ontario N2L 3G1, Canada}
\affiliation{Department of Physics and Astronomy, University of Waterloo, 200
University Avenue West, Waterloo, Ontario N2L 3G1, Canada}

\author{T. G.~McConkey}
\thanks{These two authors contributed equally to this work.}
\affiliation{Institute for Quantum Computing, University of Waterloo, 200
University Avenue West, Waterloo, Ontario N2L 3G1, Canada}
\affiliation{Department of Electrical and Computer Engineering, University of
Waterloo, 200 University Avenue West, Waterloo, Ontario N2L 3G1, Canada}

\author{J. R.~Rinehart}
\affiliation{Institute for Quantum Computing, University of Waterloo, 200
University Avenue West, Waterloo, Ontario N2L 3G1, Canada}
\affiliation{Department of Physics and Astronomy, University of Waterloo, 200
University Avenue West, Waterloo, Ontario N2L 3G1, Canada}

\author{C. T.~Earnest}
\affiliation{Institute for Quantum Computing, University of Waterloo, 200
University Avenue West, Waterloo, Ontario N2L 3G1, Canada}
\affiliation{Department of Physics and Astronomy, University of Waterloo, 200
University Avenue West, Waterloo, Ontario N2L 3G1, Canada}

\author{C.R. H.~McRae}
\affiliation{Institute for Quantum Computing, University of Waterloo, 200
University Avenue West, Waterloo, Ontario N2L 3G1, Canada}
\affiliation{Department of Physics and Astronomy, University of Waterloo, 200
University Avenue West, Waterloo, Ontario N2L 3G1, Canada}

\author{D.~Shiri}
\thanks{Department of Physics, Chalmers University of Technology, SE-412 96
G\"{o}teborg, Sweden.}
\affiliation{Institute for Quantum Computing, University of Waterloo, 200
University Avenue West, Waterloo, Ontario N2L 3G1, Canada}
\affiliation{Department of Physics and Astronomy, University of Waterloo, 200
University Avenue West, Waterloo, Ontario N2L 3G1, Canada}

\author{J. D.~Bateman}
\thanks{Present address: The Edward S. Rogers Sr. Department of Electrical and
Computer Engineering, University of Toronto, 10 King's College Road, Toronto,
Ontario M5S 3G4, Canada.}
\affiliation{Institute for Quantum Computing, University of Waterloo, 200
University Avenue West, Waterloo, Ontario N2L 3G1, Canada}
\affiliation{Department of Physics and Astronomy, University of Waterloo, 200
University Avenue West, Waterloo, Ontario N2L 3G1, Canada}

\author{Y.~Rohanizadegan}
\affiliation{Institute for Quantum Computing, University of Waterloo, 200
University Avenue West, Waterloo, Ontario N2L 3G1, Canada}
\affiliation{Department of Physics and Astronomy, University of Waterloo, 200
University Avenue West, Waterloo, Ontario N2L 3G1, Canada}

\author{B.~Penava}
\affiliation{INGUN Pr\"{u}fmittelbau GmbH, Max-Stromeyer-Stra\ss{}e 162, D-78467
Konstanz, Germany}

\author{P.~Breul}
\affiliation{INGUN Pr\"{u}fmittelbau GmbH, Max-Stromeyer-Stra\ss{}e 162, 
D-78467 Konstanz, Germany}

\author{S.~Royak}
\affiliation{INGUN Pr\"{u}fmittelbau GmbH, Max-Stromeyer-Stra\ss{}e 162, D-78467
Konstanz, Germany}

\author{M.~Zapatka}
\affiliation{INGUN USA, Inc., 252 Latitude Lane, Suite 102, Lake Wylie, South
Carolina 29710-8152, USA}

\author{A. G.~Fowler}
\affiliation{Google, Inc., Santa Barbara, California 93117, USA}

\author{M.~Mariantoni}
\email[Corresponding author: ]{matteo.mariantoni@uwaterloo.ca}
\affiliation{Institute for Quantum Computing, University of Waterloo, 200 
University Avenue West, Waterloo, Ontario N2L 3G1, Canada}
\affiliation{Department of Physics and Astronomy, University of Waterloo, 200 
University Avenue West, Waterloo, Ontario N2L 3G1, Canada}

\date{\today}

\begin{abstract}
Quantum computing architectures are on the verge of scalability, a key
requirement for the implementation of a universal quantum computer. The next
stage in this quest is the realization of quantum error correction codes, which
will mitigate the impact of faulty quantum information on a quantum computer.
Architectures with ten or more quantum bits~(qubits) have been realized using
trapped ions and superconducting circuits. While these implementations are
potentially scalable, true scalability will require systems engineering to
combine quantum and classical hardware. One technology demanding imminent
efforts is the realization of a suitable wiring method for the control and
measurement of a large number of qubits. In this work, we introduce an
interconnect solution for solid-state qubits: \textit{The quantum socket}. The
quantum socket fully exploits the third dimension to connect classical
electronics to qubits with higher density and better performance than
two-dimensional methods based on wire bonding. The quantum socket is based on
spring-mounted micro wires \textendash~\textit{the three-dimensional wires}
\textendash~that push directly on a micro-fabricated chip, making electrical
contact. A small wire cross section ($\sim$~\SI{1}{\milli\meter}), nearly
non-magnetic components, and functionality at low temperatures make the quantum
socket ideal to operate solid-state qubits. The wires have a coaxial geometry
and operate over a frequency range from DC to~\SI{8}{\giga\hertz}, with a
contact resistance of~$\sim$~\SI{150}{\milli\ohm}, an impedance mismatch 
of~$\sim$~\SI{10}{\ohm}, and minimal crosstalk. As a proof of principle, we
fabricated and used a quantum socket to measure superconducting resonators at a
temperature of~$\sim$~\SI{10}{\milli\kelvin}. Quantum error correction codes
such as the surface code will largely benefit from the quantum socket, which
will make it possible to address qubits located on a two-dimensional lattice.
The present implementation of the socket can be readily extended to accommodate
a quantum processor with a~$10 \times 10$ qubit lattice, which would allow the
realization of a simple quantum memory.
\end{abstract}

\pacs{03.67.Lx, 06.60.Ei, 85.40.Ls, 85.25.-j}
\keywords{Quantum Computing; Scalable Qubit Architectures; Quantum Socket;
Three-Dimensional Wires; Microwave Interconnects; Superconducting Quantum
Circuits; Quantum Error Correction}

\maketitle

\section{INTRODUCTION}
	\label{INTRODUCTION}

The work to be presented in this article lies at the boundary between physics
and electrical engineering and aims at the experimental implementation of
practical hardware technology for quantum computation~\cite{Gambetta:2015}. At
this point in time, one of the main objectives in the quantum computing
community is to build and prototype tools for scalable architectures that may
lead to the realization of a universal quantum computer~\cite{Deutsch:1985}. In
this project, we undertake the task of implementing an extensible wiring method
for the operation of a quantum processor based on solid-state devices, e.g.,
superconducting qubits~\cite{Clarke:2008, Devoret:2013, Martinis:2015}. Possible
experimental solutions based on wafer bonding techniques~\cite{Miller:2012,
Abraham:2014:a, Brecht:2015, Rosenberg:2016} or coaxial through-silicon
vias~\cite{Bruno:2016} as well as theoretical proposals~\cite{Gambetta:2015,
Brecht:2016} have recently addressed the wiring issue, highlighting it as a
priority for quantum computing.

Building a universal quantum computer~\cite{DiVincenzo:1995, Nielsen:2000,
DiVincenzo:2000, Mermin:2007, Ladd:2010, PerezDelgado:2011} will make it
possible to execute quantum algorithms~\cite{Montanaro:2016}, which would have
profound implications on scientific research and society. For a quantum computer
to be competitive with the most advanced classical computer, it is widely
believed that the qubit operations will require error rates on the order
of~$10^{-15}$ or less. Achieving such error rates is only possible by means of
quantum error correction (QEC) algorithms~\cite{Nielsen:2000, Mermin:2007,
Gottesman:2010}, which allow for the implementation of fault tolerant operations
between logical qubits. A logical qubit is realized as an ensemble of a large
number of physical qubits (on the order~$10^3$ or larger), where each physical
qubit behaves as an effective quantum-mechanical two-level system.

Among various QEC algorithms, the most practical at present is the surface code
algorithm~\cite{Raussendorf:2007, Fowler:2012}. The surface code requires only
nearest-neighbor interactions between physical qubits on a two-dimensional
lattice, one- and two-qubit physical gates, and qubit measurement with error
rates below approximately~$10^{-2}$~\cite{Fowler:2012}. Both fast gates and
measurements are indispensable to run quantum algorithms efficiently; execution
times on the order of tens of nanoseconds for physical operations (e.g., a gate
or measurement operation) are highly desirable and represent the
state-of-the-art in qubit technology today. Under these conditions, to factorize
a~$2000$-bit number using Shor's algorithm~\cite{Shor:1994} requires more
than~$220$ million physical qubits (i.e., approximately~$4000$ logical qubits
storing meaningful data; note that these logical qubits occupy less
than~\SI{10}{\percent} of the quantum computer, the remainder of which is used
to generate special states facilitating computation), with an overall
computation time of~$\sim$~\SI{1}{\day}~\cite{Fowler:2012}.

Quantum computing architectures can be implemented using photons~\cite{Kok:2007,
OBrien:2009}, trapped ions~\cite{Monroe:2013}, spins in
molecules~\cite{Cory:2006} and quantum dots~\cite{Cody:2012, Hanson:2007,
Maune:2012, Zwanenburg:2013, Higginbotham:2014}, spins in
silicon~\cite{Zwanenburg:2013, OGorman:2016}, and superconducting quantum
circuits~\cite{Clarke:2008, Devoret:2013, Martinis:2015}. The last are leading
the way for the realization of the first surface code logical qubit, which is
one of the priorities in the quantum computing community at present. Recently,
several experiments based on superconducting quantum circuits have demonstrated
the principles underlying the surface code. These works have shown a complete
set of physical gates with fidelities beyond the surface code
threshold~\cite{Barends:2014}, the parity measurements necessary to detect
quantum errors~\cite{Corcoles:2015, Riste:2015}, and have realized a classical
version of the surface code on a one-dimensional array of nine physical
qubits~\cite{Kelly:2015}. Notably, the planar design inherent to the
superconducting qubit platform will make it possible to implement large
two-dimensional qubit arrays, as required by the surface code.

Despite all these accomplishments, a truly scalable qubit architecture has yet
to be demonstrated. Wiring is one of the most basic unsolved scalability issues
common to most solid-state qubit implementations, where qubit arrays are
fabricated on a chip. The conventional wiring method based on wire bonding
suffers from fundamental scaling limitations as well as mechanical and
electrical restrictions. Wire bonding relies on bonding pads located at the
edges of the chip. Given a two-dimensional lattice of~$N \times N$ physical
qubits on a square chip, the number of wire bonds that can be placed scales
approximately as~$4 N$ ($N$ bonds for each chip side). Wire bonding will thus
never be able to reach the required~$N^2$ law according to which physical qubits
scale on a two-dimensional lattice. Furthermore, for large~$N$, wire bonding
precludes the possibility of accessing physical qubits in the center region of
the chip, which is unacceptable for a physical implementation of the surface
code. In the case of superconducting qubits, for example, qubit control and
measurement are typically realized by means of microwave pulses or, in general,
pulses requiring large frequency bandwidths. By their nature, these pulses
cannot be reliably transmitted through long sections of quasi-filiform wire
bonds. In fact, stray capacitances and inductances associated with wire bonds as
well as the self-inductance of the bond itself limit the available frequency
bandwidth, thus compromising the integrity of the control and measurement
signals~\cite{Mutus:2014}. Additionally, the placement of wire bonds is prone to
errors and inconsistency in spacing~\footnote{Even though state-of-the-art
bonding machines can significantly mitigate these issues.}.

In this work, we set out to solve the wiring bottleneck common to almost all
solid-state qubit implementations. Our solution is based on suitably packaged
\textit{three-dimensional micro wires} that can reach any area on a given chip
from above. We define this wiring system as the \textit{quantum socket}. The
wires are coaxial structures consisting of a spring-loaded inner and outer
conductor with diameters of~\SI{380}{\micro\meter} and \SI{1290}{\micro\meter},
respectively, at the smallest point and with a maximum outer diameter
of~\SI{2.5}{\milli\meter}. The movable section of the wire is characterized by a
maximum stroke of approximately~\SI{2.5}{\milli\meter}, allowing for a wide
range of on-chip mechanical compression. All wire components are non magnetic,
thereby minimizing any interference with the qubits. The three-dimensional wires
work both at room temperature and at cryogenic temperatures as low
as~$\sim$\SI{10}{\milli\kelvin}. The wires' test-retest reliability is
excellent, with marginal variability over hundreds of measurements. Their
electrical performance is good from DC to at least~\SI{8}{\giga\hertz}, with a
contact resistance smaller than~\SI{150}{\milli\ohm} and an instantaneous
impedance mismatch of approximately~\SI{10}{\ohm}. Notably, the coaxial design
of the wires strongly reduces unwanted crosstalk, which we measured to be at
most~\SI{-45}{\deci\bel} for a realistic quantum computing application.

In a recent work~\cite{Devoret:2013}, seven sequential stages necessary to the
development of a quantum computer were introduced. At this time, the next stage
to be reached is the implementation of a single logical qubit characterized by
an error rate that is at least one order of magnitude lower than that of the
underlying physical qubits. In order to achieve this task, a two-dimensional
lattice of~$10 \times 10$ physical qubits with an error rate of at
most~$10^{-3}$ is required~\cite{Fowler:2012}. In the case of superconducting
qubits such a lattice can be realized on a chip area of~$\SI{72}{\milli\meter}
\times \SI{72}{\milli\meter}$ (the largest square that can be diced from a
standard $4$~inch wafer) and wired by means of a quantum socket. It is feasible
to further miniaturize the three-dimensional wires so as to achieve a wire
density of~\SI{2.5e5}{\per\meter\squared}. This would allow the manipulation
of~$\sim 10^5$ physical qubits and, possibly, the realization of simple fault
tolerant operations~\cite{Fowler:2012}. Furthermore, the wires could serve as
interconnect between a quantum hardware layer fabricated on one chip and a
classical hardware layer realized on a separate chip, just above the quantum
layer. The classical hardware would be used to manipulate the qubits and could
be implemented by means of rapid single-flux-quantum~(RSFQ) digital
circuitry~\cite{Brock:2000, Mukhanov:2011}.

The implications of our cryogenic micro-wiring method go beyond quantum
computing applications, providing a useful addition to the packaging industry
for research applications at low temperatures. With this work, we demonstrate
that the laborious and error-prone wire bonding technique can be substituted by
the simple procedure of inserting the chip into a sample box equipped with
three-dimensional wires. Our concept will expedite sample packaging and, thus,
experiment turn over even for research not directly related to quantum
information.

This research article is organized as follows. In
Sec.~\ref{THE:QUANTUM:SOCKET:DESIGN}, we introduce the quantum socket design,
with special focus on the three-dimensional wires
(cf.~Subsec.~\ref{Three:dimensional:wires}), the microwave package
(cf.~Subsec.~\ref{Microwave:package}), and the package holder
(cf.~Subsec.~\ref{Package:holder}). In the same section, we show microwave
simulations of a bare three-dimensional wire, of a wire connected to a pad on a
chip, and of an entire microwave package
(cf.~Subsec.~\ref{Microwave:simulations}). In
Sec.~\ref{THE:QUANTUM:SOCKET:IMPLEMENTATION}, we show the physical
implementation of all components described in
Sec.~\ref{THE:QUANTUM:SOCKET:DESIGN}. In this section, we describe the materials
used in the quantum socket and show how the socket is assembled. In particular,
we describe the magnetic and thermal properties of the quantum socket components
(cf.~Subsec.~\ref{Magnetic:properties} and Subsec.~\ref{Thermal:properties},
respectively) as well as the spring characterization
(cf.~Subsec.~\ref{Spring:characterization}). Moreover, we discuss in detail the
quantum socket alignment procedure (cf.~Subsec.~\ref{Alignment}). In
Sec.~\ref{CHARACTERIZATION}, we present a variety of measurements used to
characterize the quantum socket operation. We first focus on a four-port
measurement used to estimate a wire contact resistance
(cf.~Subsec.~\ref{Four:point:measurements}). We then show a series of microwave
measurements on samples with different geometries and materials, both at room
temperature and at~\SI{77}{\kelvin}. These measurements comprise two-port
scattering parameter (S-parameter) experiments
(cf.~Subsec.~\ref{Two:port:scattering:parameters}), time-domain
reflectometry~(TDR) analysis (cf.~Subsec.~\ref{Time:domain:reflectometry}), and
signal crosstalk tests (cf.~Subsec.~\ref{Signal:crosstalk}). In
Sec.~\ref{APPLICATIONS:TO:SUPERCONDUCTING:RESONATORS}, we show an application of
the quantum socket relevant to superconducting quantum computing, where the
socket is used to measure aluminum~(Al) superconducting resonators at a
temperature of approximately~\SI{10}{\milli\kelvin}. Finally, in
Sec.~\ref{CONCLUSIONS}, we envision an extensible quantum computing architecture
where a quantum socket is used to connect to a~$10 \times 10$ lattice of
superconducting qubits and comment on the possibility to use the socket in
conjunction with RSFQ electronics.

\section{THE QUANTUM SOCKET DESIGN}
	\label{THE:QUANTUM:SOCKET:DESIGN}

The development of the quantum socket required a stage of meticulous
micro-mechanical and microwave design and simulations. It was determined that a
spring-loaded interconnect \textendash~the three-dimensional wire
\textendash~was the optimal method to electrically access devices
lithographically fabricated on a chip and operated in a cryogenic environment.
An on-chip contact pad geometrically and electrically matched to the bottom
interface of the wire can be placed easily at any desired location on the chip
as part of the fabrication process, thus making it possible to reach any point
on a two-dimensional lattice of qubits. The contact pad is realized as a thin
metallic film deposited on a dielectric substrate; arbitrary film thicknesses
can be deposited. The rest of the sample is fabricated on chip following a
similar process. The coaxial design of the wire provides a wide operating
frequency bandwidth, while the springs allow for mechanical stress relief during
the cooling process. The three-dimensional wires used in this work take
advantage of the knowledge in the existing field of microwave circuit
testing~\cite{Zapatka:2009}. However, reducing the wire dimensions to a few
hundred micrometers and using it to connect to quantum-mechanical
micro-fabricated circuits at low temperatures resulted in a significant
extension of existing implementations and applications.

In this section, we will describe the design of the three-dimensional wires, of
the microwave package used to house the wires, and of the microwave package
holder. Additionally, we will show a set of microwave simulations of the main
components of the quantum socket.

\subsection{Three-dimensional wires}
	\label{Three:dimensional:wires}

\begin{figure*}
	\centering
	\includegraphics[width=0.99\textwidth]{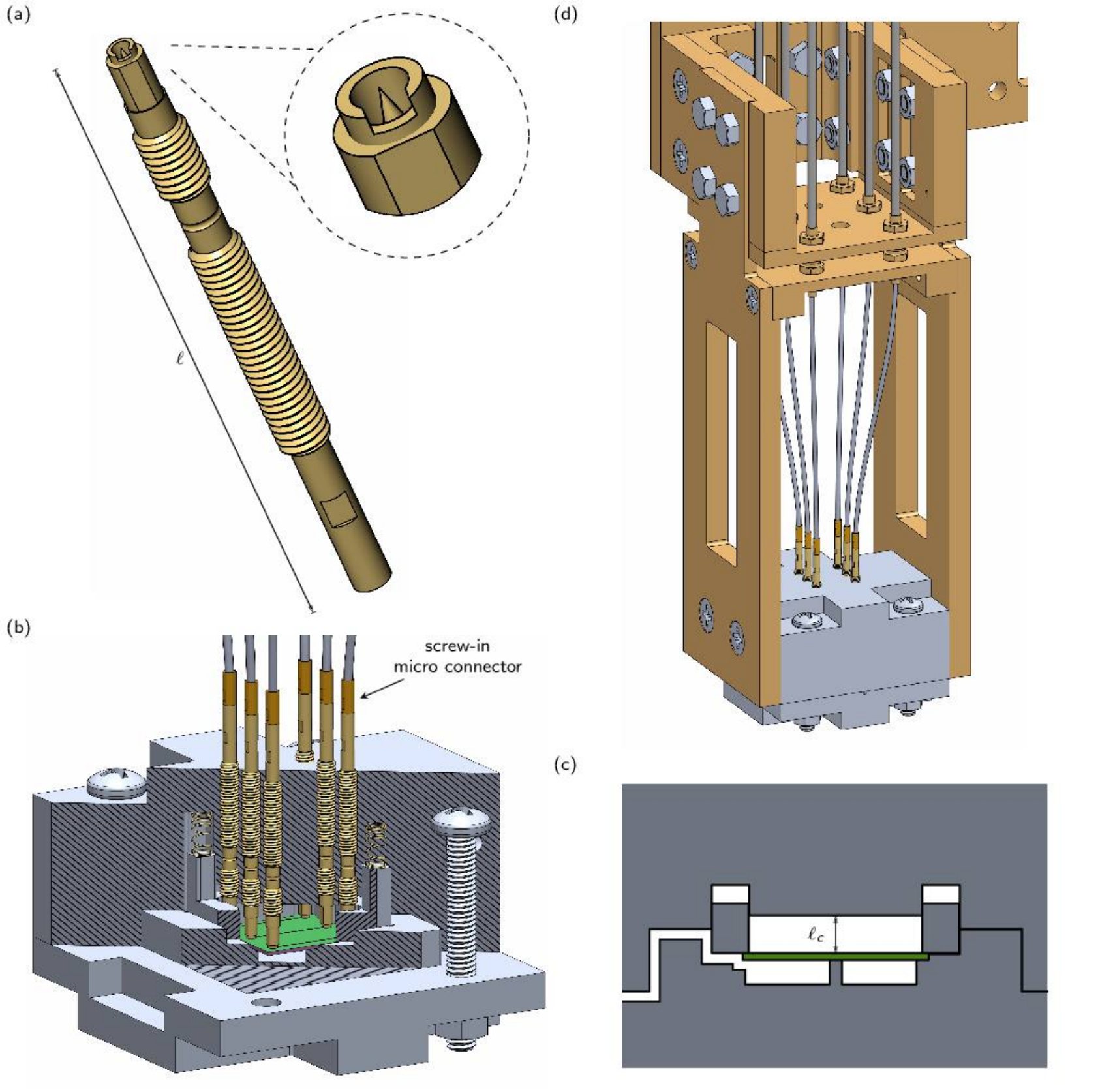}
\caption{Computer-aided designs of the three-dimensional wire, microwave
package, and package holder. (a) A wire of length~$\ell =
\SI{30.5}{\milli\meter}$ along with a detail of the contact head (inset). (b)
Assembled microwave package including six three-dimensional wires, washer,
washer springs, and chip (shown in green). The arrow indicates the screw-in
micro connector on the back end of the wire. Forward hatching indicates the
washer cutaway, whereas backward hatching indicates both lid and sample holder
cutaways. (c) Cross section of the microwave package showing the height of the
upper cavity, which coincides with the minimum compression
distance~$\ell_{\textrm{c}}$ of the three-dimensional wires
(cf.~Appendix~\ref{WIRE:COMPRESSION}). (d) Microwave package mounted to the
package holder, connected, in turn, to the mounting plate of a DR with SMP
connectors.}
	\label{Figure01:Bejanin}
\end{figure*}

\begin{figure*}[t]
	\centering
	\includegraphics[width=0.99\textwidth]{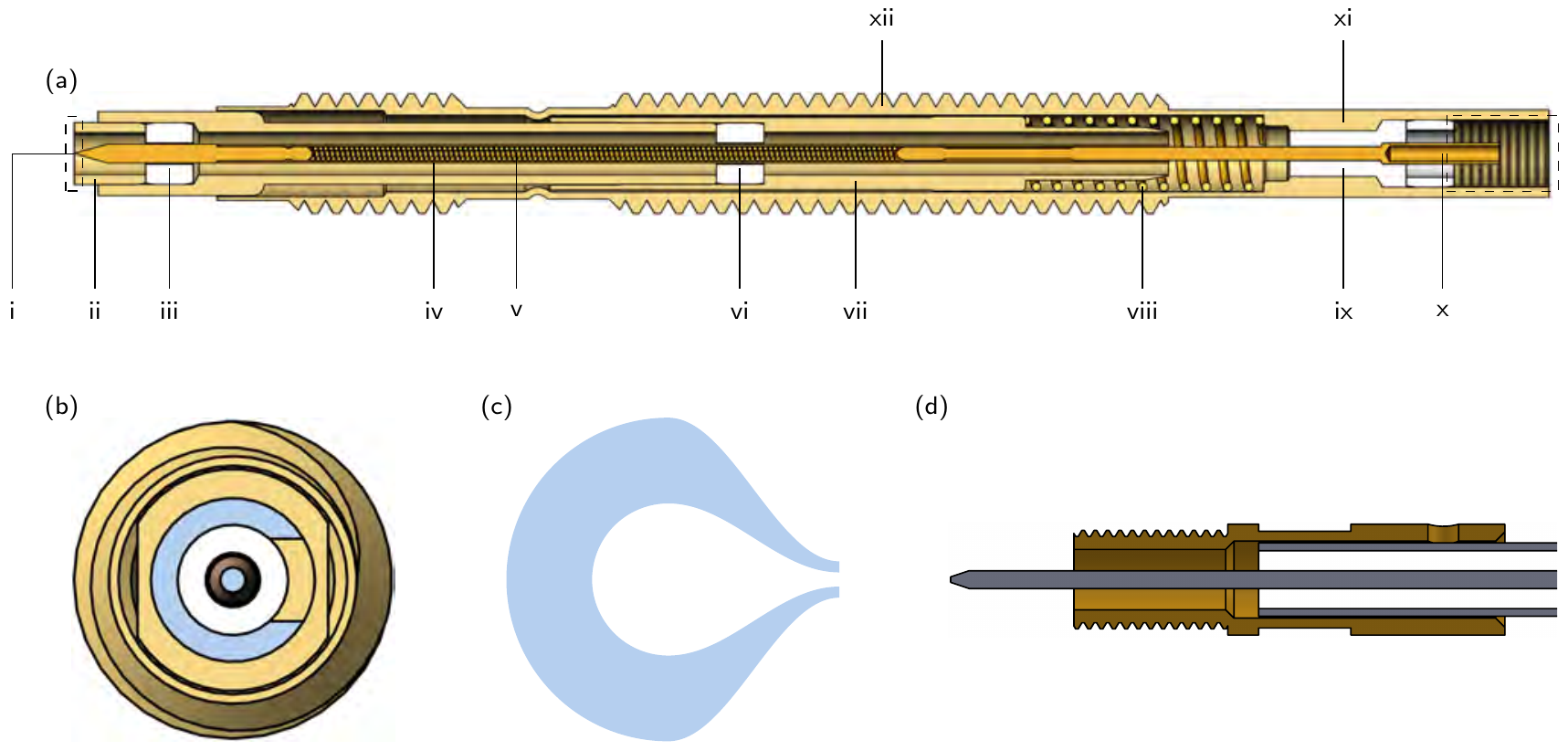}
\caption{Two-dimensional cut view of the three-dimensional wire, contact pad,
and screw-in micro connector. (a) Side view of the wire cross section. The wire
components are: i, spring-loaded center conductor of the contact head; ii,
spring-loaded outer conductor of the contact head; iii, vi, and ix, dielectric
spacers; iv, center conductor barrel; v, center conductor spring; vii, outer
conductor barrel; viii, outer conductor spring; x, center conductor tail; xi,
outer conductor tail; xii, threaded outer body. The dashed box on the left
indicates the contact head; that on the right indicates the female threads
included for use with the screw-in micro connector. (b) Front view of the wire.
The blue surface indicates the wire bottom interface. (c) On-chip contact pad.
Here, the blue surface indicates the pad dielectric gap, whereas the white
surfaces refer to conductors. (d) Screw-in micro connector. The left end of the
micro connector mates with the back end of the three-dimensional wire; the right
end is soldered to a coaxial cable, the inner conductor of which serves as the
inner conductor of the micro connector (slotting into x).}
	\label{Figure02:Bejanin}
\end{figure*}

Figure~\ref{Figure01:Bejanin} shows the design of the quantum socket components.
Figure~\ref{Figure01:Bejanin}~(a) displays a model of a three-dimensional wire.
The coaxial design of the wire is visible from the image, which features a
wire~\SI{30.5}{\milli\meter} long when uncompressed. The wire is characterized
by an inner cylindrical pin of diameter~\SI{380}{\micro\meter} and an outer
cylindrical body (the electrical ground) of diameter~\SI{1290}{\micro\meter} at
its narrowest region; this region is the bottommost section of the wire and,
hereafter, will be referred to as the wire \textit{contact head} (cf.~the inset
of Fig.~\ref{Figure01:Bejanin}~(a), as well as the dashed box on the left of
Fig.~\ref{Figure02:Bejanin}~(a)). The contact head terminates at the wire
\textit{bottom interface}; this interface is designed to mate with a pad on a
chip (cf.~Fig.~\ref{Figure02:Bejanin}~(b) and (c)). The outer body includes a
rectangular aperture, the \textit{tunnel}, to prevent shorting the inner
conductor of an on-chip coplanar waveguide~(CPW) transmission
line~\cite{Collin:2001, Pozar:2011}; the transmission line can connect the pad
with any other structure on the chip. Two different tunnel dimensions were
designed, with the largest one reducing potential alignment errors. These errors
can result in undesired short-circuit connections to ground. The tunnel height
was~\SI{300}{\micro\meter} in both cases, with a width of~\SI{500}{\micro\meter}
or \SI{650}{\micro\meter}. The internal spring mechanisms of the wire allow the
contact head to be compressed; the maximum stroke was designed to
be~\SI{2.5}{\milli\meter}, corresponding to a working stroke
of~\SI{2.0}{\milli\meter}.

The outer body of the three-dimensional wire is an~M$2.5$ male thread used to
fix the wire to the lid of the microwave package
(cf.~Fig.~\ref{Figure01:Bejanin}~(b) and (d)). The thread is split into two
segments of length~\SI{3.75}{\milli\meter} and \SI{11.75}{\milli\meter} that are
separated by a constriction with outer diameter~\SI{1.90}{\milli\meter}. The
constriction is necessary to assemble and maintain in place the inner components
of the three-dimensional wire. A laser-printed marker is engraved into the top
of the outer body. The marker is aligned with the center of the tunnel, making
it possible to mate the wire bottom interface with a pad on the underlying chip
with a high degree of angular precision. The grooves partially visible on the
bottom of Fig.~\ref{Figure01:Bejanin}~(a) are used to torque the wire into
a~\SI{2.5}{\milli\meter} thread in the package's lid.

Figure~\ref{Figure02:Bejanin}~(a) shows a lateral two-dimensional cut view of
the three-dimensional wire. Two of the main wire components are the inner and
outer barrel, which compose part of the inner and outer conductor. The inner
conductor barrel is a hollow cylinder with outer and inner diameters
of~\SI{380}{\micro\meter} and \SI{290}{\micro\meter} (indicated as part~iv in
Fig.~\ref{Figure02:Bejanin}~(a)), respectively. This barrel encapsulates the
inner conductor spring. The outer conductor barrel is a hollow cylinder as well,
in this case with an inner diameter of~\SI{870}{\micro\meter} (parts~ii and
vii). Three polytetrafluoroethylene (PTFE) disks serve as spacers between the
inner and outer conductor; such disks contribute marginally to the wire
dielectric volume, the majority of which is air or vacuum. The outer spring is
housed within the outer barrel towards its back end, just before the last PTFE
disk on the right-hand side of the wire. The \textit{back end} of the wire is a
region comprising a female thread on the outer conductor and an inner conductor
barrel (cf.~dashed box on the right-hand side of
Fig.~\ref{Figure02:Bejanin}~(a)).

The inner conductor tip is characterized by a conical geometry with an opening
angle of~$\SI{30}{\degree}$. Such a sharp design was chosen to ensure that the
tip would pierce through any possible oxide layer forming on the contact pad
metallic surface, thus allowing for a good electrical contact.

Figure~\ref{Figure02:Bejanin}~(c) shows the design of a typical on-chip pad used
to make contact with the bottom interface of a three-dimensional wire. The pad
comprises an inner and outer conductor, with the outer conductor being grounded.
The pad in the figure was designed for a silver~(Ag) film of
thickness~\SI{3}{\micro\meter}. A variety of similar pads were designed for
gold~(Au) and Al films with thickness ranging between
approximately~\SI{100}{\nano\meter} and \SI{200}{\nano\meter}. The pad inner
conductor is a circle with diameter~\SI{320}{\micro\meter} that narrows to a
linear trace (i.e., the inner conductor of a CPW transmission line) by means of
a raised-cosine taper. The raised cosine makes it possible to maximize the pad
area, while minimizing impedance mismatch. As designed, the wire and pad allow
for lateral and rotational misalignment of~$\mp\SI{140}{\micro\meter}$ and
$\mp\SI{28}{\degree}$, respectively. The substrate underneath the pad is assumed
to be silicon~(Si) with a relative permittivity~$\epsilon_\textrm{r} \simeq 11$.
The dielectric gap between the inner and outer conductor
is~\SI{180}{\micro\meter} in the circular region of the pad; the outer edge of
the dielectric gap then follows a similar raised-cosine taper as the inner
conductor. The pad characteristic impedance is designed to be~$Z_\textrm{c} =
\SI{50}{\ohm}$.

\subsection{Microwave package}
	\label{Microwave:package}

The microwave package comprises three main parts: The lid; the sample holder;
the grounding washer. The package is a parallelepiped with a height
of~\SI{30}{\milli\meter} and with a square base of side
length~\SI{50}{\milli\meter}. The chip is housed inside the sample holder.  All
these components mate as shown in Fig.~\ref{Figure01:Bejanin}~(b) and (c).

In order to connect a three-dimensional wire to a device on a chip, the wire is
screwed into an~M$2.5$ female thread that is tapped into the lid of the
microwave package, as depicted in Fig.~\ref{Figure01:Bejanin}~(b). The pressure
applied by the wire to the chip is set by the depth of the wire in the package.
The wire stroke, thread pitch, and alignment constraints impose discrete
pressure settings (cf.~Appendix~\ref{WIRE:COMPRESSION}). In the present
implementation of the quantum socket, the lid is designed to hold a set of six
three-dimensional wires, which are arranged in two parallel rows. In each row,
the wires are spaced by~\SI{5.75}{\milli\meter} from center to center, with the
two rows being separated by a distance of~\SI{11.5}{\milli\meter}.

A square chip of lateral dimensions~$\SI{15}{\milli\meter} \times
\SI{15}{\milli\meter}$ is mounted in the sample holder in a similar fashion as
in Ref.~\cite{Wenner:2011:a}. The outer edges of the chip rest on four
protruding lips, which are~\SI{1}{\milli\meter} wide. Hereafter, those lips will
be referred to as the \textit{chip recess}. For design purposes, a chip
thickness of~\SI{550}{\micro\meter} is assumed. Correspondingly, the chip recess
is designed so that the top of the chip is~\SI{100}{\micro\meter} above the
adjacent surface of the chip holder, i.e., the depth of the recess
is~\SI{450}{\micro\meter} (cf.~Fig.~\ref{Figure01:Bejanin}~(c)). The outer edges
of the chip are pushed on by a spring-loaded grounding washer.
The~\SI{100}{\micro\meter} chip protrusion ensures a good electrical connection
between chip and washer, as shown in Fig.~\ref{Figure01:Bejanin}~(c).

The grounding washer was designed to substitute the large number of lateral
bonding wires that would otherwise be required to provide a good ground to the
chip (as shown, for example, in Fig.~6 of Ref.~\cite{Wenner:2011:a}). The washer
springs are visible in Fig.~\ref{Figure01:Bejanin}~(b), which also shows a cut
view of the washer. The washer itself is electrically grounded by means of the
springs as well as through galvanic connection to the surface of the lid. The
four feet of the washer, which can be seen in the cut view of
Fig.~\ref{Figure01:Bejanin}~(b), can be designed to be shorter or longer. This
makes it possible to choose different pressure settings for the washer.

After assembling the package, there exist two electrical cavities
(cf.~Fig.~\ref{Figure01:Bejanin}~(d)): One above the chip, formed by the lid,
washer, and metallic surface of the sample (\textit{upper cavity}), and one
below the chip, formed by the sample holder and metallic surface of the sample
(\textit{lower cavity}). The hollow cavity above the sample surface has
dimensions~$\SI{14}{\milli\meter} \times \SI{14}{\milli\meter} \times
\SI{3.05}{\milli\meter}$. The dimensions of the cavity below the sample
are~$\SI{13}{\milli\meter} \times \SI{13}{\milli\meter} \times
\SI{2}{\milli\meter}$. The lower cavity helps mitigate any parasitic capacitance
between the chip and the box (ground). Additionally, it serves to lower the
effective permittivity, increasing the frequency of the substrate modes
(cf.~Subsec.~\ref{Microwave:simulations}).

A pillar of square cross section with side length of~\SI{1}{\milli\meter} is
placed right below the chip at its center; the pillar touches the bottom of the
chip, thus providing mechanical support~\footnote{The pillar was included in the
initial design as there was concern over potential damage to the substrate from
mechanical strain due to the three-dimensional wires pushing on the top of the
chip.}. The impact of such a pillar on the microwave performance of the package
will be described in Subsec.~\ref{Microwave:simulations}. A channel with a
cross-sectional area of~$\SI{800}{\micro\meter} \times \SI{800}{\micro\meter}$
connects the inner cavities of the package to the outside, thus making it
possible to evacuate the inner compartments of the package. This channel
meanders to prevent external electromagnetic radiation from interfering with the
sample.

\begin{figure*}[t!]
	\centering
	\includegraphics[width=0.99\textwidth]{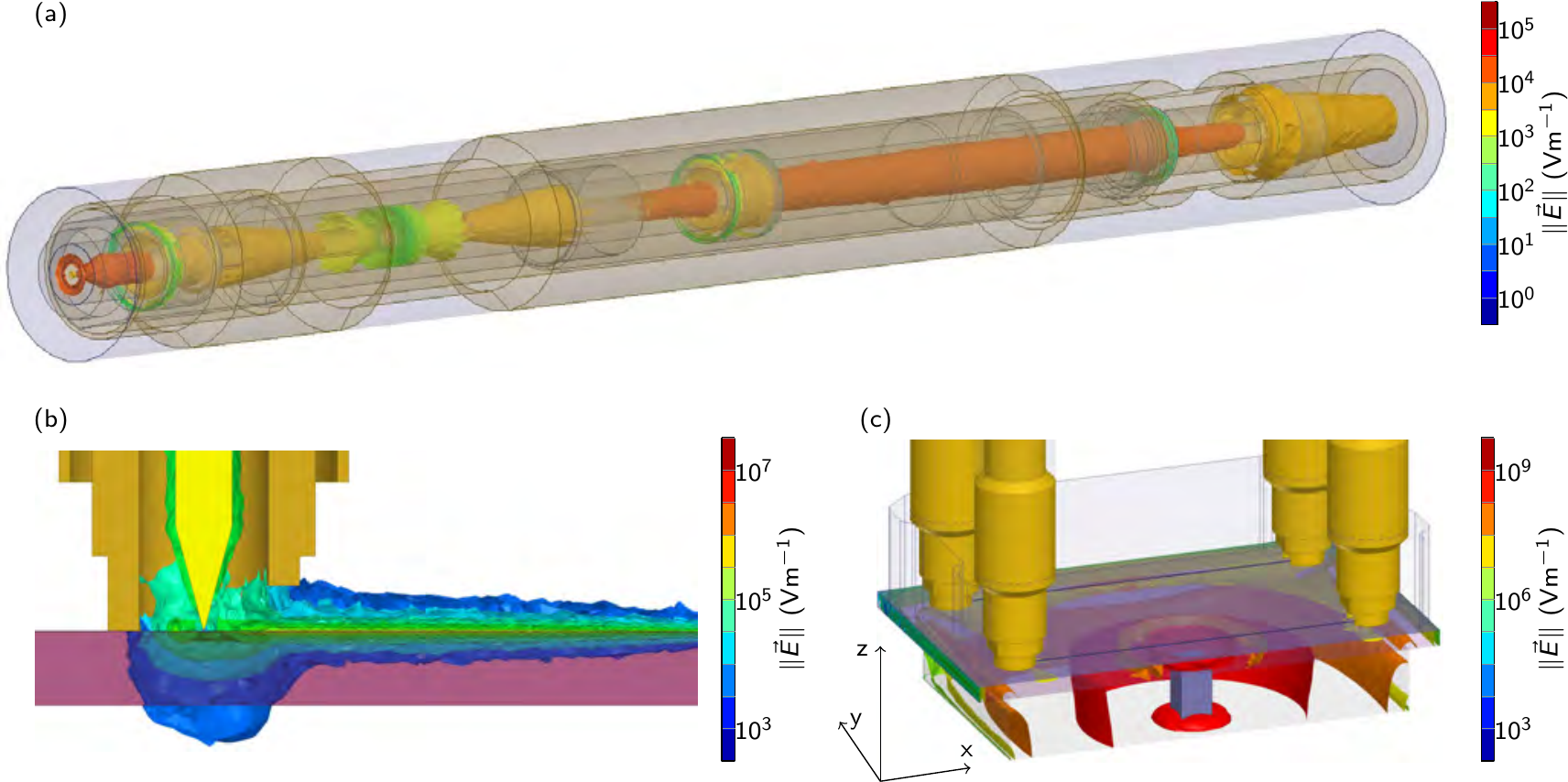}
\caption{Numerical simulations of the electric field distribution. (a) Field for
a three-dimensional wire at~\SI{6}{\giga\hertz}. (b) Field in proximity of
the~\SI{90}{\degree} transition region at~\SI{6}{\giga\hertz}. (c) Field for the
first box mode at~\SI{6.3}{\giga\hertz}. Color bar scales are indicated in their
respective panels. The~$x$, $y$, and $z$ directions of a Cartesian coordinate
system are also indicated. In~(b), the cross section of the transition region is
shown. Note the large volume occupied by the electric field beneath the contact
pad. In~(c), an offset cross section of the first box mode is shown. The field
confinement due to the pillar is clearly visible. Additionally, the simulation
shows a slight field confinement in the region surrounding the chip recess.
Time-domain animations of the simulated electric field can be found in the
Supplemental Material at~\url{http://www.Supplemental-Material-Bejanin} .}
	\label{Figure03:Bejanin}
\end{figure*}

\subsection{Package holder}
	\label{Package:holder}

The three-dimensional wires, which are screwed into the microwave package, must
be connected to the qubit control and measurement electronics. In addition, for
cryogenic applications, the package must be thermally anchored to a
refrigeration system in order to be cooled to the desired temperature.
Figure~\ref{Figure01:Bejanin}~(d) shows the mechanical module we designed to
perform both electrical and thermal connections. In this design, each
three-dimensional wire is connected to a \textit{screw-in micro connector},
which is indicated by an arrow in Fig.~\ref{Figure01:Bejanin}~(b) and is shown
in detail in Fig.~\ref{Figure02:Bejanin}~(d). One end of the micro connector
comprises a male thread and an inner conductor pin that mate with the back end
of the three-dimensional wire. The other end of the micro connector is soldered
to a coaxial cable. The micro connector is necessary because the high
temperatures generated by soldering a coaxial cable directly to the wire back
end would damage some of the inner wire components~\footnote{The breaking
temperature of those components is lower than the melting temperature of
available eutectic solders.}.

The end of each coaxial cable opposite to the three-dimensional wire is soldered
to a sub-miniature push-on~(SMP) connector. The SMP connectors are bolted to a
horizontal plate attached to the microwave package by means of two vertical
fixtures, as shown in Fig.~\ref{Figure01:Bejanin}~(d). The vertical fixtures and
the horizontal plate constitute the package holder. The package holder and
microwave package form an independent assembly. A horizontal mounting plate,
designed to interface with the package holder, houses a set of matching SMP
connectors. The mounting plate is mechanically and, thus, thermally anchored to
the mixing chamber~(MC) stage of a dilution refrigerator~(DR). This design
significantly simplifies the typical mounting procedure of a sample box to a
cryostat, since the package holder and microwave package can be conveniently
assembled remotely from the DR and attached to it just prior to commencing an
experiment.

\subsection{Microwave simulations}
	\label{Microwave:simulations}

The three-dimensional wires, the~\SI{90}{\degree} transition between the wire
and the on-chip pad as well as the inner cavities of the fully-assembled
microwave package were extensively simulated by means of the high frequency
three-dimensional full-wave electromagnetic field simulation software~(HFSS) by
Ansys, Inc. The results for the electromagnetic field distribution at a
frequency of approximately~\SI{6}{\giga\hertz}, which is a typical operation
frequency for superconducting qubits, are shown in Fig.~\ref{Figure03:Bejanin}.
Figure~\ref{Figure03:Bejanin}~(a) shows the field behavior for a bare
three-dimensional wire. The field distribution resembles that of a coaxial
transmission line except for noticeable perturbations at the dielectric PTFE
spacers. Figure~\ref{Figure03:Bejanin}~(b) shows the~\SI{90}{\degree} transition
region. This is a critical region for signal integrity since abrupt changes in
physical geometry cause electrical reflections~\cite{Simons:2001, Zapatka:2009}.
In order to minimize such reflections, an impedance-matched pad was designed.
However, this leads to a large electromagnetic volume in proximity of the pad,
as seen in Fig.~\ref{Figure03:Bejanin}~(b), possibly resulting in parasitic
capacitance and crosstalk.

\begin{table}[b!]
\caption{Simulation results for the first three box modes of the lower cavity
inside the assembled microwave package shown in Fig.~\ref{Figure01:Bejanin}~(b).
The dielectric used for these simulations was Si at room temperature with
relative permittivity~$\epsilon_{\textrm{r}} = 11.68$. ``Vacuum'' indicates that
no Si is present in the simulation. ``with pillar'' indicates that
the~$\SI{1.0}{\milli\meter} \times \SI{1.0}{\milli\meter} \times
\SI{2.0}{\milli\meter}$ support pillar is present. $\textrm{TE}_{xyz}$ indicates
the number of half-wavelengths spanned by the electric field in the~$x$, $y$,
and $z$ directions, respectively (cf.~Fig~\ref{Figure03:Bejanin}~(c)). Note that
the frequency of the first mode of the upper cavity
is~$\sim\SI{17.2}{\giga\hertz}$.}
\begin{center}
	\begin{ruledtabular}
		\begin{tabular}{cccc}
		\raisebox{0mm}[3mm][0mm]{} & $\textrm{TE}_{110}$ & 
		$\textrm{TE}_{120}$ & $\textrm{TE}_{210}$ \\
        \raisebox{0mm}[0mm][2mm]{} & \footnotesize{(\SI{}{\giga\hertz})} & 
        \footnotesize{(\SI{}{\giga\hertz})} & 
        \footnotesize{(\SI{}{\giga\hertz})} \\
        \hline
        \hline
        \raisebox{0mm}[3mm][0mm]{Vacuum} & $15.7$ & $24.2$ & $24.2$ \\
        \hline
        \raisebox{0mm}[3mm][0mm]{Vacuum with pillar} & $13.1$ & $23.6$ & $23.6$ 
        \\
        \hline
        \raisebox{0mm}[3mm][0mm]{Si} & $13.5$ & $16.8$ & $16.8$ \\
        \hline
        \raisebox{0mm}[3mm][0mm]{Si with pillar} & $6.3$ & $16.2$ & $16.9$ \\
        \vspace{-4.5mm}
		\end{tabular}
	\end{ruledtabular}
\end{center}
	\label{Table01:Bejanin}
\end{table}

In addition to considering the wire and the transition region, the electrical
behavior of the inner cavities of the package was studied analytically and
simulated numerically. As described in Subsec.~\ref{Microwave:package}, the
metallic surface of the chip effectively divides the cavity of the sample holder
into two regions: A vacuum cavity above the metal surface and a cavity partially
filled with dielectric below the metal surface. The latter is of greatest
concern as the dielectric acts as a perturbation to the cavity
vacuum~\footnote{Provided the vacuum still constitutes the majority of the
volume of the cavity.}, thus lowering the box modes. For a simple rectangular
cavity, the frequency~$f$ of the first mode due to this perturbation can be
found as~\cite{Pozar:2011},
\begin{equation}
f = f_0 - \frac{ f_0 \left( \epsilon_{\textrm{r}} - 1 \right) d_{\textrm{s}} }{ 
2 b } \quad ,
	\label{Equation:01}
\end{equation}
where~$f_0$ is the frequency of the unperturbed mode, $\epsilon_{\textrm{r}}$
the relative permittivity of the dielectric, $d_{\textrm{s}}$ the substrate
thickness, and $b$ the cavity height. From Eq.~(\ref{Equation:01}), we estimated
this box mode to be~\SI{12.8}{\giga\hertz}. However, considering the presence of
the pillar, the three-dimensional wires, etc., we had to use numerical
simulations to obtain a more accurate estimate of the lowest box modes. The
results for the first three modes are reported in Table~\ref{Table01:Bejanin}.
Discounting the pillar, the analytical and simulated values are in good
agreement with each other. The addition of the support pillar significantly
lowers the frequency of the modes. In fact, it increases the relative filling
factor of the cavity by confining more of the electromagnetic field to the
dielectric than to vacuum. Given the dimensions of this design, the pillar leads
to a first mode which could interfere with typical qubit frequencies. In spite
of this, the pillar was included in the final design in order provide a degree
of mechanical support. Note that the pillar can alternatively be realized as a
dielectric material, e.g., PTFE; a dielectric pillar would no longer cause field
confinement between the top surface of the pillar and the metallic surface of
the chip.

\section{THE QUANTUM SOCKET IMPLEMENTATION}
	\label{THE:QUANTUM:SOCKET:IMPLEMENTATION}

\begin{figure*}[t!]
	\centering
	\includegraphics[width=0.99\textwidth]{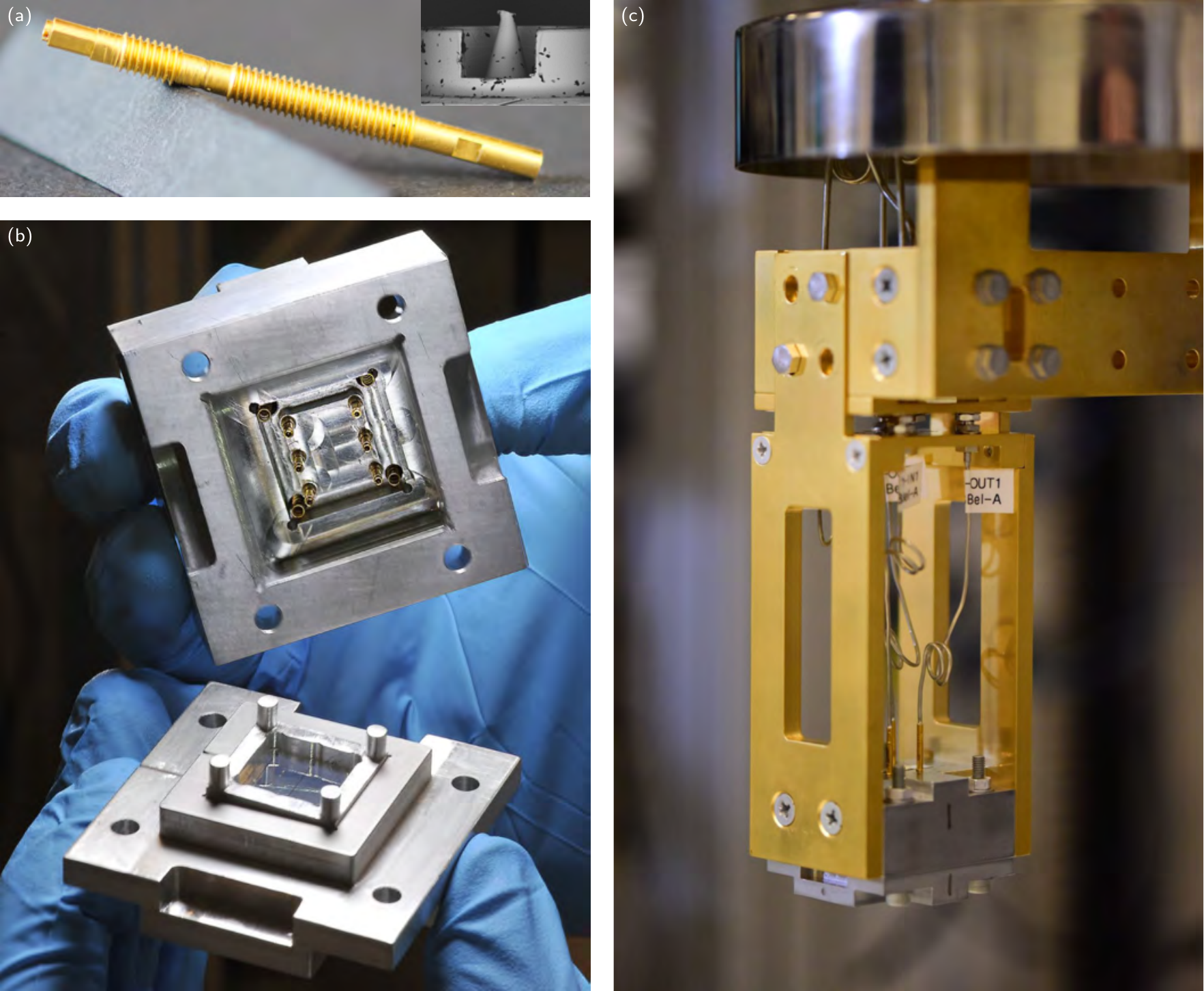}
\caption{Images of the quantum socket as implemented. (a) Macro photograph of a
three-dimensional wire; inset: SEM image of the contact head. Note that the tip
of the inner conductor retained small metallic flakes that were scraped off the
on chip pads. (b) Microwave package lid with six three-dimensional wires and
four washer springs, washer, and sample holder with chip installed. (c) Package
holder with attached microwave package mounted to the MC stage of a DR. On the
top of the panel is visible the lid of a custom-made magnetic shield from the
Amuneal Manufacturing Corporation, type Amumetal 4K~(A4K).}
	\label{Figure04:Bejanin}
\end{figure*}

The physical implementation of the main components of the quantum socket is
displayed in Fig.~\ref{Figure04:Bejanin}. In particular,
Fig.~\ref{Figure04:Bejanin}~(a) shows a macro photograph of a three-dimensional
wire. The inset shows a scanning electron microscope~(SEM) image of the wire
contact head, featuring the~\SI{500}{\micro\meter} version of the tunnel. This
wire was cycled approximately ten times; as a consequence, the center conductor
of the contact head, which had a conical, sharp shape originally, flattened at
the top. The metallic components of the wire were made from bronze and brass
(cf.~Subsec.~\ref{Magnetic:properties}), and all springs from hardened beryllium
copper~(BeCu). Except for the springs, all components were gold plated without
any nickel~(Ni) adhesion underlayer. The estimated mean number of cycles before
failure for these wires is approximately~$100000$.

Figure~\ref{Figure04:Bejanin}~(b) displays the entire microwave package in the
process of locking the package lid and sample holder together, with a chip and
grounding washer already installed. As shown in the figure, two rows of
three-dimensional wires, for a total number of six wires, are screwed into the
lid with pressure settings as described in Appendix~\ref{WIRE:COMPRESSION}; each
wire is associated with one on-chip CPW pad. The four springs that mate with the
grounding washer feet are embedded in corresponding recesses in the lid; the
springs are glued in these recesses by way of Vibra-Tite~$12110$ (from ND
Industries, Inc.), a medium-strength thread locker that works well at low
temperatures. All package components were made from high-purity Al.

Figure~\ref{Figure04:Bejanin}~(c) shows a picture of the assembled microwave
package attached to the package holder; the entire structure is attached to the
MC stage of a DR. All parts of the assembly were made from high thermal
conductivity~C10100 oxygen-free electrolytic~(OFE) copper
alloy~\footnote{Residual-resistivity ratio larger than~$300$.}. The parts were
polished to a mirror finish before being gold plated~\footnote{No Ni adhesion
underlayer and an Au bath with minimum magnetic impurities were used. The
deposited Au was hard Au, type~I, grade C (MIL-G-45204C), with a total measured
quantity of Ni, iron, and cobalt of~\SI{0.204}{\percent}.}. The coaxial cables
between the screw-in micro connectors and the SMP connectors are from the EZ
Form Cable Corporation, model EZ 47-CU-TP~(EZ~47). The SMP connectors, also from
EZ Form, are models SMP bulkhead jack for $0.047$~inch coaxial cables (SMP~047;
installed in the package holder horizontal plate) and SMP bulkhead plug with
limited detent for $0.086$~inch cables (SMP~086; installed in the mounting plate
attached to the MC stage of the DR). All SMP connectors were custom-made
non-magnetic connectors.

In the remainder of this section, we will discuss the magnetic and thermal
properties of the materials used to implement the quantum socket as well as the
spring characterization and the alignment procedure.

\subsection{Magnetic properties}
	\label{Magnetic:properties}

An important stage in the physical implementation of the quantum socket was the
choice of materials to be used for the three-dimensional wires, the microwave
package, and the package holder. In fact, it has been shown that non-magnetic
components in proximity of superconducting qubits are critical to preserve long
qubit coherence~\cite{Frunzio:2005, Song:2009:a, Song:2009:b, Megrant:2012}.

The three-dimensional wires are the closest devices to the qubits. For this
reason, all their components should be made using non-magnetic materials. Due to
machining constraints, however, alloys containing ferromagnetic impurities
(iron~(Fe), cobalt~(Co), and Ni) had to be used. For the outer conductor
components we used brass, which is easy to thread; the chosen grade was ISO
CuZn21Si3P (EN CW724R)~\footnote{Alloy~430, UNS C69300.}. For the inner
conductor components, brass CW724R did not meet the machining requirements.
Consequently, we decided to use copper alloy (phosphor bronze) grade DIN 2.1030
- CuSn8 (EN CW453K)~\footnote{UNS C52100.}. The chemical composition for these
two materials is reported in Table~\ref{Table04:Bejanin} of
Appendix~\ref{THE:QUANTUM:SOCKET:MAGNETISM}. The dielectric spacers were made
from PTFE and the rest of the components from hardened BeCu; both materials are
non-magnetic. The weight percentage of ferromagnetic materials is non-negligible
for both CW453K and CW724R. Thus, we performed a series of tests using a zero
Gauss chamber~(ZGC) in order to ensure both materials were sufficiently
non-magnetic. The results are given in
Appendix~\ref{THE:QUANTUM:SOCKET:MAGNETISM} and show that the magnetic
impurities should be small enough not to disturb the operation of
superconducting quantum devices.

The microwave package and grounding washer were made from high-purity Al alloy
5N5 (\SI{99.9995}{\percent} purity) provided by Laurand Associates, Inc. The
very low level of impurities in this alloy assures minimal stray magnetic fields
generated by the package itself, as confirmed by the magnetic tests discussed in
Appendix~\ref{THE:QUANTUM:SOCKET:MAGNETISM}.

\subsection{Thermal properties}
	\label{Thermal:properties}

The thermal conductance of the three-dimensional wires is a critical parameter
to be analyzed for the interconnection with devices at cryogenic temperatures.
Low thermal conductivity would result in poor cooling of the devices, which, in
the case of qubits, may lead to an incoherent thermal mixture of the qubit
ground state~$\ket{\textrm{g}}$ and excited
state~$\ket{\textrm{e}}$~\cite{Corcoles:2011}. Even a slightly mixed state would
significantly deteriorate the fidelity of the operations required for
QEC~\cite{Fowler:2014}. It has been estimated that some of the qubits in the
experiment of Ref.~\cite{Kelly:2015}, which relies solely on Al wire bonds as a
means of thermalization, were characterized by an excited state
population~$P_{\textrm{e}} \simeq 0.04$. Among other possible factors, it is
believed that this population was due to the poor thermal conductance of the Al
wire bonds. In fact, these bonds become superconductive at the desired qubit
operation temperature of~$\sim \SI{10}{\milli\kelvin}$, preventing the qubits
from thermalizing and, thus, from being initialized in~$\ket{\textrm{g}}$ with
high fidelity.

In order to compare the thermal performance of an Al wire bond with that of a
three-dimensional wire, we estimated the heat transfer rate per kelvin of a
wire, $\Pi_{\textrm{t}}$, using a simplified coaxial geometry. At a temperature
of~\SI{25}{\milli\kelvin}, we calculated~$\Pi_{\textrm{t}} \simeq
\SI{6e-7}{\watt\per\kelvin}$. At the same temperature, the heat transfer rate
per kelvin of a typical Al wire bound was estimated to be~$\Pi_{\textrm{b}}
\simeq \SI{4e-12}{\watt\per\kelvin}$
(cf.~Appendix~\ref{THERMAL:CONDUCTANCE:OF:A:THREE:DIMENSIONAL:WIRE} for more
details). A very large number of Al wire bonds would thus be required to obtain
a thermal performance comparable to that of a single three-dimensional wire.

\subsection{Spring characterization}
	\label{Spring:characterization}

Another critical step in the physical implementation of the quantum socket was
to select springs that work at cryogenic temperatures. In fact, the force that a
wire applies to a chip depends on these springs. This force, in turn, determines
the wire-chip contact resistance, which impacts the socket's DC and microwave
performance. Among various options, we chose custom springs made from hardened
BeCu.

To characterize the springs, their compression was assessed at room temperature,
in liquid nitrogen (i.e., at a temperature~$T \simeq \SI{77}{\kelvin}$), and in
liquid helium ($T \simeq \SI{4.2}{\kelvin}$). Note that a spring working
at~\SI{4.2}{\kelvin} is expected to perform similarly at a temperature
of~\SI{10}{\milli\kelvin}. A summary of the thermo-mechanical tests is reported
in Appendix~\ref{THERMO:MECHANICAL:TESTS}. The main conclusion of the tests is
that the springs do not break (even after numerous temperature cycles) and have
similar spring constants at all measured temperatures.

\begin{figure}[t!]
	\centering
	\includegraphics[width=0.49\textwidth]{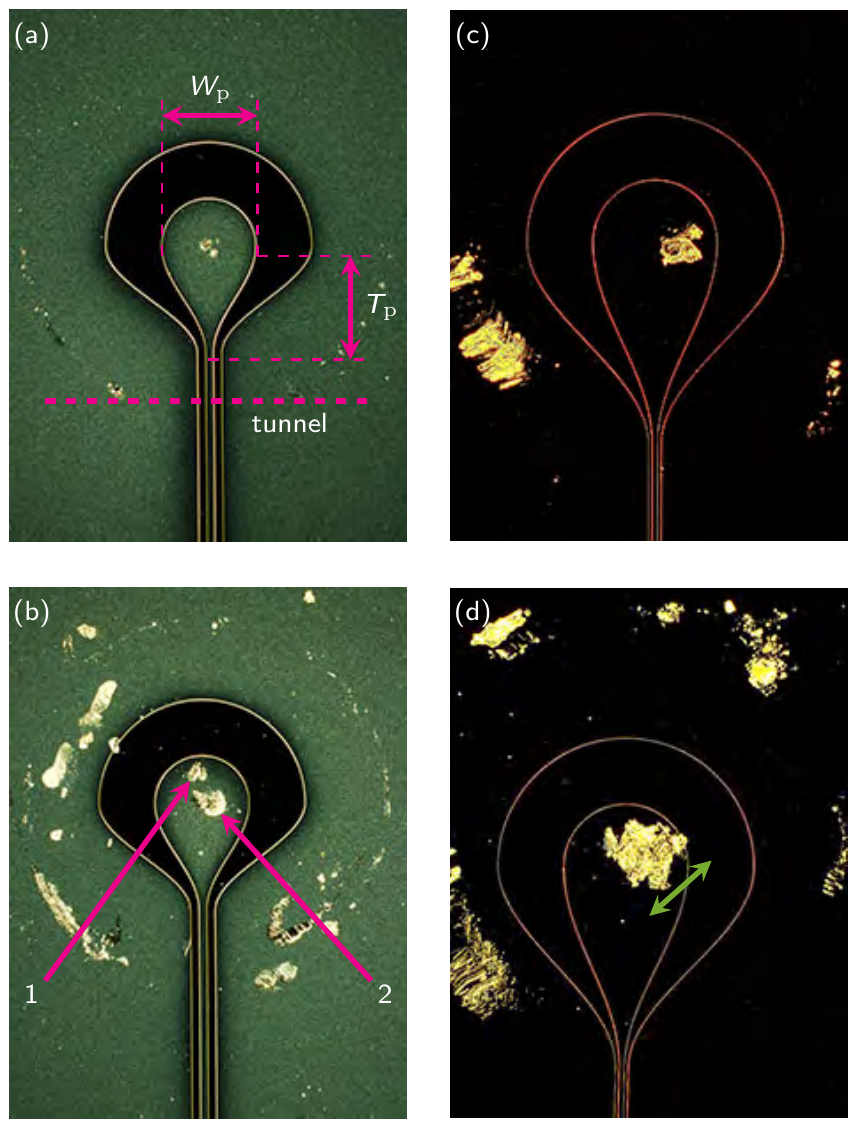}
\caption{Micro images used to evaluate the alignment procedure of the
three-dimensional wires. (a)-(b) Ag pads. The magenta arrows indicate the
first~(1) and second~(2) mating instance. The lengths~$W_{\textrm{p}}$ and
$T_{\textrm{p}}$ are indicated in~(a) by means of magenta bars. (c)-(d) Al pad
before and after a cooling cycle to~$\sim~\SI{10}{\milli\kelvin}$. Center
conductor dragging due to cooling is indicated by a green bar. The magenta
dashed line in~(a) indicates tunnel (i.e., rotational) alignment for the~Ag pad.
The images (dark field) were taken with an Olympus~MX61 microscope at~$\times 5$
magnification, manual exposure, and $\sim~\SI{500}{\milli\second}$ exposure
time. Note that the geometries for the pads in panels~(a) and (b) are optimized
for a~\SI{3}{\micro\meter} Ag film and, thus, are slightly different than those
for the pads in panels~(c) and (d), which are designed for
a~\SI{120}{\nano\meter} Al film.}
	\label{Figure05:Bejanin}
\end{figure}

\subsection{Alignment}
	\label{Alignment}

In order to implement a quantum socket with excellent interconnectivity
properties, it was imperative to minimize machining errors and mitigate the
effects of any residual errors. These errors are mainly due to: Dicing
tolerances; tapping tolerances of the~M$2.5$-threaded holes of the lid;
tolerances of the mating parts for the inner cavities of the lid and sample
holder; tolerances of the chip recess. These errors can cause both lateral and
rotational misalignment and become likely worse when cooling the quantum socket
to low temperatures. More details on alignment errors can be found in
Appendix~\ref{ALIGNMENT:ERRORS}.

The procedure to obtain an ideal and repeatable alignment comprises three main
steps: Optimization of the contact pad and tunnel geometry; accurate and precise
chip dicing; accurate and precise package machining. For the quantum socket
described in this work, the optimal tunnel width was found to
be~\SI{650}{\micro\meter}. This maintained reasonable impedance matching, while
allowing greater CPW contact pad and tapering dimensions. The contact pad
width~$W_{\textrm{p}}$ and taper length~$T_{\textrm{p}}$ were chosen to
be~$W_{\textrm{p}} = \SI{320}{\micro\meter}$ and $T_{\textrm{p}} =
\SI{360}{\micro\meter}$. These are the maximum dimensions allowable that
accommodate the geometry of the wire bottom interface for a nominal lateral and
rotational misalignment of~$\mp~\SI{140}{\micro\meter}$ and
$\mp~\SI{28}{\degree}$, respectively. In order to select the given pad
dimensions, we had to resort to a~$50~\Omega$ matched raised-cosine tapering.

The majority of the chips used in the experiments presented here was diced with
a dicing saw from the DISCO Corporation, model~DAD3240. To obtain a desired die
length, both the precision of the saw stage movement and the blade's kerf had to
be considered. For the DAD3240 saw, the former is~$\sim\SI{4}{\micro\meter}$,
whereas the latter changes with usage and materials. For the highest accuracy
cut, we measured the kerf on the same type of wafer just prior to cutting the
actual die. In order to achieve maximum benefit from the saw, rotational and
lateral alignment dicing markers were incorporated on the wafer. Such a
meticulous chip dicing procedure is only effective in conjunction with a
correspondingly high level of machining accuracy and precision. We used standard
computer numerical control~(CNC) machining with a tolerance of
$1$~thou~(\SI{25.4}{\micro\meter}), although electrical discharge machining can
be pursued if more stringent tolerances are required.

Following the aforementioned procedures we were able to achieve the desired
wire-pad matching accuracy and precision, which resulted in a test-retest
reliability (repeatability) of~\SI{100}{\percent} over~$94$ instances. These
figures of merit were tested in two steps: First, by micro imaging several pads
that were mated to a three-dimensional wire
(cf.~\ref{On:chip:pad:micro:imaging}); second, by means of DC resistance tests
(cf.~\ref{DC:resistance:tests}).

\subsubsection{On-chip pad micro imaging}
	\label{On:chip:pad:micro:imaging}

Micro imaging was performed on a variety of different samples, four of which are
exemplified in Fig.~\ref{Figure05:Bejanin}. The figure shows a set of micro
images for Ag and Al pads (details regarding the fabrication of these samples
are available in Appendix~\ref{SAMPLE:FABRICATION}).
Figure~\ref{Figure05:Bejanin}~(a) and (b) show two Ag pads that were mated with
the three-dimensional wires at room temperature. Panel~(a) shows a mating
instance where the wire bottom interface perfectly matched the on chip pad.
Panel~(b) shows two mating instances that, even though not perfectly matched,
remained within the designed tolerances. Notably, simulations of imperfect
mating instances revealed that an off-centered wire does not significantly
affect the microwave performance of the quantum socket. Finally, panels~(c) and
(d) display two Al pads which were both mated with a wire one time. While the
pad in~(c) was operated only at room temperature, the pad in~(d) was part of an
assembly that was cooled to~$\sim~\SI{10}{\milli\kelvin}$ for approximately
three months. The image was taken after the assembly was cycled back to room
temperature and shows dragging of the wire by a few tens of micrometers. Such a
displacement can likely be attributed to the difference in the thermal expansion
of Si and Al (cf.~Appendix~\ref{ALIGNMENT:ERRORS}).

As a diagnostic tool, micro images of a sample already mounted in the sample
holder after a mating cycle can be obtained readily by means of a handheld
digital microscope.

\subsubsection{DC resistance tests}
	\label{DC:resistance:tests}

\begin{table*}[ht]
\caption{DC resistance tests. Multiple Au samples were measured. For all samples
the length from the center of one pad to that of the opposite pad of the CPW
center conductor is~$L_{\textrm{pp}} = \SI{11.5}{\milli\meter}$. In the table
are reported: The width~$W$ of each CPW transmission line; the thickness~$d$ of
the metal; the metal volume resistivity~$\rho$ at room temperature or
at~\SI{77}{\kelvin}; the input and output wire pressure
settings~$\ell^{\textrm{i}}_{\textrm{p}}$ and $\ell^{\textrm{o}}_{\textrm{p}}$,
respectively; the operating temperature~$T$; the number of measurements~$N$; the
estimated trace resistance~$R^{\textrm{t}}$ (for the Au samples, the very large
parallel resistance~$\sim \SI{46}{\kilo\ohm}$ at room temperature due to the
titanium~(Ti) adhesion layer was neglected); the measured resistances
$R_{\textrm{io}}$, $R_{\textrm{ig}}$, and $R_{\textrm{og}}$. For a given chip,
each resistance was measured independently~$N$ times under similar measurement
conditions. The mean values and standard deviations of~$R_{\textrm{io}}$ are
provided; the minimum values of~$R_{\textrm{ig}}$ and $R_{\textrm{og}}$ are
given. Note that because $R^{\textrm{c}} + R^{\textrm{wc}} \ll R^{\textrm{t}}$,
we expect $R_{\textrm{io}} \approx R^{\textrm{t}}$. The discrepancy between the
estimated and measured values ($R^{\textrm{t}}$ and $R_{\textrm{io}}$) for the
Au and Al samples is mainly due to uncertainties associated with the metal
thickness~$d$. The inaccuracies are smaller for thicker films, as in the case of
the~\SI{3}{\micro\meter} Ag samples.}
\begin{center}
	\begin{ruledtabular}
		\begin{tabular}{cccccccccccc}
		\raisebox{0mm}[3mm][0mm]{Metal} & $W$ & $d$ & $\rho$ & 
		$\ell^{\textrm{i}}_{\textrm{p}}$ & $\ell^{\textrm{o}}_{\textrm{p}}$ & 
		$T$ & $N$ & $R^{\textrm{t}}$ & $R_{\textrm{io}}$ & $R_{\textrm{ig}}$ & 
		$R_{\textrm{og}}$ \\
		\raisebox{0mm}[0mm][2mm]{-} & \footnotesize{(\SI{}{\micro\meter})} & 
		\footnotesize{(\SI{}{\nano\meter})} & 
		\footnotesize{(\SI{1e-9}{\ohm\meter})} & 
		\footnotesize{(\SI{}{\milli\meter})} & 
		\footnotesize{(\SI{}{\milli\meter})} & \footnotesize{(\SI{}{\kelvin})} 
		& \footnotesize {(-)} & \footnotesize{(\SI{}{\ohm})} & 
		\footnotesize{(\SI{}{\ohm})} & \footnotesize{(\SI{}{\mega\ohm})} & 
		\footnotesize{(\SI{}{\mega\ohm})} \\
		\hline
		\hline
		\raisebox{0mm}[3mm][0mm]{Au} & $10$ & $100$ & $22$ & $4.52$ & $4.44$ & 
		$300$ & $30$ & $253$ & $218(3)$ & $31$ & $31$ \\
		\hline
		\raisebox{0mm}[3mm][0mm]{Au} & $10$ & $100$ & $22$ & $4.97$ & $4.89$ & 
		$300$ & $2$ & $253$ & $223(0)$ & $38$ & $38$ \\
		\hline
		\raisebox{0mm}[3mm][0mm]{Au} & $10$ & $100$ & $22$ & $4.18$ & $4.11$ & 
		$300$ & $2$ & $253$ & $217(0)$ & $39$ & $39$ \\
		\hline
		\raisebox{0mm}[3mm][0mm]{Au} & $10$ & $100$ & $22$ & $4.57$ & $4.45$ & 
		$300$ & $2$ & $253$ & $229(0)$ & $28.8$ & $28.6$ \\
		\hline
		\raisebox{0mm}[3mm][0mm]{Au} & $10$ & $200$ & $22$ & $4.60$ & $4.70$ & 
		$300$ & $10$ & $126.5$ & $98.0(7)$ & $50$ & $50$ \\
		\hline
		\raisebox{0mm}[3mm][0mm]{Au} & $10$ & $200$ & $4.55$ & 
		$4.60$\footnote{\label{ActualT}At~\SI{300}{\kelvin}.} & 
		$4.70$\footref{ActualT} & $77$ & $6$ & $26.16$ & $36.02(2)$ & $77.3$ & 
		$81.8$ \\
		\hline
		\raisebox{0mm}[3mm][0mm]{Ag} & $30$ & $3000$ & $16$ & $4.60$ & $4.70$ & 
		$300$ & $6$ & $2.04$ & $2.71(4)$ & $0.0043$ & $0.0043$ \\
		\hline
		\raisebox{0mm}[3mm][0mm]{Al} & $15$ & $120$ & $26$ & $4.25$ & $4.07$ & 
		$300$ & $24$ & $166.1$ & $171(1)$ & $0.0042$ & $0.0042$ \\
		\vspace{-4.5mm}
		\end{tabular}
	\end{ruledtabular}
\end{center}
	\label{Table02:Bejanin}
	\vspace{-3.5mm}
\end{table*}

In contrast to the micro imaging tests, which require the removal of the
microwave package's lid, DC resistance tests can be performed \textit{in situ}
at room temperature after the package and package holder have been fully
assembled. These tests were performed on all devices presented in this work,
including Au, Ag, and Al samples.

The typical test setup comprises a microwave package with two three-dimensional
wires each mating with an on-chip pad. The two pads are connected by means of a
CPW transmission line with series resistance~$R^{\textrm{t}}$. The back end of
the wires is connected to a coaxial cable ending in a microwave connector,
similar to the setup in Figs.~\ref{Figure01:Bejanin}~(d) and
\ref{Figure04:Bejanin}~(c). The DC equivalent circuit of this setup can be
represented by way of a four-terminal Pi network. The circuit comprises an input
``$\textrm{i}$'' and output ``$\textrm{o}$'' terminal, two terminals connected
to a common ground ``$\textrm{g}$,'' an input-output resistor with
resistance~$R_{\textrm{io}}$, and two resistors to ground with
resistance~$R_{\textrm{ig}}$ and $R_{ \textrm{og}}$. The~$\textrm{i}$ and
$\textrm{o}$ terminals correspond to the inner conductor of the two microwave
connectors. The outer conductor of both connectors is grounded.

The resistance~$R_{\textrm{io}}$ is that of the center conductor of the CPW
transmission line, including the contact resistance~$R^{\textrm{c}}$ for each
wire-pad interface and the series resistance~$R^{\textrm{wc}}$ of the wire's and
coaxial cable's inner conductor, $R_{\textrm{io}} = R^{\textrm{t}} + 2 (
R^{\textrm{c}} + R^{\textrm{wc}} )$. The resistances~$R_{\textrm{ig}}$ and
$R_{\textrm{og}}$ are those of the path between each center conductor and ground
and include the resistance of the inner and outer conductor of the various
coaxial cables and wires as well as any wire-pad contact resistance. Ideally,
these ground resistances should be open circuits. In reality, they are expected
to have a finite but large value because of the intrinsic resistance of the Si
wafers used as a substrate.

The design parameters, electrical properties, measurement conditions as well as
the measured values of~$R_{\textrm{io}}$, $R_{\textrm{ig}}$, and $R_{
\textrm{og}}$ for various Au, Ag, and Al samples are reported in
Table~\ref{Table02:Bejanin}. The resistances were probed by means of a
multimeter from the Fluke Corporation, model 289~\footnote{In a few instances,
the values were confirmed using a precision source-measure unit from Keysight
Technologies Inc., model~B2911A.}. Measuring resistances significantly different
from the expected values means that either a lateral or rotational misalignment
occurred.

The resistances for some Au samples were also measured at~\SI{77}{\kelvin} to
verify whether a good room temperature alignment persisted in cryogenic
conditions. The cold measurements were realized by dunking the package holder
into liquid nitrogen.

The measured value of~$R_{\textrm{io}}$ for the Ag samples is larger than the
estimated trace resistance by~$\sim~\SI{650}{\milli\ohm}$. This simple result
makes it possible to find an upper bound value for the contact resistance,
$R^{\textrm{c}} \lesssim \SI{325}{\milli\ohm}$. A more accurate estimate of the
contact resistance based on four-point measurements will be described
in~Subsec.\ref{Four:point:measurements}.

The DC resistance testing procedure presented here will be useful in
integrated-circuit quantum information processing, where, for example, CPW
transmission lines can serve as qubit readout lines~\cite{Corcoles:2015,
Riste:2015, Kelly:2015}. These tests can be expanded to encompass different
circuit structures such as the qubit control lines utilized in
Ref.~\cite{Kelly:2015}.

\section{CHARACTERIZATION}
	\label{CHARACTERIZATION}

The three-dimensional wires are multipurpose interconnects that can be used to
transmit signals over a wide frequency range, from DC to~$\SI{10}{\giga\hertz}$.
These signals can be: The current bias used to tune the transition frequency of
a superconducting qubit; the Gaussian-modulated sinusoidal or the rectangular
pulses that, respectively, make it possible to perform XY and Z control on a
qubit; the continuous monochromatic microwave tones used to read out a qubit
state or to populate and measure a superconducting resonator~\cite{Clarke:2008,
Corcoles:2015, Kelly:2015, Megrant:2012}. In general, the wires can be used to
transmit any baseband modulated carrier signal within the specified frequency
spectrum, at room and cryogenic temperatures.

In this section, we report experimental results for a series of measurements
aiming at a complete electrical characterization of the quantum socket at room
temperature and at approximately~\SI{77}{\kelvin} (i.e., in liquid nitrogen).
First, we performed four-point measurements to estimate the contact resistance
of a three-dimensional wire. Second, we measured the S-parameters of a wire at
room temperature. Third, we measured the S-parameters of the quantum socket with
an Au sample at room temperature and at~\SI{77}{\kelvin} and an Ag sample at
room temperature. Fourth, we realized time-domain measurements of the quantum
socket. Last, we performed four-port S-parameter measurements in order to assess
the socket crosstalk properties.

\subsection{Four-point measurements}
	\label{Four:point:measurements}

The wire-pad contact resistance~$R^{\textrm{c}}$ is an important property of the
quantum socket. In fact, a large~$R^{\textrm{c}}$ would result in significant
heating when applying DC bias signals and rectangular pulses, thus deteriorating
qubit performance.

\begin{figure}[t!]
	\centering
	\includegraphics[width=0.49\textwidth]{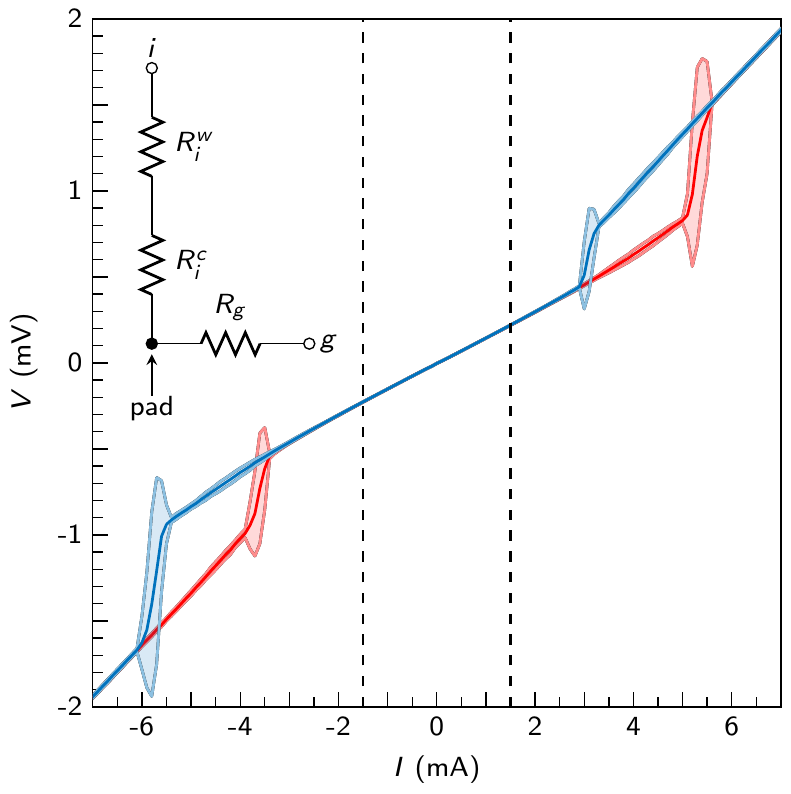}
\caption{I-V characteristic curve for~$R_{\textrm{ig}}$. The sweeps were
conducted by both increasing~(red) and decreasing~(blue) the applied current
between~$-\SI{7}{\milli\ampere}$ and $+\SI{7}{\milli\ampere}$. The voltage
measurements were delayed by~\SI{5}{\milli\second} and averaged
over~\SI{50}{\milli\second}. The displayed data is averaged over~$100$~sweeps.
The shaded region indicates two standard deviations. The dashed black lines
indicate the region ($\mp~\SI{1.5}{\milli\ampere}$) for which the resistance
value was found using linear regression. The origin of the hystereses is
explained in the main text. The inset shows the circuit diagram of the device
under test, including all resistors measured by means of the four-point
measurement. The position of the pad is indicated by an arrow.}
	\label{Figure06:Bejanin}
\end{figure}

In order to assess~$R^{\textrm{c}}$ for the inner and outer conductor of a
three-dimensional wire, we performed four-point measurements using the setup
shown in the inset of Fig.~\ref{Figure06:Bejanin}. Using this setup, we
were able to measure both the series resistance of the wire~$R^{\textrm{w}}$ and
the contact resistance~$R^{\textrm{c}}$. This allows us to estimate the overall
heating that could be generated during a qubit experiment.

The setup comprises a microwave package with a chip entirely coated with
a~\SI{120}{\nano\meter}~thick Al film; no grounding washer was used. The package
featured three three-dimensional wires, of which two were actually measured; the
third wire was included to provide mechanical stability. The package was
attached to the MC stage of a DR and connected to a set of phosphor bronze
twisted pairs. The twisted pairs were thermally anchored at all DR stages and
connected at room temperature to a precision source-measure unit~(SMU) from
Keysight Technologies Inc., model~B2911A.

We measured the resistance between the inner conductor of a wire and ground,
$R_{\textrm{ig}}$. This resistance comprises the inner conductor wire
resistance~$R_{\textrm{i}}^{\textrm{w}}$ in series with the inner conductor
contact resistance~$R_{\textrm{i}}^{\textrm{c}}$ and any resistance to ground,
$R_{\textrm{g}}$. Note that, at the operation temperature of the experiment
($\sim~\SI{10}{\milli\kelvin}$), Al is superconducting and, thus, the metal
resistance can be neglected.

Figure~\ref{Figure06:Bejanin} shows the current-voltage~(I-V) characteristic
curve for~$R_{\textrm{ig}}$. With increasing bias currents, the contact
resistance results in hot-spot generation leading to a local breakdown of
superconductivity. For sufficiently high bias currents, superconductivity breaks
down completely. At such currents, the observed hysteretic behavior indicates
the thermal limitations of our setup. Note, however, that these currents are at
least one order of magnitude larger than the largest bias current required in
typical superconducting qubit experiments~\cite{Barends:2013}.

In order to estimate~$R_{\textrm{ig}}$ from the I-V characteristic curve, we
selected the bias current region from~$-\SI{1.5}{\milli\ampere}$
to~$+\SI{1.5}{\milli\ampere}$ and fitted the corresponding slope. We
obtained~$R_{\textrm{ig}} \simeq \SI{148}{\milli\ohm}$. This value, which
represents an upper bound for the wire resistance and the wire-pad contact
resistance, $(R_{\textrm{i}}^{\textrm{w}} + R_{\textrm{i}}^{\textrm{c}})$, is
significantly larger than that associated with Al wire
bonds~\cite{Teverovsky:2004}. In future versions of the three-dimensional wires
we will attempt to reduce the wire-pad contact resistance by rounding the tip of
the center conductor, stiffening the wire springs, using a thicker metal film
for the pads, and plating the contact pads with Au or titanium nitride. We note,
however, that even a large value of the wire and/or wire-pad contact resistance
will not significantly impair the quantum socket microwave performance.

\subsection{Two-port scattering parameters}
	\label{Two:port:scattering:parameters}

\begin{figure*}[t!]
	\centering
	\includegraphics[width=0.99\textwidth]{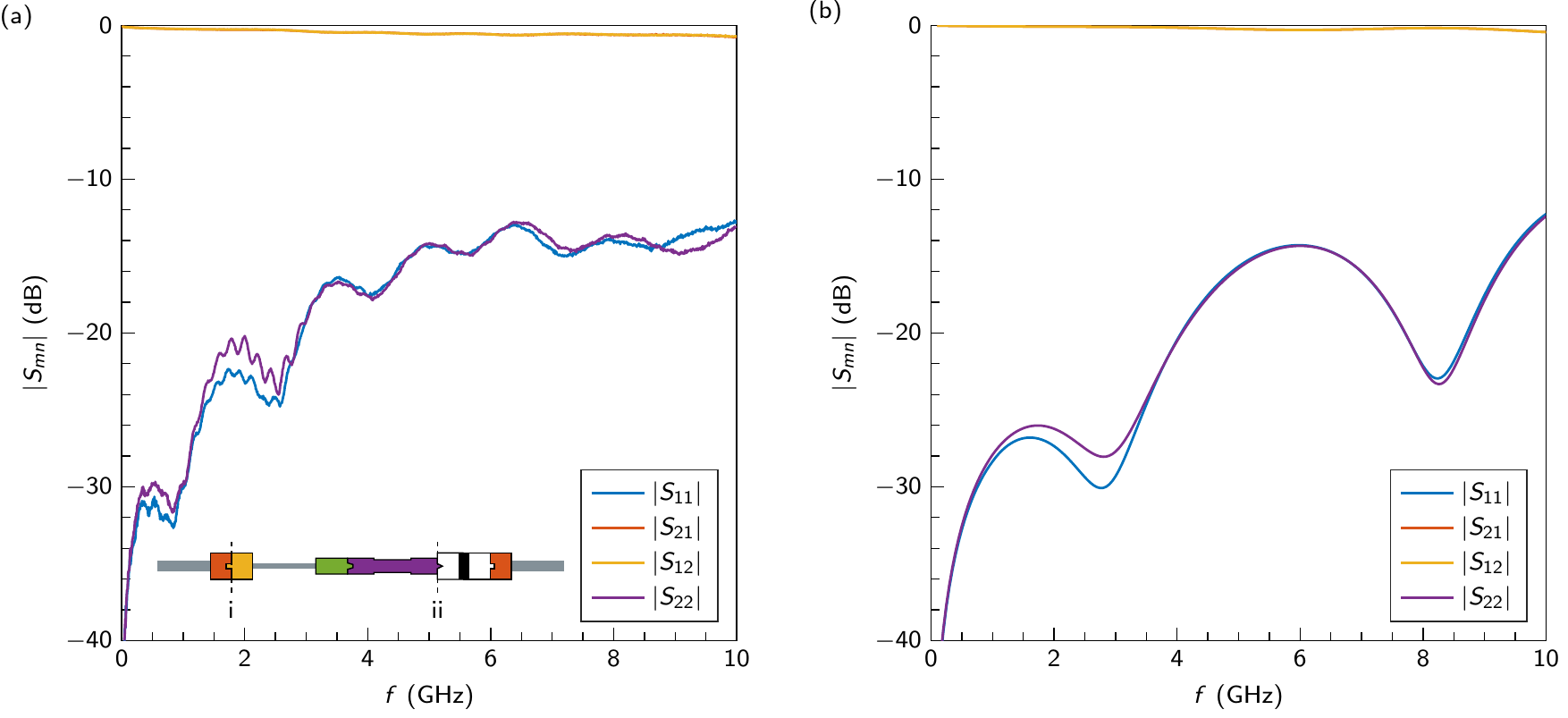}
\caption{S-parameter measurements and simulations of a three-dimensional wire at
room temperature. (a) Magnitude of the measured S-parameters~$\abs{S_{mn}}$,
with~$m , n = \{ 1 , 2 \}$. Inset: Image of the measurement setup. From left to
right: Segment of flexible coaxial cable (grey); SMA female connector (red);
after plane~i, SMA male connector (orange); segment of EZ~47 cable (grey);
screw-in micro connector (green); three-dimensional wire (purple); after
plane~ii, 2.92~mm end launch connector (white and black); SMA female connector
(red); segment of flexible coaxial cable (grey). (b) S-parameter simulations.
The lower attenuation is due to idealized material properties and connections.}
	\label{Figure07:Bejanin}
\end{figure*}

The S-parameter measurements of a bare three-dimensional wire were realized by
means of the setup shown in the inset of Fig.~\ref{Figure07:Bejanin}~(a). The
device under test~(DUT) comprises a cable assembly attached to a
three-dimensional wire by means of a screw-in micro connector. The cable
assembly is made of an approximately~\SI{230}{\milli\meter} long semi-rigid
coaxial cable EZ~47, which is soldered to an EZ Form custom-made Sub-miniature
type A~(SMA) male connector, model~705538-347. The other end of the coaxial
cable is soldered to the screw-in micro connector. The SMA connector of the DUT
is connected to one port of a vector network analyzer~(VNA) from Keysight,
model~PNA-L N5230A by means of a flexible coaxial cable. The bottom interface of
the wire is connected to a 2.92~mm end launch connector from Southwest
Microwave, Inc., model~1092-01A-5, which then connects to the other port of the
VNA through a second flexible coaxial cable. The 2.92~mm adapter is
characterized by a flush coaxial back plane, which mates with the wire bottom
interface well enough to allow for S-parameter measurements up
to~\SI{10}{\giga\hertz}.

In order to measure the S-parameters of the DUT, a two-tier calibration was
performed. First, a two-port electronic calibration module~(ECal) from Keysight,
model~N4691B, with 2.92~mm male connectors was used to set the measurement
planes to the end of the flexible cables closer to the DUT. Second, a
port-extension routine was performed to correct for the insertion loss, phase,
and delay of the 2.92~mm adapter. This made it possible to set the measurement
planes to the ports of the DUT.

The magnitudes of the measured reflection and transmission S-parameters are
displayed in Fig.~\ref{Figure07:Bejanin}~(a). We performed microwave simulations
of a three-dimensional wire for the same S-parameters
(cf.~Subsec.~\ref{Microwave:simulations} for the electric field distribution),
the results of which are plotted in Fig.~\ref{Figure07:Bejanin}~(b). The
S-parameters were measured and simulated between \SI{10}{\mega\hertz} and
\SI{10}{\giga\hertz}. The S-parameters~$\abs{S_{21}}$ and $\abs{S_{12}}$ show a
featureless microwave response, similar to that of a coaxial transmission line.
The attenuation at~\SI{6}{\giga\hertz} is~$\abs{S_{21}} \simeq
\SI{-0.58}{\deci\bel}$ and the magnitude of the reflection coefficients at the
same frequency is~$\abs{S_{11}} \simeq \SI{-13.8}{\deci\bel}$ and $\abs{S_{22}}
\simeq \SI{-14.0}{\deci\bel}$. The phase of the various S-parameters (not shown)
behaves as expected for a coaxial transmission line. All measurements were
performed at room temperature.

The S-parameter measurements of a three-dimensional wire indicate a very good
microwave performance. However, these measurements alone are insufficient to
fully characterize the quantum socket operation. A critical feature that
deserves special attention is the~\SI{90}{\degree} transition region between the
wire bottom interface and the on-chip CPW pad. It is well-known
that~\SI{90}{\degree} transitions can cause significant impedance mismatch and,
thus, signal reflection~\cite{Simons:2001}. In quantum computing applications,
these reflections could degrade both the qubit control and readout fidelity.

\begin{figure*}[t]
	\centering
	\includegraphics[width=0.99\textwidth]{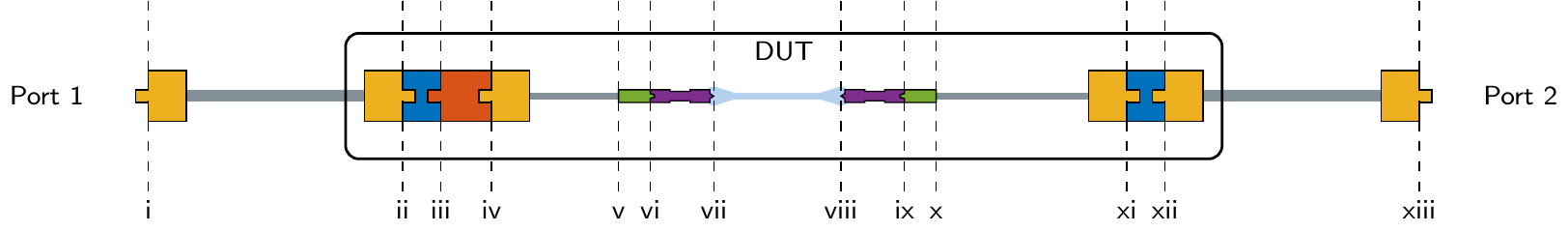}
\caption{Microwave characterization setup. The vertical black dashed lines
indicate main reflection planes. The yellow terminations correspond to SMA male
connectors at the end of each cable. The input(output) flexible cable
corresponds to the region in between planes~i and ii(xii and xiii), in gray; the
blue blocks correspond to SMA female bulkhead adapters; the plane~ii(xii)
correspond to the input(output) port of the DUT; the orange block corresponds to
an SMA male to SMA female adapter; the EZ 47 input(output) cable corresponds to
the region in between planes~iv and v(x and xi), in gray; the plane~v(x)
corresponds to the solder connection on the three-dimensional wire; the
plane~vi(ix) is associated with the screw-in micro connector; the
plane~vii(viii) corresponds to the~\SI{90}{\degree} interface connecting each
three-dimensional wire to the input(output) of the CPW transmission line~(pale
blue). The three-dimensional wires are indicated in purple.}
	\label{Figure08:Bejanin}
\end{figure*}

Figure~\ref{Figure08:Bejanin} shows a typical setup for the characterization of
a wiring configuration analogous to that used for qubit operations. The setup
comprises a VNA from Keysight, model~PNA-X N5242A, with ports~$1$ and $2$
connected to a pair of flexible coaxial cables from Huber+Suhner AG,
model~SucoFlex 104-PE (with SMA male connectors). In order to calibrate the
measurement, the flexible cables were first connected to a two-port ECal module
from Keysight, model~N4691-60006, featuring 3.5~mm female
connectors~\footnote{Unless specified, all calibrations were performed using SMA
female to 3.5~mm male adapters. These adapters introduced negligible calibration
errors from DC to~\SI{10}{\giga\hertz}.}. These cables were then connected to
the SMA female bulkhead adapter at the input and output ports of the DUT shown
in Fig.~\ref{Figure08:Bejanin}. The DUT incorporates a microwave package with a
pair of three-dimensional wires, which address one CPW transmission line on an
Au or Ag chip. The microwave package was attached to the package holder, as
described in Subsec.~\ref{Package:holder} and
Sec.~\ref{THE:QUANTUM:SOCKET:IMPLEMENTATION} (cf. also
Figs.~\ref{Figure01:Bejanin}~(d) and \ref{Figure04:Bejanin}~(c)). For the
measurements in this section, however, the SMP adapters were substituted by SMA
female-female bulkhead adapters. The transmission line geometrical dimensions
and wire pressure settings are reported in Table~\ref{Table02:Bejanin}; only
the~\SI{200}{\nano\meter} Au samples and the Ag samples were characterized at
microwave frequencies. The back end of each three-dimensional wire is connected
to one end of an EZ~47 cable by means of the screw-in micro connector described
in Subsec.~\ref{Package:holder}; the other end of the EZ~47 cable is soldered to
an SMA male connector. One of the EZ cables is~\SI{228}{\milli\meter} long and
the other is~\SI{232}{\milli\meter} long; the longer cable (output) is connected
directly to the SMA bulkhead adapter, whereas the shorter cable (input) is
prolonged using an SMA female-male adapter and, then, connected to the bulkhead
adapter. These bulkhead adapters are the reference planes~ii and xii associated
with the input and output ports of the DUT, respectively, as shown in
Fig.~\ref{Figure08:Bejanin}.

\begin{figure*}[ht]
	\centering
	\includegraphics[width=0.99\textwidth]{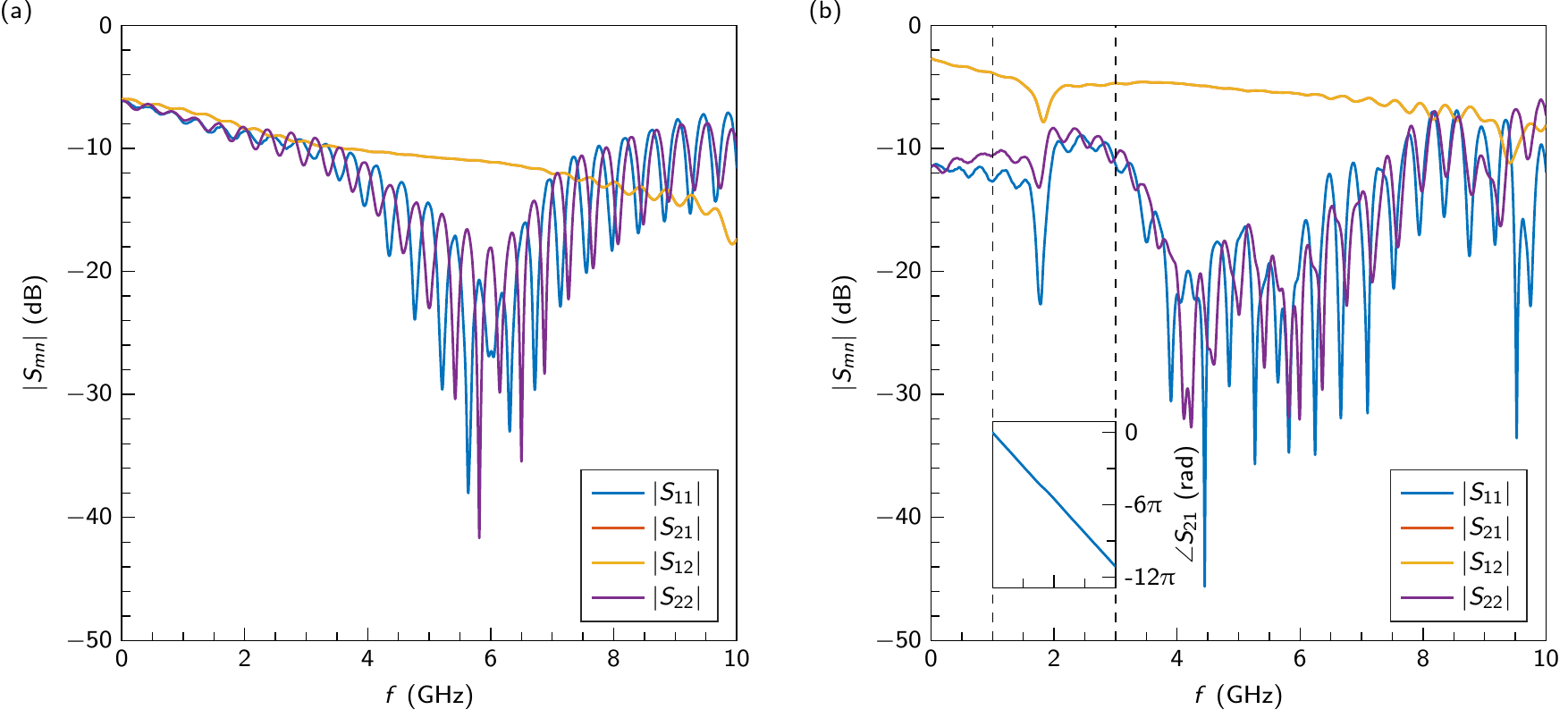}
\caption{S-parameter measurements for the Au sample. (a) $\abs{S_{mn}}$ at~room
temperature. (b) $\abs{S_{mn}}$ at~\SI{77}{\kelvin}. The transmission
coefficients show that the DUT is a reciprocal device (i.e., $S_{21} \simeq
S_{12}$), as expected for a passive structure. The inset in (b) shows the
unwrapped phase angle~$\angle S_{21}$; the black dashed lines delimit the
frequency region between~\SI{1}{\giga\hertz} and \SI{3}{\giga\hertz}. Note that
the reflection coefficients~$S_{11}$ and $S_{22}$ are relatively large at very
low frequency. This is expected for a very lossy transmission line. In fact, the
center conductor for the Au sample is characterized by a series resistance
(cf.~Table~\ref{Table02:Bejanin}) $R_{\textrm{io}} \simeq \SI{98}{\ohm}$ at room
temperature, which corresponds to~$S_{11} \sim S_{22} \simeq -
\SI{6}{\deci\bel}$ at~\SI{10}{\mega\hertz}, and $R_{\textrm{io}} \simeq
\SI{36}{\ohm}$ at~\SI{77}{\kelvin}, which corresponds to~$S_{11} \sim S_{22}
\simeq - \SI{12}{\deci\bel}$ at~\SI{10}{\mega\hertz}. These findings are
consistent with the TDR results to be shown in Fig.~\ref{Figure11:Bejanin},
where the large impedance steps are also due to the large series resistance
(cf.~Subsec.~\ref{Time:domain:reflectometry}). The low-loss Ag sample shows much
lower reflection coefficients at low frequency
(cf.~Fig.~\ref{Figure10:Bejanin}), whereas the lossy Al sample shows high
reflections at low frequency and room temperature
(cf.~Fig.~\ref{Figure15:Bejanin}~(a) in
Sec.~\ref{APPLICATIONS:TO:SUPERCONDUCTING:RESONATORS}).}
	\label{Figure09:Bejanin}
\end{figure*}

\begin{figure*}[ht]
	\centering
	\includegraphics[width=0.99\textwidth]{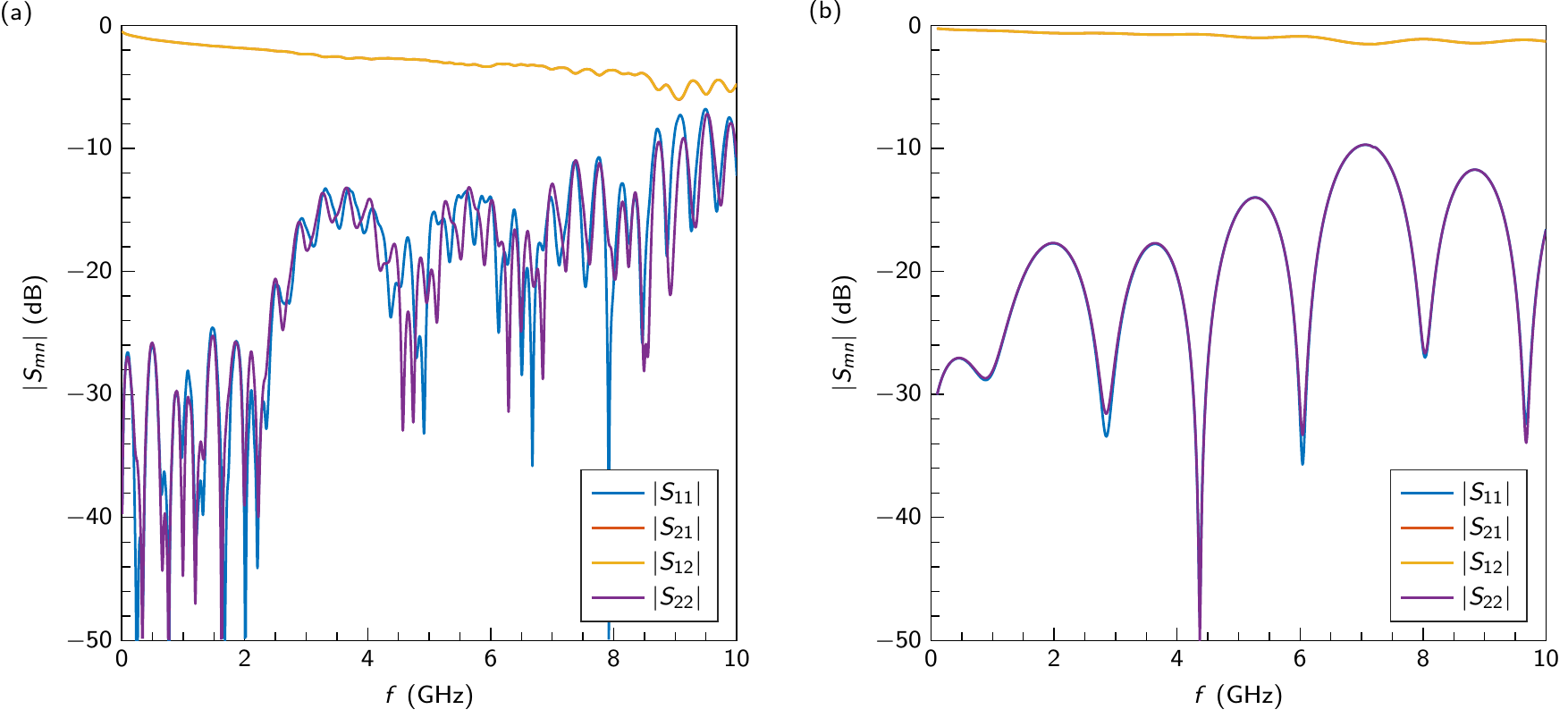}
\caption{S-parameters of the Ag sample. (a) $\abs{S_{mn}}$ measurement at~room
temperature. (b) $\abs{S_{mn}}$ microwave simulation. Note that the attenuation
for the measured data is larger than that in the simulation because the latter
does not include the EZ~47 cables used in the DUT of
Fig.~\ref{Figure08:Bejanin}.}
	\label{Figure10:Bejanin}
\end{figure*}

We performed a two-port S-parameter measurement of the DUT
from~\SI{10}{\mega\hertz} to \SI{10}{\giga\hertz}. We selected an intermediate
frequency~(IF) bandwidth~$\Delta f_{\textrm{IF}} = \SI{500}{\hertz}$, a constant
excitation power~$P_{\textrm{RF}} = 0~\textrm{dB-milliwatts}$~(dBm), and
$N_{\textrm{RF}} = 12001$ or $N_{\textrm{RF}} = 100001$ measurement points for
the Au and Ag samples, respectively. The measurement results at room temperature
for the Au and Ag samples are shown in Figs.~\ref{Figure09:Bejanin}~(a) and
\ref{Figure10:Bejanin}~(a), respectively~\footnote{Note that the Ag measurements
were performed without using the SMA female to 3.5~mm male adapters.}. The
results for the Au sample at~\SI{77}{\kelvin} are shown in
Fig.~\ref{Figure09:Bejanin}~(b).

The S-parameter measurements of the Au sample show that the quantum socket
functions well at microwave frequencies, both at room temperature and
at~\SI{77}{\kelvin}. Since most of the mechanical shifts have already occurred
when cooling to~\SI{77}{\kelvin}~\cite{Marquardt:2000}, this measurement allows
us to deduce that the socket will continue functioning even at lower
temperatures, e.g., $\sim~\SI{10}{\milli\kelvin}$. The Au sample, however, is
characterized by a large value of~$R_{\textrm{io}}$, which may conceal unwanted
features both in the transmission and reflection measurements. Therefore, we
prepared an Ag sample that exhibits a much lower resistance even at room
temperature. The behavior of the Ag S-parameters is similar to that of a
transmission line or coaxial connector. For example, $\abs{S_{11}}$ is
approximately~\SI{-15}{\deci\bel}; as a reference, for a high-precision SMA
connector at the same frequency~$\abs{S_{11}} \simeq \SI{-30}{\deci\bel}$.

The presence of the screw-in micro connector can occasionally deteriorate the
microwave performance of the quantum socket. In fact, if the micro connector is
not firmly tightened, a dip in the microwave transmission is observed. At room
temperature, it is straightforward to remove the dip by simply re-tightening the
connector when required. On the contrary, for the measurements
at~\SI{77}{\kelvin} and for any other application in a cryogenic environment
assuring that the micro connector is properly torqued at all times can be
challenging. Figure~\ref{Figure09:Bejanin}~(b), for example, shows the
S-parameters for an Au sample measured at~\SI{77}{\kelvin}. A microwave dip
appeared at approximately~\SI{1.8}{\giga\hertz}, with a~\SI{3}{\deci\bel}
bandwidth of approximately~$\SI{200}{\mega\hertz}$. The inset in
Fig.~\ref{Figure09:Bejanin}~(b) displays the phase angle of~$S_{21}$
between~\SI{1}{\giga\hertz} and \SI{3}{\giga\hertz}, showing that the dip is
unlikely a Lorentzian-type resonance (more details in the Supplemental Material
at~\url{http://www.Supplemental-Material-Bejanin}). Note that the dip is far
from the typical operation frequencies for superconducting qubits. Additionally,
as briefly described in Sec.~\ref{CONCLUSIONS}, we will remove the screw-in
micro connector from future generations of the three-dimensional wires.

Figure~\ref{Figure10:Bejanin}~(b) shows a simulation of the S-parameters for the
Ag sample, for the same frequency range as the actual measurements. While there
are visible discrepancies between the measured and simulated S-parameters, the
latter capture well some of the characteristic features of the microwave
response of the DUT. In particular, the measured and simulated reflection
coefficients display a similar frequency dependence. It is worth mentioning that
we also simulated the case where the wire bottom interface is not perfectly
aligned with the on-chip pad (results not shown). We considered lateral
misalignments of~\SI{100}{\micro\meter} and rotational misalignments
of~$\sim~\SI{20}{\degree}$. This allowed us to study more realistic scenarios,
such as those shown in Fig.~\ref{Figure05:Bejanin}. We found that the departure
between the misaligned and the perfectly aligned simulations was marginal. For
example, the transmission S-parameters varied only by
approximately~$\mp~\SI{0.5}{\deci\bel}$.

In Appendix~\ref{MICROWAVE:PARAMETERS}, we show a set of microwave parameters
obtained from the measured S-parameters for the Au sample at room temperature
and at~\SI{77}{\kelvin} and for the Ag sample at room temperature. These
parameters make it possible to characterize the input and output impedance as
well as the dispersion properties of the quantum socket.

\subsection{Time-domain reflectometry}
	\label{Time:domain:reflectometry}

In TDR measurements, a rectangular pulse with fast rise time and fixed length is
applied to a DUT; the reflections (and all re-reflections) due to all reflection
planes in the system (i.e., connectors, geometrical changes, etc.) are then
measured by way of a fast electrical sampling module. The reflections are, in
turn, related to the impedances of all of the system components. Thus, TDR makes
it possible to estimate any impedance mismatch and its approximate spatial
location in the system.

TDR measurements were performed on the DUT shown in Fig.~\ref{Figure08:Bejanin},
with the same Au or Ag sample as for the measurements in
Subsec.~\ref{Two:port:scattering:parameters}. As always, the Au sample was
measured both at room temperature and at~\SI{77}{\kelvin}, whereas the Ag sample
was measured only at room temperature. The TDR setup is analogous to that used
for the S-parameter measurements, with the following differences: The DUT input
and output reference planes were extended to include the SucoFlex flexible
coaxial cables (i.e., these cables were not calibrated out); when testing the
DUT input port, the output port was terminated in a load with
impedance~$Z_{\textrm{L}} = Z_{\textrm{c}}$ and vice versa when testing the DUT
output port. The TDR measurements were realized by means of a sampling
oscilloscope from Teledyne LeCroy, model~WaveExpert 100H; the oscilloscope
features an electrical sampling module with~\SI{20}{\giga\hertz} bandwidth and a
TDR step generator, model~ST-20. The generated signal is a voltage square wave
characterized by a nominal pulse rise time of~\SI{20}{\pico\second}, amplitude
of~\SI{250}{\milli\volt}, pulse width of~\SI{300}{\nano\second}, and pulse
repetition rate of~\SI{1}{\mega\hertz}. The voltage reflected by the DUT,
$V^{-}$, is acquired as a function of time~$t$ by means of the sampling module.
This time is the round-trip interval necessary for the voltage pulse to reach a
DUT reflection plane and return back to the sampling module. The measured
quantity is given by~\footnote{Note that the DUT is a piecewise transmission
line inhomogeneously filled with dielectric materials. Transforming the time~$t$
into distance is only possible with detailed knowledge of geometries and
materials for all regions of the DUT. Since this information is not known to a
high degree of accuracy, we prefer to express all measured quantities as a
function of~$t$.}
\begin{equation}
V_{\textrm{meas}} ( t ) = V^+ ( t ) + V^- ( t ) \quad ,
	\label{Equation:02}
\end{equation}
where~$V^+$ is the amplitude of the incident voltage square wave. From
Eq.~(\ref{Equation:02}), we can obtain the first-order instantaneous impedance
as~\cite{Bogatin:2003}
\begin{equation}
Z ( t ) = Z_{\textrm{c}} \frac{1 + \xi( t )}{1 - \xi( t )} \quad ,
	\label{Equation:03}
\end{equation}
where~$\xi( t ) = \left( V_{\textrm{meas}} ( t ) - V^+ \right) / V^+$.

\begin{figure}[t!]
	\centering
	\includegraphics[width=0.49\textwidth]{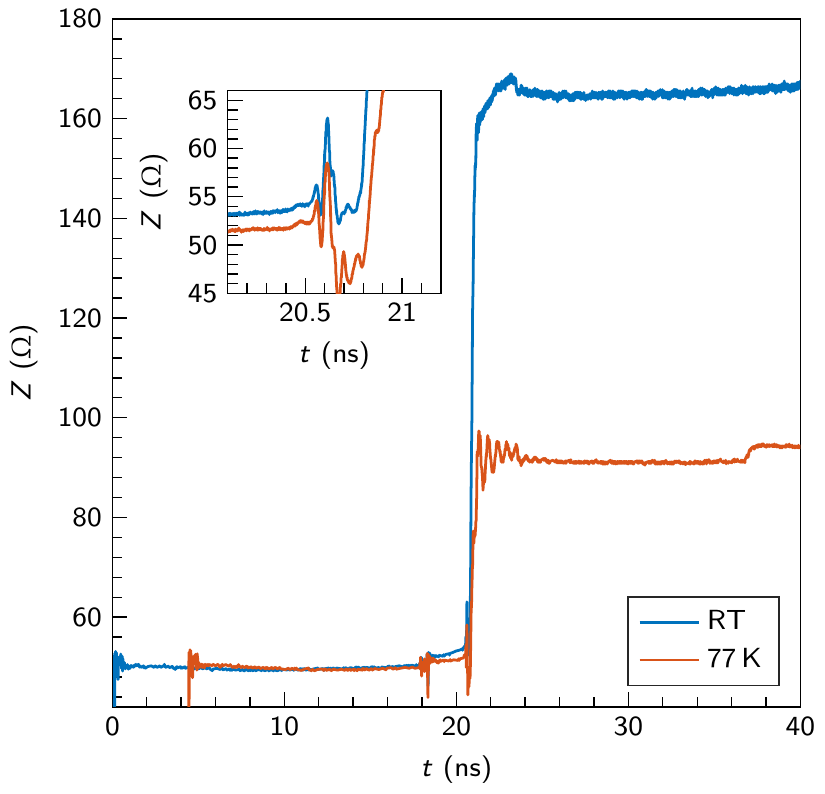}
\caption{TDR measurements of the Au sample at room temperature~(blue) and
\SI{77}{\kelvin}~(red). The inset shows the room temperature data associated
with part of the EZ~47 cable, the input three-dimensional wire, and part of the
CPW transmission line.}
	\label{Figure11:Bejanin}
\end{figure}

Figure~\ref{Figure11:Bejanin} shows~$Z ( t )$ for the DUT with the Au sample at
room temperature and at~\SI{77}{\kelvin}; the measurement refers to the input
port of the DUT, including a~\SI{2.0}{\meter} flexible cable. The figure inset
shows the room temperature data for a shorter time interval. This corresponds to
a space interval beginning at a point between planes~iv and v and ending at a
point between planes~vii and viii (cf.~Fig.~\ref{Figure08:Bejanin}).

\begin{figure}[h]
	\centering
	\includegraphics[width=0.49\textwidth]{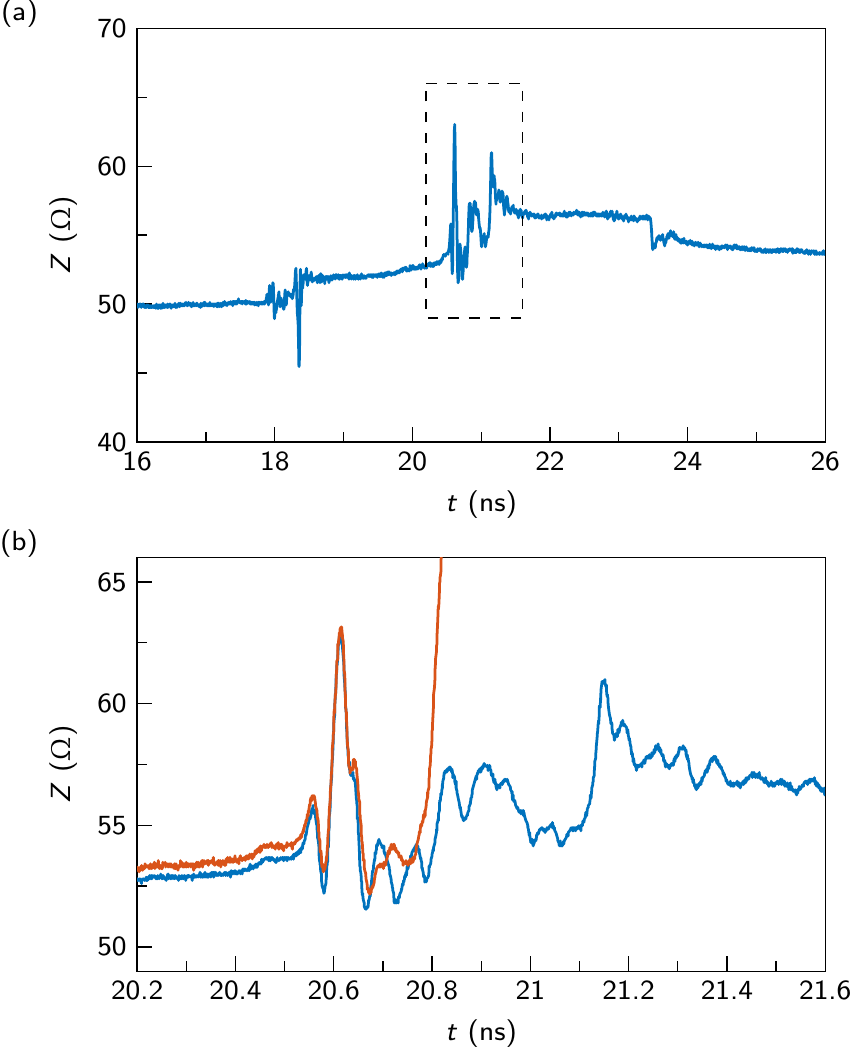}
\caption{TDR measurement of the Ag sample at room temperature. (a) Measurement
of port~$1$ of the setup in Fig.~\ref{Figure08:Bejanin}. (b) Zoomin of~(a)
addressing the three-dimensional wire and the~\SI{90}{\degree} transition region
between the wire and the CPW transmission line~(blue). The room temperature Au
data~(red) is also displayed as a reference.}
	\label{Figure12:Bejanin}
\end{figure}

Figure~\ref{Figure12:Bejanin}~(a) shows~$Z ( t )$ for the Ag sample at room
temperature. Figure~\ref{Figure12:Bejanin}~(b) displays the data in~(a) for a
time interval corresponding to a space interval beginning at a point between
planes~iv and v and ending at a point between planes~x and xi; as a reference,
the Au data are overlaid with the Ag data.

\begin{figure*}[ht]
	\centering
	\includegraphics[width=0.99\textwidth]{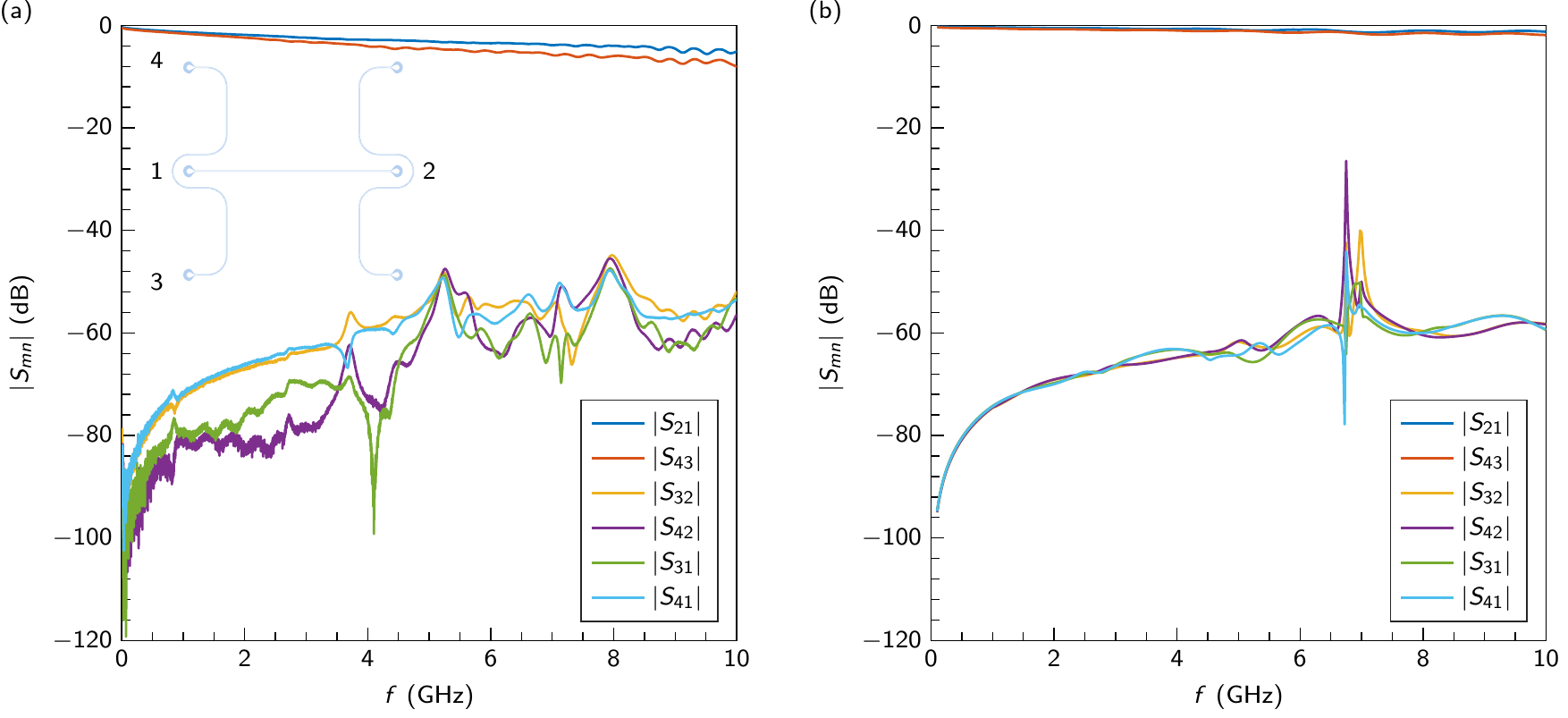}
\caption{Signal crosstalk. (a) Transmission and crosstalk coefficients for the
Ag sample shown in the inset. The numbers adjacent to the pads in the inset
correspond to the device ports. Reciprocal and reflection S-parameters are not
shown. (b) Microwave simulation of the same device. The origin of the peaks at
approximately~\SI{7}{\giga\hertz} is explained in the main text.}
	\label{Figure13:Bejanin}
\end{figure*}

For the Au sample, the first main reflection plane (plane~ii) is encountered
at~$t \simeq \SI{18}{\nano\second}$. The second main reflection plane (plane~v)
appears after~$\sim~\SI{2.5}{\nano\second}$ relative to the first plane, at~$t
\simeq \SI{20.5}{\nano\second}$. From that time instant and for a span of
approximately~\SI{250}{\pico\second}, the TDR measurement corresponds to~$Z ( t
)$ of the three-dimensional wire itself. The maximum impedance mismatch between
the EZ Form cable and the three-dimensional wire is approximately~\SI{10}{\ohm}.
The third main reflection plane (plane~vii) corresponds to the~\SI{90}{\degree}
transition region; for the Au sample, it is impossible to identify features
beyond this plane owing to the large series resistance of the on-chip CPW
transmission line. From empirical evidence, the impedance~$Z ( t )$ of a lossy
line with series resistivity~$\rho$ increases linearly with the length of the
line~$L$ as~$\rho L / ( W d )$. In fact, for the Au sample we measured an
impedance step across the CPW transmission line of
approximately~$\SI{100}{\ohm}$ at room temperature and \SI{40}{\ohm}
at~\SI{77}{\kelvin}. These steps are approximately the~$R_\textrm{io}$ values
reported in Table~\ref{Table02:Bejanin}.

In order to obtain a detailed measurement of the impedance mismatch beyond
the~\SI{90}{\degree} transition region, we resorted to the TDR measurements of
the DUT with the much less resistive Ag sample. First, we confirmed that~$Z ( t
)$ of the input three-dimensional wire for the Ag sample is consistent with the
TDR measurements of the Au sample; this is readily verified by inspecting
Fig.~\ref{Figure12:Bejanin}~(b). The three-dimensional wire is the structure
ending at the onset of the large impedance step shown by the Au overlaid data.
The structure spanning the time interval from~$t \simeq
\SI{20.75}{\nano\second}$ to~$t \simeq \SI{21}{\nano\second}$ is associated with
the input transition region, the CPW transmission line, and the output
transition region. The output three-dimensional wire starts at~$t \simeq
\SI{21}{\nano\second}$, followed by the EZ Form coaxial cable, which finally
ends at the SMA bulkhead adapter at~$t \simeq \SI{23.5}{\nano\second}$. The
maximum impedance mismatch associated with the transition regions and the CPW
transmission line is~$\sim~\SI{5}{\ohm}$. Notably, this mismatch is smaller than
the mismatch between the three-dimensional wire and the coaxial cable. This is
an important result. In fact, while it would be hard to diminish the impedance
mismatch due to the transition region, it is feasible to further minimize the
wire mismatch by creating accurate lumped-element models of the wire and use
them to minimize stray capacitances and/or inductances~\footnote{Work in
progress.}.

It is worth comparing~$Z ( t )$ of the quantum socket with that of a standard
package for superconducting qubits, where wire bonds are used to make
interconnections between a printed circuit board and the control and measurement
lines of a qubit on a chip. A detailed study of the impedance mismatch
associated with wire bonds is found in Ref.~\cite{Mutus:2014}, where the authors
have shown that a long wire bond (of length between~$\sim~\SI{1}{\milli\meter}$
and $\SI{1.5}{\milli\meter}$; typical length in most applications) can lead to
an impedance mismatch larger than~\SI{15}{\ohm} (cf.~Fig.~S3 in the
supplementary information of Ref.~\cite{Mutus:2014}); on the contrary, a short
wire bond (between~$\sim~\SI{0.3}{\milli\meter}$ and $\SI{0.5}{\milli\meter}$;
less typical) results in a much smaller mismatch, approximately~\SI{2}{\ohm}. In
terms of impedance mismatch the current implementation of the quantum socket,
which is limited by the mismatch of the three-dimensional wires, lies in between
these two extreme scenarios.

\subsection{Signal crosstalk}
	\label{Signal:crosstalk}

Crosstalk is a phenomenon where a signal being transmitted through a channel
generates an undesired signal in a different channel. Inter-channel isolation is
the figure of merit that quantifies signal crosstalk and that has to be
maximized to improve signal integrity. Crosstalk can be particularly large in
systems operating at microwave frequencies, where, if not properly designed,
physically adjacent channels can be significantly affected by coupling
capacitances and/or inductances. In quantum computing implementations based on
superconducting quantum circuits, signal crosstalk due to wire bonds has been
identified to be an important source of errors and methods to mitigate it have
been developed~\cite{Wenner:2011:a, Abraham:2014:b, Abraham:2015}. However,
crosstalk remains an open challenge and isolations (opposite of crosstalk) lower
than~\SI{20}{\deci\bel} are routinely observed when using wire
bonds~\footnote{Daniel T.~Sank (private communication).}. The coaxial design of
the three-dimensional wires represents an advantage over wire bonds. The latter,
being open structures, radiate more electromagnetic energy that is transferred
to adjacent circuits. The former, being enclosed by the outer conductor, limit
crosstalk due to electromagnetic radiation.

In realistic applications of the quantum socket, the three-dimensional wires
must land in close proximity of several on-chip transmission lines. In order to
study inter-channel isolation in such scenarios, we designed a special device
comprising a pair of CPW transmission lines, as shown in the inset of
Fig.~\ref{Figure13:Bejanin}~(a). One transmission line connects two
three-dimensional wires (ports~$1$ and $2$), exactly as for the devices studied
in Subsecs.~\ref{Two:port:scattering:parameters} and
\ref{Time:domain:reflectometry}; the other line, which also connects two
three-dimensional wires (ports~$3$ and $4$), circumvents the wire at port~$1$ by
means of a CPW semicircle. The distance between the semicircle and the wire
outer conductor is designed to be as short as possible,
$\sim~\SI{100}{\micro\meter}$.

The chip employed for the crosstalk tests is similar to the Ag sample used for
the socket microwave characterization and was part of a DUT analogous to that
shown in Fig.~\ref{Figure08:Bejanin}. The DC resistances of the center trace of
the~$1-2$ and $3-4$ transmission lines were measured and found to
be~$\sim~\SI{2.8}{\ohm}$ and $\sim~\SI{4.5}{\ohm}$, respectively (note that
the~$3-4$ transmission line is~$\sim~\SI{18.0}{\milli\meter}$ long). All DC
resistances to ground and between the two transmission lines were found to be on
the order of a few kilohms, demonstrating the absence of undesired short circuit
paths. A four-port calibration and measurement of the DUT were conducted by
means of the ECal module and PNA-X. We selected a frequency range
from~\SI{10}{\mega\hertz} to \SI{10}{\giga\hertz}, $\Delta f_{\textrm{IF}} =
\SI{1}{\kilo\hertz}$, $P_{\textrm{RF}} = 0$~dBm, and $N_{\textrm{RF}} = 64001$.
Among the $16$ S-parameters, Fig.~\ref{Figure13:Bejanin}~(a) shows the magnitude
of the transmission coefficients~$S_{21}$ and $S_{43}$, along with the magnitude
of the crosstalk coefficients~$S_{31}, S_{41}, S_{32}$, and $S_{42}$.

The results show that the isolation in the typical qubit operation bandwidth,
between~\SI{4}{\giga\hertz} and \SI{8}{\giga\hertz}, is larger
than~$\sim~\SI{45}{\deci\bel}$. Note that the crosstalk coefficients shown in
Fig.~\ref{Figure13:Bejanin}~(a) include attenuation owing to the series
resistance of the Ag transmission lines. The actual isolation, due only to
spurious coupling, would thus be smaller by a few decibels.

Figure~\ref{Figure13:Bejanin}~(b) shows the microwave simulations of the
crosstalk coefficients, which agree reasonably well with the experimental
results. These simulations are based on the models explained in
Subsec.~\ref{Microwave:simulations}. From simulations, we believe the isolation
is limited by the crosstalk between the CPW transmission lines, instead of the
three-dimensional wires. Note that the peaks at
approximately~\SI{7}{\giga\hertz} correspond to an enhanced crosstalk due to a
box mode in the microwave package. The peaks appear in the simulations, which
are made for a highly conductive package, and may appear in measurements
performed below~$\sim~\SI{1}{\kelvin}$, when the Al package becomes
superconductive. For the room temperature measurements shown in
Fig.~\ref{Figure13:Bejanin}~(a), these peaks are smeared out due to the highly
lossy Al package.

\section{APPLICATIONS TO SUPERCONDUCTING RESONATORS}
	\label{APPLICATIONS:TO:SUPERCONDUCTING:RESONATORS}

\begin{figure}[b!]
	\centering
	\includegraphics[width=0.99\columnwidth]{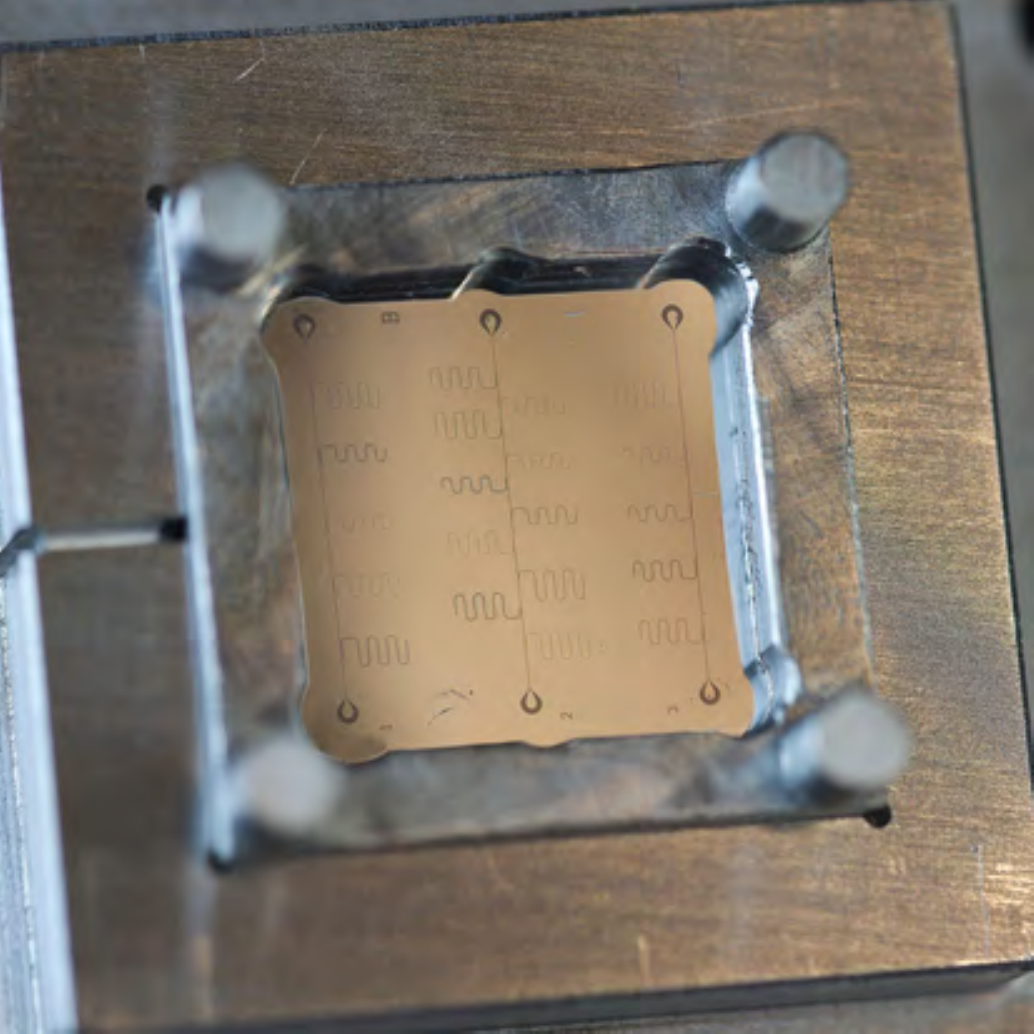}
\caption{Macro photograph of an Al sample mounted in a sample holder with
grounding washer. The image shows three CPW transmission lines each coupled to a
set of~$\lambda / 4 $-wave resonators. The grounding washer, with its four
protruding feet, is placed above the chip covering the chip edges. The marks
imprinted by the bottom interface of the three-dimensional wires on the Al pads
are noticeable. More detailed images of these marks are shown in
Fig.~\ref{Figure05:Bejanin}.}
	\label{Figure14:Bejanin}
\end{figure}

Thus far, we have shown a detailed characterization of the quantum socket in DC
and at microwave frequencies, both at room temperature and at~\SI{77}{\kelvin}.
In order to demonstrate the quantum socket operation in a realistic quantum
computing scenario, we used a socket to wire a set of superconducting CPW
resonators cooled to approximately~\SI{10}{\milli\kelvin} in a DR. We were able
to show an excellent performance in the frequency range from~\SI{4}{\giga\hertz}
to~\SI{8}{\giga\hertz}, which is the bandwidth of our measurement apparatus.

\begin{figure*}[t!]
	\centering
	\includegraphics[width=0.99\textwidth]{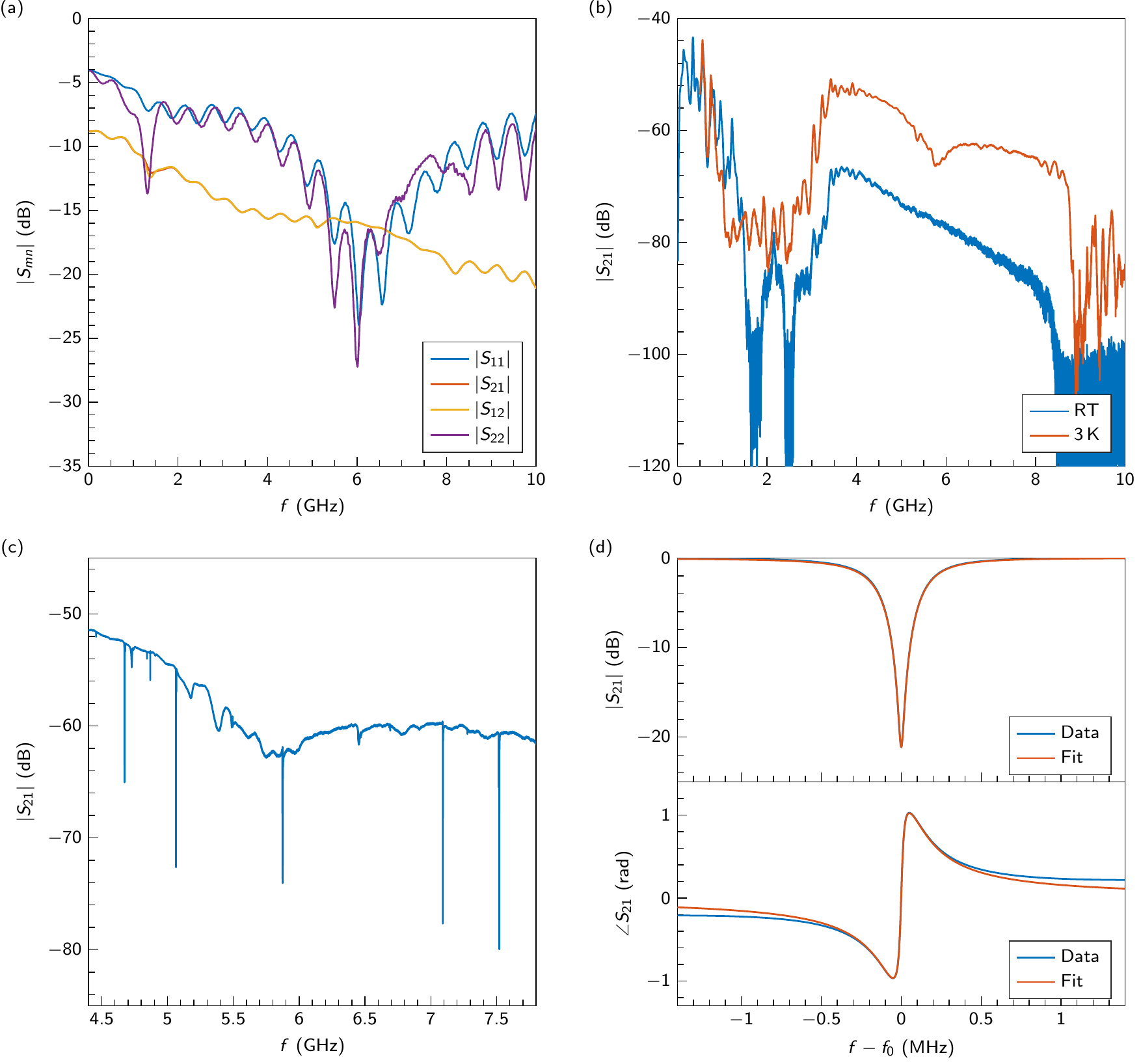}
\caption{Measurements of Al resonators. (a) Benchtop measurement of the
S-parameters of the Al sample conducted at room temperature. (b) $\abs{S_{21}}$
measurement of the sample mounted on the MC stage of the DR at room temperature
(blue) and $\sim~\SI{3}{\kelvin}$ (orange). (c) $\abs{S_{21}}$ measurement of
the sample at approximately~\SI{10}{\milli\kelvin}. The five dips correspond to
$\lambda / 4 $-wave resonators. (d) Magnitude and phase of $S_{21}$ for
resonator~$3$.}
	\label{Figure15:Bejanin}
\end{figure*}

The experimental setup is described in
Appendix~\ref{DILUTION:REFRIGERATOR:SETUP} and shown in
Fig.~\ref{Figure19:Bejanin}. Figure~\ref{Figure14:Bejanin} shows a macro
photograph of a~$\SI{15}{\milli\meter} \times \SI{15}{\milli\meter}$ chip housed
in the sample holder; the chip is the Al sample described in
Subsec.~\ref{Alignment}, with geometrical and DC electrical parameters reported
in Table~\ref{Table02:Bejanin}. The sample comprises a set of three CPW
transmission lines, each connecting a pair of three-dimensional wire pads;
multiple shunted CPW resonators are coupled to each transmission line. In this
section, we will focus only on transmission line three and its five resonators.
The transmission line has a center conductor width of~\SI{15}{\micro\meter} and
gap width of~\SI{9}{\micro\meter}, resulting in a characteristic impedance of
approximately~\SI{50}{\ohm}. The resonators are~$\lambda / 4$-wave resonators,
each characterized by a center conductor of width~$W$ and a dielectric gap of
width~$G$. The open end of the resonators runs parallel to the transmission line
for a length~$\ell_{\kappa}$, providing a capacitive coupling;
a~\SI{5}{\micro\meter} ground section separates the gaps of the transmission
line and resonators (cf.~Fig.~S2 of the Supplemental Material
at~\url{http://www.Supplemental-Material-Bejanin}). The nominal resonance
frequency~$\tilde{f}_0$ as well as all the other resonator parameters are
reported in Table~\ref{Table03:Bejanin}.

\begin{table}[t!]
\caption{Resonator parameters. The measured resonance frequency is~$f_0$. The
rescaled coupling and internal quality factors~$Q_{\textrm{c}}^*$ and
$Q_{\textrm{i}}$, respectively, are obtained from the fits of the measured
transmission coefficients (cf.~main text for details).}
\begin{center}
	\begin{ruledtabular}
		\begin{tabular}{cccccccccccccc}
		\raisebox{0mm}[3mm][0mm]{$i$} & $\tilde{f}_0$ & $f_0$ & $W$ & $G$ & 
		$\ell_{\textrm{c}}$ & $Q_{\textrm{c}}^*$ & $Q_{\textrm{i}}$ \\
		\raisebox{0mm}[0mm][2mm]{(-)} & \footnotesize{(\SI{}{\mega\hertz})} & 
		\footnotesize{(\SI{}{\mega\hertz})} & 
		\footnotesize{(\SI{}{\micro\meter})} & 
		\footnotesize{(\SI{}{\micro\meter})} & 
		\footnotesize{(\SI{}{\micro\meter})} & \footnotesize{(-)} & 
		\footnotesize{(-)} \\
		\hline
		\hline
		\raisebox{0mm}[3mm][0mm]{1} & $4600.0$ & $4673.2$ & $8$ & $5$ & $400$ & 
		$5012$ & $21243$ \\
		\hline
		\raisebox{0mm}[3mm][0mm]{2} & $5000.0$ & $5064.5$ & $15$ & $9$ & $300$ 
		& $14567$ & $165559$ \\
		\hline
		\raisebox{0mm}[3mm][0mm]{3} & $5800.0$ & $5872.9$ & $25$ & $15$ & $400$ 
		& $10269$ & $47165$ \\
		\hline
		\raisebox{0mm}[3mm][0mm]{4} & $7000.0$ & $7091.7$ & $15$ & $9$ & $300$ 
		& $6230$ & $54894$ \\
		\hline
		\raisebox{0mm}[3mm][0mm]{5} & $7400.0$ & $7520.1$ & $8$ & $5$ & $400$ & 
		$4173$ & $28353$ \\
		\vspace{-4.5mm}
		\end{tabular}
	\end{ruledtabular}
\end{center}
	\label{Table03:Bejanin}
\end{table}

A typical DR experiment employing the quantum socket consists of the following
steps. First, the sample is mounted in the microwave package, which has already
been attached to the package holder (cf.~Subsec.~\ref{Package:holder} and
Sec.~\ref{THE:QUANTUM:SOCKET:IMPLEMENTATION}). Second, a series of DC tests is
performed at room temperature. The results for a few Al samples are reported in
Table~\ref{Table02:Bejanin}. Third, the package holder assembly is characterized
at room temperature by measuring its S-parameters. The results of such a
measurement are shown in Fig.~\ref{Figure15:Bejanin}~(a). Fourth, the package
holder is mounted by means of the SMP connectors to the MC stage of the DR and
an~$S_{21}$ measurement is performed. The results (magnitude only) are shown in
Fig.~\ref{Figure15:Bejanin}~(b) in the frequency range
between~\SI{10}{\mega\hertz} and \SI{10}{\giga\hertz}. Fifth, the various
magnetic and radiation shields of the DR are closed and the DR is cooled down.
Sixth, during cooldown the~$S_{21}$ measurement is repeated first
at~$\sim~\SI{3}{\kelvin}$ and, then, at the DR base temperature of
approximately~\SI{10}{\milli\kelvin}. The results are shown in
Figs.~\ref{Figure15:Bejanin}~(b) and (c), respectively.
At~$\sim~\SI{3}{\kelvin}$ we note the appearance of a shallow dip at
approximately~\SI{5.7}{\giga\hertz}, probably due to a screw-in micro connector
becoming sightly loose while cooling
(cf.~Subsec.~\ref{Two:port:scattering:parameters}). It is important to mention
that in the next generation of three-dimensional wires we will eliminate the
screw-in micro connector, since we believe we found a technique to overcome the
soldering issues detailed in~Subsec.~\ref{Package:holder}
(cf.~Sec.~\ref{CONCLUSIONS} for a brief description). At the base temperature,
all five resonators are clearly distinguishable as sharp dips on the relatively
flat microwave background of the measurement network. We then select a narrower
frequency range around each resonator and make a finer~$S_{21}$ measurement. For
example, Fig.~\ref{Figure15:Bejanin}~(d) shows the magnitude and phase of the
resonance dip associated with resonator number~$3$.

The S-parameters of each resonator were measured with the PNA-X power set
to~\SI{10}{\deci\bel}m. Considering that the total input channel attenuation at
room temperature is~$\sim~\SI{76}{\deci\bel}$ at~\SI{5}{\giga\hertz}, the power
at the resonator input is approximately~\SI{-66}{\deci\bel}m (a
few~\SI{}{\deci\bel} higher when cold).

The normalized inverse transmission coefficient~$\tilde{S}_{21}^{-1}$ was fitted
as in Ref.~\cite{Megrant:2012}. This procedure makes it possible to accurately
estimate both the internal~$Q_{\textrm{i}}$ and the rescaled
coupling~$Q_{\textrm{c}}^*$ quality factors of a resonator. The fit results are
shown in Table~\ref{Table03:Bejanin}. The plot of the fits for the magnitude and
phase of~$S_{21}$ for resonator~$3$ are overlaid with the measured data in
Fig.~\ref{Figure15:Bejanin}~(d). The real and imaginary parts
of~$\tilde{S}_{21}^{-1}$ for the same resonator, as well as the associated fit,
are shown in Fig.~S3 in the Supplemental Material
at~\url{http://www.Supplemental-Material-Bejanin}.

\begin{figure*}[t!]
	\centering
	\includegraphics[width=0.99\textwidth]{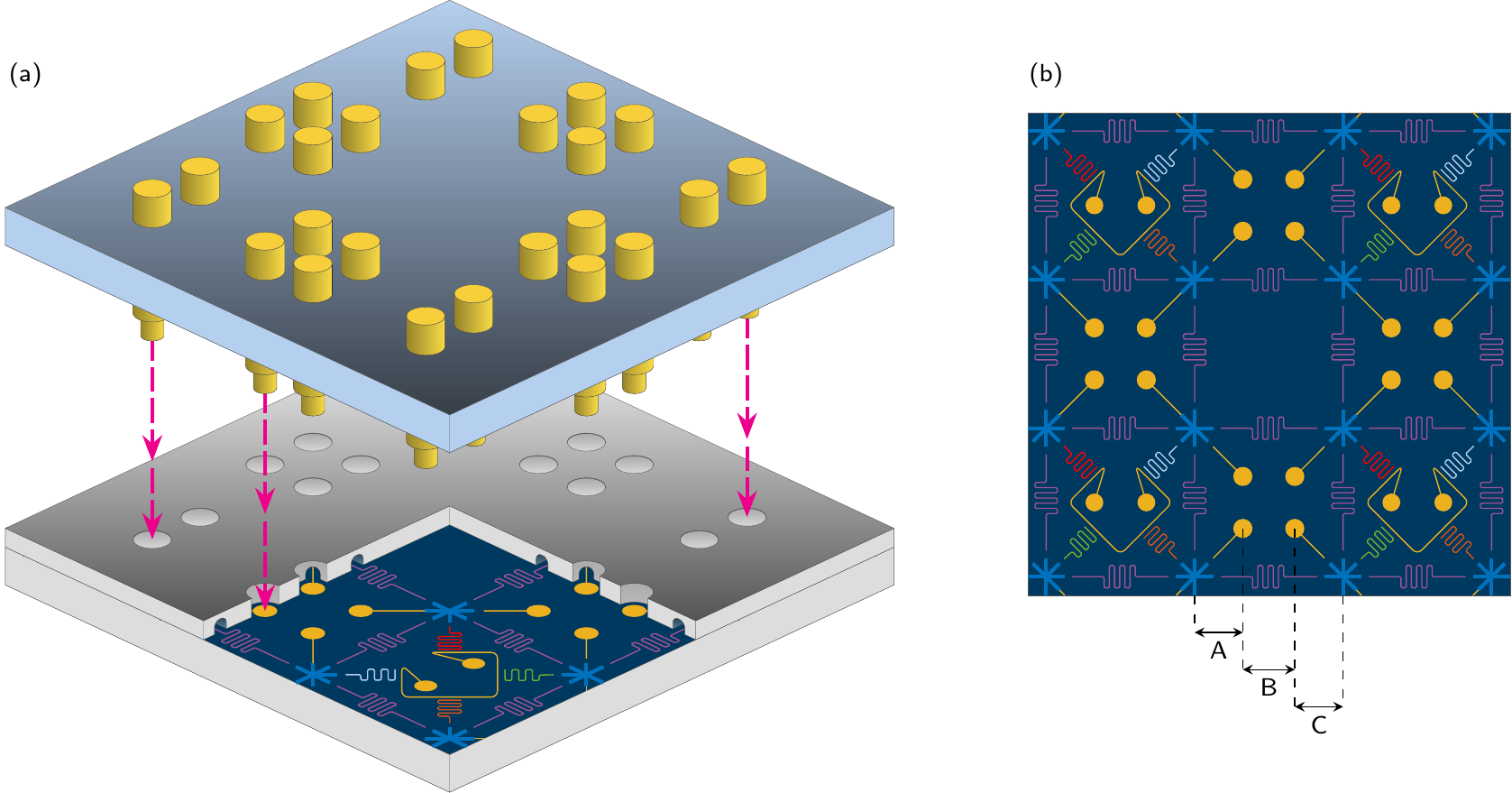}
\caption{Extensible quantum computing architecture. (a) Architecture main three
layers: The quantum hardware~(bottom); the shielding interlayer~(middle); the
three-dimensional wiring mesh~(top). The vertical magenta dashed lines with
double arrows show the mounting procedure to be used to prepare the assembly.
The middle layer~(thinner) will be metalized on the bottom and wafer bonded to
the quantum layer beneath. The back end of the wires~(top) will be connected to
SMPS connectors~(not shown). (b) Two-dimensional view of the quantum hardware.
The distances between coupling resonators and wire pdas are~$A = C =
\SI{2.25}{\milli\meter}$ and $B = \SI{3.5}{\milli\meter}$. Note that the Z
control lines are represented as galvanically connected to the heptatons,
similar to Ref.~\cite{Barends:2013}. The measurement can be multiplexed so that
four qubits are readout by one line only.}
	\label{Figure16:Bejanin}
\end{figure*}

\section{CONCLUSIONS}
	\label{CONCLUSIONS}

Figure~\ref{Figure16:Bejanin} shows an extensible quantum computing architecture
where a two-dimensional square lattice of superconducting qubits is wired by
means of a quantum socket analogous to that introduced in this work. The
architecture comprises three main layers: The quantum hardware; the shielding
interlayer; the three-dimensional wiring mesh.

As shown in Fig.~\ref{Figure16:Bejanin}~(a), the quantum hardware is realized as
a two-dimensional lattice of superconducting qubits with nearest neighbor
interactions. The qubits are a modified version of the Xmon presented in
Ref.~\cite{Barends:2013}. Each qubit is characterized by seven arms that make it
possible to connect it to one XY and one Z control line as well as one
measurement resonator and four inter-qubit coupling resonators. We name this
type of qubit the~\textit{heptaton}. The inter-qubit coupling is mediated by
means of superconducting CPW resonators that allow the implementation of
control~Z~(CZ) gates between two neighboring qubits~\cite{Haack:2010,
Mariantoni:2011}. A set of four heptatons can be readout by way of a single CPW
transmission line connected to four CPW resonators, each with a different
resonant frequency. Figure~\ref{Figure16:Bejanin} also shows the on-chip pads
associated with each three-dimensional wire. In the Supplemental Material
at~\url{http://www.Supplemental-Material-Bejanin}, we propose a more general
surface code architecture where each qubit can be measured by means of two
different resonators, one with frequency above and the other with frequency
below all coupling resonator frequencies.

Assuming a pitch between two adjacent three-dimensional wires
of~\SI{1}{\milli\meter}, the lateral dimension of one square cell having four
heptatons at its edges is~\SI{8}{\milli\meter}. The three distances~$A$, $B$,
and $C$ between wire pads and resonators leading to this quantity are indicated
in Fig.~\ref{Figure16:Bejanin}~(b). It is thus possible to construct a
two-dimensional lattice of~$10 \times 10$ heptatons on a square chip with
lateral dimension~$9 \times \SI{8}{\milli\meter} = \SI{72}{\milli\meter}$.
A~$\SI{72}{\milli\meter} \times \SI{72}{\milli\meter}$ square chip is the
largest chip that can be diced from a standard~$4$~inch wafer. This will allow
the implementation of a logical qubit based on the surface code, with at least
distance five~\cite{Fowler:2012}. In this architecture, the coupling resonators
act as a \textit{coherent spacer} between pairs of qubits, i.e., they allow a
sufficient separation to accommodate the three-dimensional wires, while
maintaining qubit coherence during the CZ gates. Additionally, these resonators
will help mitigate qubit crosstalk compared to architectures based on direct
capacitive coupling between adjacent qubits (cf.~Ref.~\cite{Kelly:2015}). In
fact, they will suppress qubit-mediated coupling between neighboring control
lines~\cite{Mariantoni:2008}. It is worth noting that adjacent coupling
resonators can be suitably designed to be at different frequencies, thus further
diminishing qubit-mediated crosstalk.

Implementing a large qubit chip with a lateral dimension
of~\SI{72}{\milli\meter} presents significant challenges to the qubit operation
at microwave frequencies. A large chip must be housed in a large microwave
package, causing the appearance of box modes that can interfere with the qubit
control and measurement sequences~\cite{Wenner:2011:a}. Moreover, a large chip
will inevitably lead to floating ground planes that can generate unwanted
slotline modes~\cite{Wenner:2011:a}. All these parasitic effects can be
suppressed by means of the shielding interlayer, as shown in
Fig.~\ref{Figure16:Bejanin}~(a). This layer can be wafer
bonded~\cite{Miller:2012, Abraham:2014:a, Brecht:2015, Rosenberg:2016} to the
quantum layer. Through holes and cavities on the bottom part of the layer can be
readily fabricated using standard Si etching techniques. The holes will house
the three-dimensional wires whereas the cavities will accommodate the qubit and
resonator structures on the quantum hardware. Large substrates also generate
chip modes that, however, can be mitigated using buried metal layers and through
vias~\cite{Abraham:2014:b}.

The three-dimensional wires to be used for the~$10 \times 10$ qubit architecture
will be an upgraded version of the wires used in this work. In particular,
the~M$2.5$ thread will be removed and the wires will be inserted in a dedicated
substrate (cf.~Fig.~\ref{Figure16:Bejanin}~(a)); additionally, the screw-in
micro connector will be substituted by a direct connection to a subminiature
push-on sub-micro~(SMPS) connector (not shown in the figure).

In future applications of the quantum socket, we envision an architecture where
the three-dimensional wires will be used as interconnect between the quantum
layer and a classical control/measurement layer. The classical layer could be
realized using RSFQ digital circuitry~\cite{Brock:2000, Mukhanov:2011}. For
example, high-sensitivity digital down-converters~(DDCs) have been fabricated
based on RSFQ electronics~\cite{Gupta:2011}. Such circuitry is operated at very
low temperatures and can substitute the room temperature electronics used for
qubit readout. Note that cryogenic DDC chips with dimensions of less
than~$\SI{5}{\milli\meter} \times \SI{5}{\milli\meter}$ can perform the same
operations presently carried out by room temperature microwave equipment with an
overall footprint of~$\sim~\SI{50}{\centi\meter} \times \SI{50}{\centi\meter}$.
Recent interest in reducing dissipation in RSFQ
electronics~\footnote{Confer~\url{https://www.iarpa.gov/index.php/research-programs/c3}
 .} will possibly enable the operation of the classical electronics in close 
proximity to the quantum hardware. We also believe it is feasible to further 
miniaturize the three-dimensional wires so that the wire outer diameter would 
be on the order of~\SI{500}{\micro\meter}. Assuming a wire-wire pitch also 
of~\SI{500}{\micro\meter}, it will therefore be possible to realize a lattice 
of~$250000$ wires connecting to~$\sim~10^5$ qubits arranged on a~$315 \times 
315$ two-dimensional qubit grid with dimensions of~$\SI{1}{\meter} \times 
\SI{1}{\meter}$. This will allow the implementation of simple fault tolerant 
operations between a few tens of logical qubits~\cite{Fowler:2012}.

\begin{acknowledgments}
Matteo Mariantoni and the Digital Quantum Matter Laboratory acknowledge the
Alfred P.~Sloan Foundation as well as the Natural Sciences and Engineering
Research Council of Canada~(NSERC), the Ministry of Research and
Innovation~(MRI) of Ontario, and the Canadian Microelectronics Corporation~(CMC)
Microsystems. Corey Rae H.~McRae acknowledges a Waterloo Institute for
Nanotechnology~(WIN) Nanofellowship-2013 and J\'{e}r\'{e}my H.~B\'{e}janin a
Provost's Entrance Award of the University of Waterloo~(UW). We also acknowledge
fruitful discussions with John M.~Martinis and the Martinis Group, Eric Bogatin,
and William D.~Oliver, as well as the group of Adrian Lupa\cb{s}cu for their
assistance in the deposition of Al films and Nathan Nelson-Fitzpatrick and the
UW's Quantum NanoFab Facility for their support. We thank Dario Mariantoni and
Alexandra V.~Bardysheva for help with a few figures.
\end{acknowledgments}

\appendix

\section{WIRE COMPRESSION}
	\label{WIRE:COMPRESSION}

In this appendix, we discuss the pressure settings of the three-dimensional
wires. In the current implementation of the quantum socket, the pressure exerted
by the three-dimensional wires on the chip is controlled by the installation
depth of the wire in the lid. This depth depends on the number of rotations used
to screw the wire into the~M$2.5$-threaded hole of the lid. Since the wire's
tunnel has to be aligned with the corresponding on-chip pad, a discrete number
of wire pressure settings is allowed. For the package shown in
Fig.~\ref{Figure01:Bejanin}~(b) and Fig.~\ref{Figure04:Bejanin}~(b), the minimum
length an unloaded wire has to protrude from the ceiling of the lid's internal
cavity to touch the chip surface is~$\ell_{\textrm{c}} =
\SI{3.05}{\milli\meter}$ (cf.~Fig.~\ref{Figure01:Bejanin}~(c)). For a maximum
wire stroke~$\Delta L = \SI{2.5}{\milli\meter}$, the maximum length an unloaded
wire can protrude from the cavity ceiling without breaking when loaded
is~$\ell_{\textrm{c}} + \Delta L = \SI{5.55}{\milli\meter}$. The first allowed
pressure setting, with wire and pad perfectly aligned, is for~$\ell_{\textrm{p}}
= \SI{3.10}{\milli\meter}$. The pitch for an~M$2.5$ screw
is~\SI{0.45}{\milli\meter}. Hence, five pressure settings are nominally
possible, for~$\ell^{}_{\textrm{p}} = 3.10 + 0.45 \, k$~mm, with~$k = 1, \, 2,
\, \ldots, \, 5$. We found the ideal pressure setting to be for~$k = 3$,
corresponding to a nominal~$\ell_{\textrm{p}} {} = {} 4.45$~mm; the actual
average setting for $12$~wires was measured to be~$\ell_{\textrm{p}} =
\SI{4.48}{\milli\meter} \mp \SI{0.28}{\milli\meter}$, with standard deviation
due to the machining tolerances. For greater depths we experienced occasional
wire damage; lesser depths were not investigated. Possible effects on the
electrical properties of the three-dimensional wires due to different pressure
settings will be studied in a future work.

\section{THE QUANTUM SOCKET MAGNETISM}
	\label{THE:QUANTUM:SOCKET:MAGNETISM}

\begin{table*}[ht]
\caption{Chemical composition (weight~\SI{}{\percent}) of the two main materials
used in the three-dimensional wires. Copper: Cu; tin: Sn; zinc: Zn; lead: Pb;
phosphorus: P.}
\begin{center}
	\begin{ruledtabular}
		\begin{tabular}{cccccccccc}
		\raisebox{0mm}[3mm][0mm]{Material} & Cu & Sn & Zn & Fe & Ni & Pb & P & 
		Si & others \\
		\hline
		\hline
		\raisebox{0mm}[3mm][0mm]{CW724R~\footnote{Confer~\url{http://www.diehl.com/en/diehl-metall/company/brands/diehl-metall-messing/ecomerica/alloys.html}
		 and 
		\url{http://www.otto-fuchs-duelken.de/fileadmin/user_upload/Downloads/OF2285_2014-02_EN.pdf}
		 .}} & $73-77$ & $0.3$ & rest & $0.3$ & $0.2$ & $\leq 0.09$ & 
		$0.04-0.10$ & $2.7-3.4$ & Al~$= 0.05$, manganese~$= 0.05$ \\
		\hline
		\raisebox{0mm}[3mm][0mm]{CW453K~\footnote{Confer~\url{http://www.steelnumber.com/en/steel_alloy_composition_eu.php?name_id=1310}
		 .}} & rest & $7.5-8.5$ & $\leq 0.2$ & $\leq 0.1$ & $\leq 0.2$ & $\leq 
		0.02$ & $0.01-0.4$ & - & $0.2$ \\
		\vspace{-4.5mm}
		\end{tabular}
	\end{ruledtabular}
\end{center}
	\label{Table04:Bejanin}
	\vspace{-3.5mm}
\end{table*}

In this appendix, we describe the measurement setup used to characterize the
magnetic properties of the materials used in the quantum socket and present the
main results. Additionally, we give an estimate of the strength of the magnetic
field caused by one three-dimensional wire inside the microwave package.

The ZGC used in our tests comprises three nested cylinders, each with a lid with
a central circular hole; the hole in the outermost lid is extended into a
chimney that provides further magnetic shielding. The walls of the ZGC are made
of an alloy of Ni and Fe (or mu-metal alloy) with a high relative
permeability~$\mu$. The alloy used for the chamber is a
CO-NETIC\textregistered~AA alloy and is characterized by a DC magnetic
permeability at~\SI{40}{\gauss}, $\mu^{40} _{\textrm{DC}} = 80000$, and an AC
magnetic permeability at~\SI{60}{\hertz} and at~\SI{40}{\gauss},
$\mu^{40}_{\textrm{AC}} = 65000$. As a consequence, the nominal magnetic field
attenuation lies between~$1000$ and $1500$. The ZGC used in our tests was
manufactured by the Magnetic Shield Corporation, model~ZG-209.

The flux gate magnetometer used to measure the magnetic field~$\vec{B}$ is a
three-axis DC milligauss meter from AlphaLab, Inc., model~MGM3AXIS. Its sensor
is a~$\SI{38}{\milli\meter} \times \SI{25}{\milli\meter} \times
\SI{25}{\milli\meter}$ parallelepiped at the end of
a~$\sim~\SI{1.2}{\meter}$~long cable; the orientation of the sensor is
calibrated to within~\SI{0.1}{\degree} and has a resolution
of~\SI{0.01}{\milli\gauss} (i.e., \SI{1}{\nano\tesla}) over a range
of~$\mp~\SI{2000}{\milli\gauss}$ (i.e., $\mp~\SI{200}{\micro\tesla}$).

The actual attenuation of the chamber was tested by measuring the value of the
Earth's magnetic field with and without the chamber in two positions, vertical
and horizontal; inside the chamber the measurements were performed a few
centimeters from the chamber base, approximately on the axis of the inner
cylinder. In these and all subsequent tests, the magnetic sensor was kept in the
same orientation and position. The results are reported in
Table~\ref{Table05:Bejanin}, which shows the type of measurement performed, the
magnitude of the measured magnetic field~$\| \vec{B} \|$, and the attenuation
ratio~$\alpha$. The maximum measured attenuation was~$\alpha \simeq 917$ in the
horizontal position.

\begin{table}[b!]
\caption{ZGC calibration. The margins of error indicated in parentheses were
estimated from the fluctuation of the magnetic sensor.}
\begin{center}
	\begin{ruledtabular}
		\begin{tabular}{ccc}
		\raisebox{0mm}[3mm][0mm]{Measurement} & $\| \vec{B} \|$ & $\alpha$ \\
		\raisebox{0mm}[0mm][2mm]{\footnotesize{(-)}} & 
		\footnotesize{(\SI{}{\milli\gauss})} & \footnotesize{(-)} \\
		\hline
		\hline
		\raisebox{0mm}[3mm][0mm]{Vertical position, background field} & 
		$554(20)$ & - \\
		\hline
		\raisebox{0mm}[3mm][0mm]{Vertical position, with ZGC} & $0.66(5)$ & 
		$842(34)$ \\
		\hline
		\raisebox{0mm}[3mm][0mm]{Horizontal position, background field} & 
		$539(20)$ & - \\
		\hline
		\raisebox{0mm}[3mm][0mm]{Horizontal position, with ZGC} & $0.59(5)$ & 
		$917(44)$ \\
		\vspace{-4.5mm}
		\end{tabular}
	\end{ruledtabular}
\end{center}
	\label{Table05:Bejanin}
\end{table}

The ZGC characterization of Table~\ref{Table05:Bejanin} also serves as a
calibration for the measurements on the materials used for the quantum socket.
In these measurements, each test sample was positioned
approximately~\SI{1}{\centi\meter} away from the magnetic sensor. The results,
which are reported in Table~\ref{Table06:Bejanin}, were obtained by taking the
magnitude of the calibrated field of each sample. The calibrated field itself
was calculated by subtracting the background field from the sample field,
component by component. Note that the background and sample fields were on the
same order of magnitude (between~\SI{0.10}{\milli\gauss} and
\SI{0.80}{\milli\gauss}), with background fluctuations on the order
of~\SI{0.10}{\milli\gauss}. Thus, we recorded the maximum value of each~$x$,
$y$, and $z$ component. Considering that the volume of the measured samples is
significantly larger than that of the actual quantum socket components, we are
confident that the measured magnetic fields of the materials should be small
enough not to significantly disturb the operation of superconducting quantum
devices. As part of our magnetism tests, we measured a block of
approximately~\SI{200}{\gram} of 5N5 Al in the ZGC; as shown in
Table~\ref{Table06:Bejanin}, the magnitude of the magnetic field was found to be
within the noise floor of the measurement apparatus~\footnote{Note that we also
performed magnetic tests by exposing all samples to a ultra-high pull neodymium
rectangular magnet, with dimensions~$\SI{25.4}{\milli\meter} {} \times {}
\SI{25.4}{\milli\meter} {} \times {} \SI{9.5}{\milli\meter}$ and a pull
of~\SI{10.4}{\kilo\gram}. We found magnetic fields with the same order of
magnitude as in Table~\ref{Table06:Bejanin}.}.

\begin{table}[b]
\caption{Magnetic field measurements of the materials used for the main
components of the quantum socket. The tested samples are significantly larger
than any component used in the actual implementation of the three-dimensional
wires and microwave package. The margins of error indicated in parentheses were
estimated from the fluctuation of the magnetic sensor.}
\begin{center}
	\begin{ruledtabular}
		\begin{tabular}{cc}
		\raisebox{0mm}[3mm][0mm]{Material} & $\| \vec{B} \|$ \\
		\raisebox{0mm}[0mm][2mm]{} & \footnotesize{(\SI{}{\milli\gauss})} \\
		\hline
		\hline
		\raisebox{0mm}[3mm][0mm]{CW724R} & $0.21(5)$ \\
		\hline
		\raisebox{0mm}[3mm][0mm]{CW453K} & $0.25(5)$ \\
		\hline
		\raisebox{0mm}[3mm][0mm]{Al 5N5} & $0.02(5)$ \\
		\vspace{-4.5mm}
		\end{tabular}
	\end{ruledtabular}
\end{center}
	\label{Table06:Bejanin}
\end{table}

A simple geometric argument allows us to estimate the actual magnetic field due
to one three-dimensional wire, without taking into account effects due to
superconductivity (most of the wire is embedded in an Al package, which is
superconductive at qubit operation temperatures). We assume that one wire
generates a magnetic field of~$0.25$~mG (i.e., the maximum field value in
Table~\ref{Table06:Bejanin}; this is a large overestimate considering the tested
samples had volumes much larger than any component in the wires) and is a
magnetic dipole positioned~$15$~mm away from a qubit. The field generated by the
wire at the qubit will then be~$B_{\textrm{q}} \simeq 0.25 \, r^{3}_{0} /
0.015^{3}~\SI{}{\milli\gauss}$, where~$r_{0} \simeq \SI{10}{\milli\meter}$ is
the distance at which the field was measured in the ZGC; thus, $B_{\textrm{q}}
\simeq 0.075~\SI{}{\milli\gauss}$. Assuming an Xmon qubit~\cite{Barends:2013}
with a superconducting quantum interference device~(SQUID) of
dimensions~$\SI{40}{\micro\meter} \times \SI{10}{\micro\meter}$, the estimated
magnetic flux due to the wire threading the SQUID is~$\Phi_{\textrm{q}} \simeq
\SI{4e-18}{\weber}$. This is approximately three orders of magnitude smaller
than a flux quantum~$\Phi_0 \simeq \SI{2.07e-15}{\weber}$; typical flux values
for the Xmon operation are on the order of~$0.5 \Phi_0$.

\section{THERMAL CONDUCTANCE OF A THREE-DIMENSIONAL WIRE}
	\label{THERMAL:CONDUCTANCE:OF:A:THREE:DIMENSIONAL:WIRE}

In this appendix, we describe the method used to estimate the thermal
performance of a three-dimensional wire and compare it to that of an Al wire
bond. Note that at very low temperature, thermal conductivities can vary by
orders of magnitude between two different alloys of the same material. The
following estimate can thus only be considered correct to within approximately
one order of magnitude. Thermal conductivity is a property intrinsic to a
material. To characterize the cooling performance of a three-dimensional wire,
we instead use the heat transfer rate (power) per kelvin difference, which
depends on the conductivity.

The power transferred across an object with its two extremities at different
temperatures depends on the cross-sectional area of the object, its length, and
the temperature difference between the extremities. Since the cross section of a
three-dimensional wire is not uniform, we assume the wire is made of two
concentric hollow cylinders. The cross-sectional area of the two cylinders is
calculated by using dimensions consistent with those of a three-dimensional
wire. The inner and outer hollow cylinders are assumed to be made of phosphor
bronze and brass alloys, respectively. The thermal conductivities of these
materials at low temperatures are determined by extrapolating measured data
to~\SI{25}{\milli\kelvin}~\footnote{Confer~\url{http://www.lakeshore.com/Documents/LSTC_appendixI_l.pdf}
 .}.

The Al wire bonds are assumed to be solid cylinders with
diameter~\SI{50}{\micro\meter}. In the superconducting state, the thermal
conductivity of Al can be estimated by extrapolating literature
values~\cite{Mueller:1978}.

The heat transfer rate per kelvin difference is calculated by multiplying the
thermal conductivity~$k_{\textrm{t}}$ with the cross-sectional area~$A$ and
dividing by the length of the thermal conductor~$\ell$. The heat transfer rate
per kelvin difference of a three-dimensional wire is calculated by summing the
heat transfer rate per kelvin difference of the inner conductor to that of the
outer conductor and is found to be~$\Pi_{\textrm{t}} \simeq
\SI{6e-7}{\watt\per\kelvin}$ at~\SI{25}{\milli\kelvin}. At the same temperature,
the heat transfer rate per kelvin of a typical Al wire bond is estimated to
be~$\Pi_{\textrm{b}} \simeq \SI{4e-12}{\watt\per\kelvin}$
(cf.~Table~\ref{Table07:Bejanin}), much lower than for a single
three-dimensional wire. Note that, instead of Al wire bonds, gold wire bonds can
be used. These are characterized by a higher thermal conductivity because they
remain normal conductive also at very low temperatures. However, Al wire bonds
remain the most common choice because easier to use.

\begin{table}[t!]
\caption{Parameters used in the estimate of the heat transfer rate per kelvin
difference for a three-dimensional wire and Al wire bond. In the table are
reported: The hollow cylinder inner diameter~$d_{\textrm{i}}$; the hollow
cylinder outer diameter and wire bond diameter~$d_{\textrm{o}}$; the hollow
cylinder and wire bond cross-sectional area~$A$; the thermal
conductivity~$k_{\textrm{t}}$. Conductor: Cond.; phosphor: Phos.}
\begin{center}
	\begin{ruledtabular}
		\begin{tabular}{ccccc}
		\raisebox{0mm}[3mm][0mm]{} & $d_{\textrm{i}}$ & $d_{\textrm{o}}$ & 
		$A$ 
		& $k_{\textrm{t}}$ \\
		\raisebox{0mm}[0mm][2mm]{} & \footnotesize{(\SI{}{\micro\meter})} & 
		\footnotesize{(\SI{}{\micro\meter})} & 
		\footnotesize{(\SI{}{\meter\squared})} & 
		\footnotesize{(\SI{}{\milli\watt\per\kelvin\per\meter})} \\
		\hline
		\hline
		\raisebox{0mm}[3.5mm][0mm]{Inner cond.} & $290$ & $380$ & $4.74 
		\times 10^{-8}$ & $3.7$ \\
		\raisebox{0mm}[3.5mm][0mm]{(Phos bronze)} & & & & \\
		\hline
		\raisebox{0mm}[3.5mm][0mm]{Outer cond.} & $870$ & $1290$ & $7.13 
		\times 10^{-7}$ & $24.1$ \\
		\raisebox{0mm}[3.5mm][0mm]{(brass)} & & & & \\
		\hline
		\raisebox{0mm}[3.5mm][0mm]{Wire bond} & - & $50$ & $1.96 \times 
		10^{-9}$ 
		& $0.01$ \\
		\raisebox{0mm}[3.5mm][0mm]{(Al)} & & & & \\
		\vspace{-4mm}
		\end{tabular}
	\end{ruledtabular}
\end{center}
	\label{Table07:Bejanin}
\end{table}

\section{THERMO-MECHANICAL TESTS}
	\label{THERMO:MECHANICAL:TESTS}

\begin{table}[b!]
\caption{Thermo-mechanical tests on hardened BeCu springs. In the table are
reported: The outer diameter~$D$ of the coil forming the helix structure of the
spring; the diameter~$d$ of the circular cross section of the spring (note that
the smallest wire diameter is~\SI{150}{\micro\meter}); the spring free
length~$L_{ \textrm{f}}$, i.e., the spring length at its relaxed position; the
number of coils~$N_{\textrm{c}}$; the spring force~$F_{\text{c}}$ (estimated at
all operating temperatures).}
\begin{center}
	\begin{ruledtabular}
		\begin{tabular}{cccccc}
		\raisebox{0mm}[3mm][0mm]{Spring type} & $D$ & $d$ & $L_{\textrm{f}}$ & 
		$N_{\textrm{c}}$ & $F_{\textrm{c}}$ \\
		\raisebox{0mm}[0mm][2mm]{} & \footnotesize{(\SI{}{\milli\meter})} & 
		\footnotesize{(\SI{}{\milli\meter})} & 
		\footnotesize{(\SI{}{\milli\meter})} & \footnotesize{(-)} & 
		\footnotesize{(\SI{}{\newton})} \\
		\hline
		\hline
		\raisebox{0mm}[3mm][0mm]{FE-113 225} & $2.30$ & $0.26$ & $11.55$ & 
		$11.25$ & $\sim 1.0$ \\
		\hline
		\raisebox{0mm}[3mm][0mm]{FE-112 157} & $1.30$ & $0.22$ & $18.00$ & 
		$42.00$ & $\sim 1.0$ \\
		\hline
		\raisebox{0mm}[3mm][0mm]{FE-50 15} & $0.60$ & $0.15$ & $31.75$ & 
		$150.00$ & $\sim 0.5$ \\
		\vspace{-4.5mm}
		\end{tabular}
	\end{ruledtabular}
\end{center}
	\label{Table08:Bejanin}
\end{table}

In this appendix, we discuss the performance of the springs used in
three-dimensional wires at various temperatures. The three types of tested
springs are called FE-113 225, FE-112 157, and FE-50 15 and their geometric
characteristics are reported in Table~\ref{Table08:Bejanin}. We ran temperature
cycle tests by dunking the springs repeatedly in liquid nitrogen and then in
liquid helium without any load. At the end of each cycle, we attempted to
compress them at room temperature. We found no noticeable changes in mechanical
performance after many cooling cycles. Subsequently, the springs were tested
mechanically by compressing them while submerged in liquid nitrogen or helium.
The setup used for the compressive loading test of the springs is shown in
Movie~4 of the Supplemental Material
at~\url{http://www.Supplemental-Material-Bejanin}, which also shows a properly
functioning spring immediately after being cooled in liquid helium. In these
tests, we only studied compression forces because in the actual experiments the
three-dimensional wires are compressed and not elongated.

The compression force was assessed by means of loading the springs with a mass.
The weight of the mass that fully compressed the spring determined the spring
compression force~$F_{\textrm{c}}$. The compression force of each spring is
reported in Table~\ref{Table08:Bejanin}. We observed through these tests that
the compression force is nearly independent of the spring temperature,
increasing only slightly when submerged in liquid helium. Assuming an operating
compression~$\Delta L = \SI{2.0}{\milli\meter}$, we expect a force
between~\SI{0.5}{\newton} and \SI{2.0}{\newton} for the inner conductor and
between \SI{2.0}{\newton} and \SI{4.0}{\newton} for the outer conductor of a
three-dimensional wire at a temperature of~\SI{10}{\milli\kelvin}. Note that we
chose spring model FE-113 225 for use with the grounding washer.

\section{ALIGNMENT ERRORS}
	\label{ALIGNMENT:ERRORS}

\begin{figure}[ht]
\centering
	\includegraphics[width=0.49\textwidth]{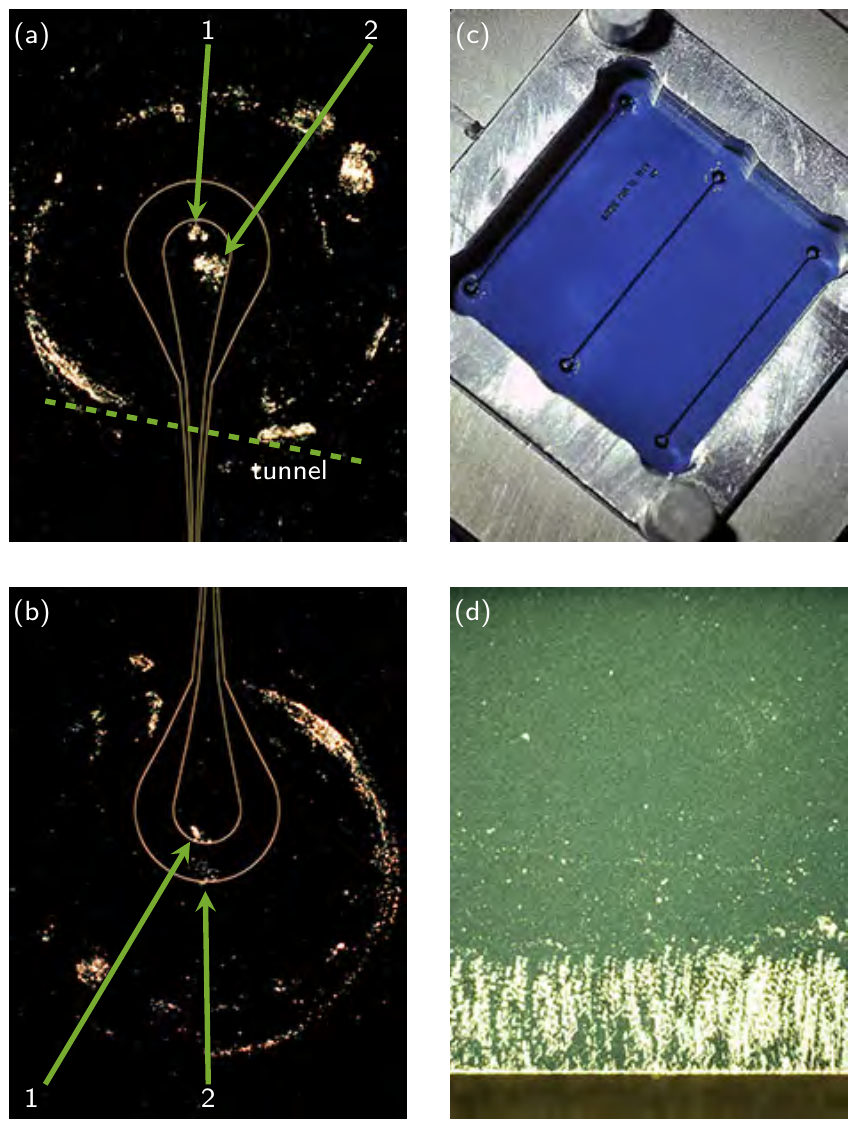}
\caption{Micro images showing three-dimensional wire alignment errors. (a)-(b)
Au pads. The pad displayed in~(a) is connected to that in~(b) by way of a CPW
transmission line approximately~\SI{11.5}{\milli\meter}~long. The die shifted
upward between the first~(green arrow~(1)) and second~(green arrow~(2)) mating
instance, resulting in a lateral misalignment for the bottom pad. The rotational
misalignment for the pad in~(a) is indicated by a dashed green line. (c)
Successful alignment for six Ag pads on the same chip. (d) Peripheral area of an
Ag sample (ground plane). The marks are due to contact with the grounding
washer.}
	\label{Figure17:Bejanin}
\end{figure}

In this appendix, we provide more details about alignment errors.
Figure~\ref{Figure17:Bejanin} shows a set of micro images for Au and Ag samples.
The Au pads in panels~(a) and (b) were mated two times at room temperature; the
three-dimensional wires used to mate these pads featured the smaller
tunnel~(\SI{500}{\micro\meter} width). The pad dimensions were~$W_{\textrm{p}} =
\SI{230}{\micro\meter}$ and $T_{\textrm{p}} = \SI{1000}{\micro\meter}$.
Noticeably, in panel~(a) the wire bottom interface matched the contact pad in
both mating instances, even though the matching was affected by a rotational
misalignment of approximately~$\SI{15}{\degree}$ with respect to the
transmission line longitudinal axis. In panel~(b) the inner conductor landed on
the dielectric gap in the second mating instance.

In our initial design, a perfect match required that the die dimensions should
be at most $1$~thou smaller than the dimensions of the chip recess, as machined.
In the case of the sample holder used to house the Au samples, the chip recess
side lengths were~\SI{15.028(5)}{\milli\meter}, \SI{15.030(5)}{\milli\meter},
\SI{15.013(5)}{\milli\meter}, and \SI{15.026(5)}{\milli\meter}. The Au samples
were diced from a Si~wafer using a dicing saw from DISCO, model DAD-2H/6, set to
obtain a~$\SI{15}{\milli\meter} \times \SI{15}{\milli\meter}$ die. Due to the
saw inaccuracies, the actual die dimensions were~$\SI{14.96(1)}{\milli\meter}
\times \SI{14.96(1)}{\milli\meter}$, significantly smaller than the chip recess
dimensions. This caused the die to shift randomly between different mating
instances, causing alignment errors.

As described in the main text, in order to minimize such errors a superior DISCO
saw was used, in combination with a DISCO electroformed bond hub diamond blade
model~ZH05-SD 2000-N1-50-F E; this blade corresponds to a nominal kerf
between~\SI{35}{\micro\meter} and \SI{40}{\micro\meter}. Additionally, we used
lateral markers spaced with increments of~\SI{10}{\micro\meter} that allowed us
to cut dies with dimensions ranging from~\SI{14.97}{\milli\meter} to
\SI{15.03}{\milli\meter}, well within the machining tolerances of the sample
holder. After machining, the actual inner dimensions of each sample holder were
measured by means of a measuring microscope. The wafers were then cut by
selecting the lateral dicing markers associated with the die dimensions that fit
best the holder being used.

Figure~\ref{Figure17:Bejanin}~(c) shows a successful alignment for six Ag pads
on the same chip; the chip is mounted in a sample holder with grounding washer.
All three main steps for an ideal and repeatable alignment
(cf.~Subsec.~\ref{Alignment}) were followed. Figure~\ref{Figure17:Bejanin}~(d)
shows the distinctive marks left by the grounding washer on an Ag film. The
marks are localized towards the edge of the die; the washer covered
approximately~\SI{500}{\micro\meter} of Ag film. This indicates a good
electrical contact at the washer-film interface.

In conclusion, it is worth commenting some of the features in
Fig.~\ref{Figure05:Bejanin}~(d) in the main text. The figure clearly shows
dragging of a three-dimensional wire due to cooling contractions. In fact, for
the Al chip recess an estimate of the lateral contraction length from room
temperature to~$\sim~\SI{4}{\kelvin}$ can be obtained as~$\Delta L_{\textrm{Al}}
= \alpha ( 4 ) L_{\textrm{Al}} \simeq \left( 4.15 \times 10^{-3} \right) \left(
\SI{15e-3}{\meter} \right) \simeq \SI{62}{\micro\meter}$, where~$\alpha ( 4 )$
is the integrated linear thermal expansion coefficient for
Al~6061-T6~\footnote{ISO~AlMg1SiCu; UNS~A96061.} at~\SI{4}{\kelvin} from
Refs.~\cite{Marquardt:2000} and $L_{\textrm{Al}}$ is the room temperature length
of the recess side~\footnote{Note that~$\alpha ( 4 )$ can be accurately
estimated from the data
at~\url{http://www.cryogenics.nist.gov/MPropsMAY/6061\%20Aluminum/6061_T6Aluminum_rev.htm}
 .}. Note that the sample holder is actually made from Al alloy 5N5; however, 
different Al alloys contract by approximately the same quantity. For the Si 
sample substrate, the lateral contraction length from room temperature 
to~$\sim~\SI{4}{\kelvin}$ is approximately given by~$\Delta L_{\textrm{Si}} 
\simeq \SI{3.2}{\micro\meter}$, where the integrated linear thermal expansion 
coefficient at~\SI{4}{\kelvin} was found in Table~2 of 
Ref.~\cite{Swenson:1983}. Below \SI{4}{\kelvin}, the thermal expansion of both 
materials is negligible for our purposes and, thus, the~\SI{4}{\kelvin} 
estimate can also be considered to be valid at~$\sim~\SI{10}{\milli\kelvin}$.

\begin{figure*}[ht]
	\centering
	\includegraphics[width=0.99\textwidth]{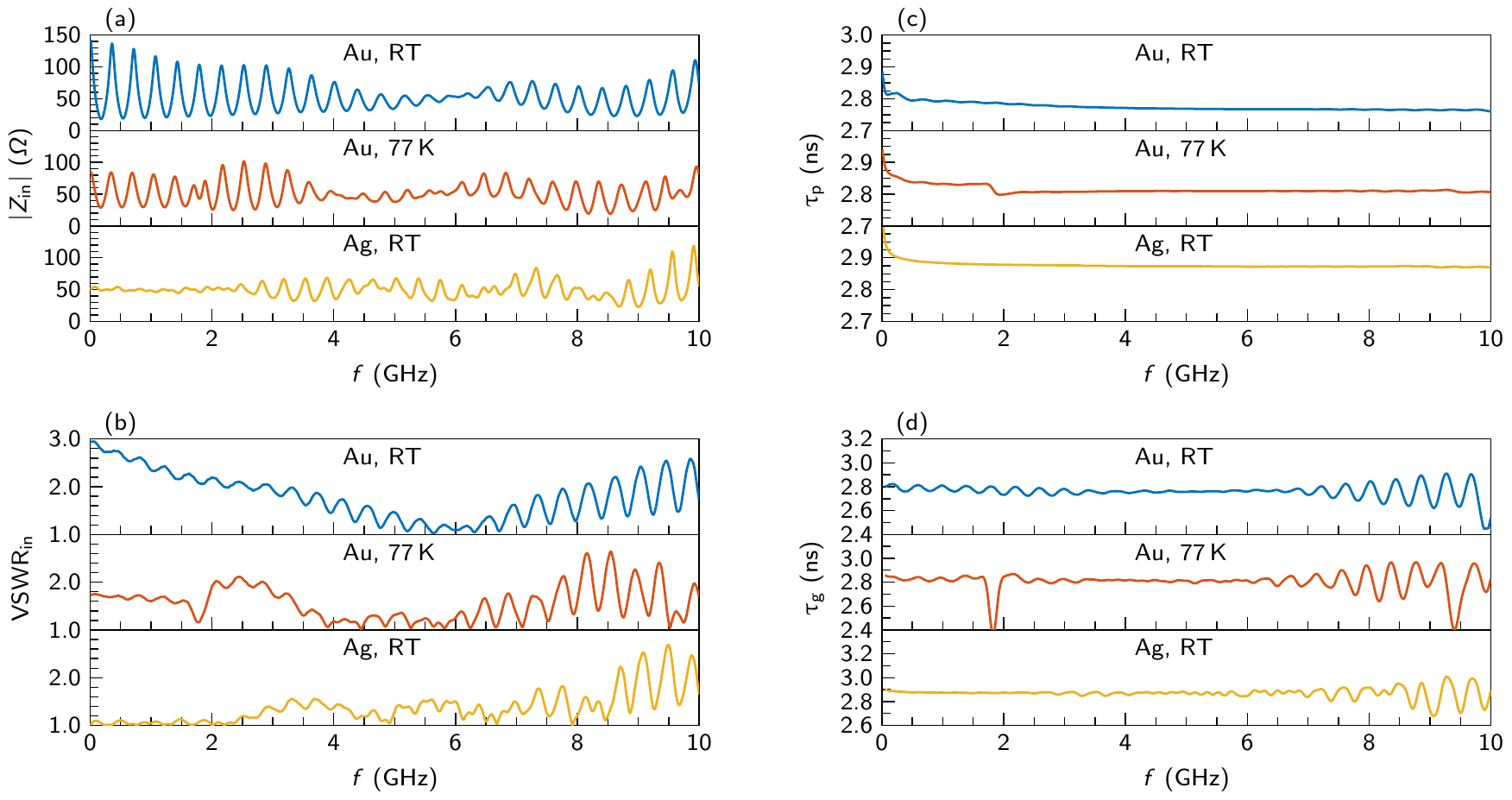}
\caption{Quantum socket microwave parameters. (a) Input impedance
magnitude~$\abs{Z_{\textrm{in}}}$. (b) Input VSWR,
$\textrm{VSWR}_{\text{\textrm{in}}}$. (c) Phase delay~$\tau^{}_{\phi}$. (d)
Group delay~$\tau^{}_{\textrm{g}}$. Blue corresponds to the Au sample at room
temperature~(RT), red to the Au sample at~\SI{77}{\kelvin}, and orange to the Ag
sample at room temperature.}
	\label{Figure18:Bejanin}
\end{figure*}

\section{SAMPLE FABRICATION}
	\label{SAMPLE:FABRICATION}

In this appendix, we outline the fabrication processes for the samples used to
test the quantum socket. A set of samples was made by liftoff of
a~$\sim~\SI{3}{\micro\meter}$ Ag film, which was grown by means of electron beam
physical vapor deposition~(EBPVD; from Intlvac Canada Inc., model~Nanochrome~II)
on a $3$~inch float-zone~(FZ) Si~$(100)$ wafer of
thickness~\SI{500}{\micro\meter}. The superconducting samples were made by
etching a~$\sim~\SI{120}{\nano\meter}$ Al film that was deposited by EBPVD on
a~\SI{500}{\micro\meter} FZ Si wafer. Last, two sets of test samples were made
by etching Au films of thickness~\SI{100}{\nano\meter} and \SI{200}{\nano\meter}
with a~\SI{10}{\nano\meter} Ti adhesion underlayer in both sets. The films were
grown by EBPVD on a $3$~inch Czochralski~(CZ) undoped Si~$(100)$ wafer of
thickness~\SI{500}{\micro\meter}.

The~\SI{3}{\micro\meter} Ag samples were required to reduce the series
resistance of the CPW transmission lines
(cf.~Subsecs.~\ref{Two:port:scattering:parameters},
\ref{Time:domain:reflectometry}, and \ref{Signal:crosstalk}). Fabricating such a
relatively thick film necessitated a more complex process as compared to that
used for the Au and Al samples. The Ag samples were fabricated with a thick
resist tone reversal process. The wafer was spun with an~AZ~P4620 positive tone
resist to create a resist thickness of~$\sim~\SI{14}{\micro\meter}$, then soft
baked for~\SI{4}{\minute} at~\SI{110}{\celsius}. Because the resist layer is so
thick, a rehydration step of~\SI{30}{\minute} was necessary before exposure.
Optical exposure was performed for~\SI{30}{\second} in a mask aligner from
S{\"U}SS MicroTec AG, model MA6, in soft contact with a photomask. After
exposure the sample was left resting for at least~\SI{3}{\hour} so that any
nitrogen created by the exposure could dissipate. The tone reversal bake was
done for~\SI{45}{\minute} in an oven set to~\SI{90}{\celsius}, filled with
ammonia gas. The sample then underwent a flood exposure for~\SI{60}{\second} and
was developed in AZ\textregistered~400K for~\SI{15}{\minute}. Subsequently,
\SI{3}{\micro\meter} of Ag was deposited and liftoff of the resist was performed
in acetone for~\SI{5}{\minute} with ultrasounds.

\section{MICROWAVE PARAMETERS}
	\label{MICROWAVE:PARAMETERS}

In this appendix, we present a set of microwave parameters that help further
analyze the performance of the quantum socket. These parameters were obtained
from the measured S-parameter data of Figs.~\ref{Figure09:Bejanin} and
\ref{Figure10:Bejanin}~(a) and are shown in Fig.~\ref{Figure18:Bejanin}. The
complex input impedance can be obtained from the frequency dependent impedance
matrix~$\textbf{Z} = [ Z_{mn}
]$as~\footnote{Confer~\url{http://www.ece.rutgers.edu/~orfanidi/ewa/} .}
\begin{equation}
Z_\textrm{{in}} = Z_{11} - \frac{Z_{12} Z_{21}}{Z_{22} - Z_\textrm{L}} \quad ,
	\label{Equation:04}
\end{equation}
where~$Z_{\textrm{L}} = Z_{\textrm{c}} = \SI{50}{\ohm}$ is the load impedance.
The impedance matrix was obtained using the measured complex S-parameter
matrix~$\textbf{S} = [ S_{mn} ]$ from
\begin{equation}
\textbf{Z} = \sqrt{\textrm{Z}_{\textrm{c}}} \left( \begin{bmatrix}
1 & 0 \\
0 & 1
\end{bmatrix} + \textbf{S} \right) \left( \begin{bmatrix}
1 & 0 \\
0 & 1
\end{bmatrix} - \textbf{S} \right)^{-1} \sqrt{\textrm{Z}_{\textrm{c}}} \quad .
	\label{Equation:05}
\end{equation}
The magnitude of~$Z_{\textrm{in}}$ is shown in Fig.~\ref{Figure18:Bejanin}~(a).

The input voltage standing wave ratio~(VSWR) was obtained from~\cite{Pozar:2011}
\begin{equation}
\textrm{VSWR}_{\text{\textrm{in}}} = \dfrac{1 + \left| S_{11} \right|}{1 - 
\left| S_{11} \right|}
	\label{Equation:06}
\end{equation}
and is displayed in Fig.~\ref{Figure18:Bejanin}~(b).

The phase delay was calculated
as~\footnote{Confer~\url{http://www.ece.rutgers.edu/~orfanidi/ewa/} .}
\begin{equation}
\tau^{}_{\phi} = -\dfrac{1}{2 \pi} \, \dfrac{\angle S_{21}}{f}
	\label{Equation:07}
\end{equation}
and is displayed Fig.~\ref{Figure18:Bejanin}~(c).

Finally, the group delay was obtained from~\cite{Pozar:2011}
\begin{equation}
\tau^{}_{\textrm{g}} = -\dfrac{1}{2 \pi} \, \dfrac{\partial}{\partial f} ( 
\angle S_{21} )
	\label{Equation:08}
\end{equation}
and is displayed in Fig.~\ref{Figure18:Bejanin}~(d). The derivative in
Eq.~(\ref{Equation:08}) was evaluated numerically by means of central finite
differences with 6th~order accuracy. The data in Fig.~\ref{Figure18:Bejanin}~(d)
were post-processed using~$1\%$ smoothing. Note that the output impedance and
VSWR were also evaluated and resembled the corresponding input parameters.

The input and output impedances as well as the VSWRs indicate a good impedance
matching up to approximately~\SI{8}{\giga\hertz}. The phase and group delays,
which are directly related to the frequency dispersion associated with the
quantum socket, indicate minimal dispersion. This is expected for a combination
of coaxial structures (the three-dimensional wires) and a CPW transmission line.
Thus, we expect wideband control pulses to be transmitted without significant
distortion in applications with superconducting qubits (cf.~Supplemental
Material at~\url{http://www.Supplemental-Material-Bejanin} for further details
about microwave pulse transmission).

\section{DILUTION REFRIGERATOR SETUP}
	\label{DILUTION:REFRIGERATOR:SETUP}

\begin{figure}[t!]
	\centering
	\includegraphics[width=0.36\textwidth]{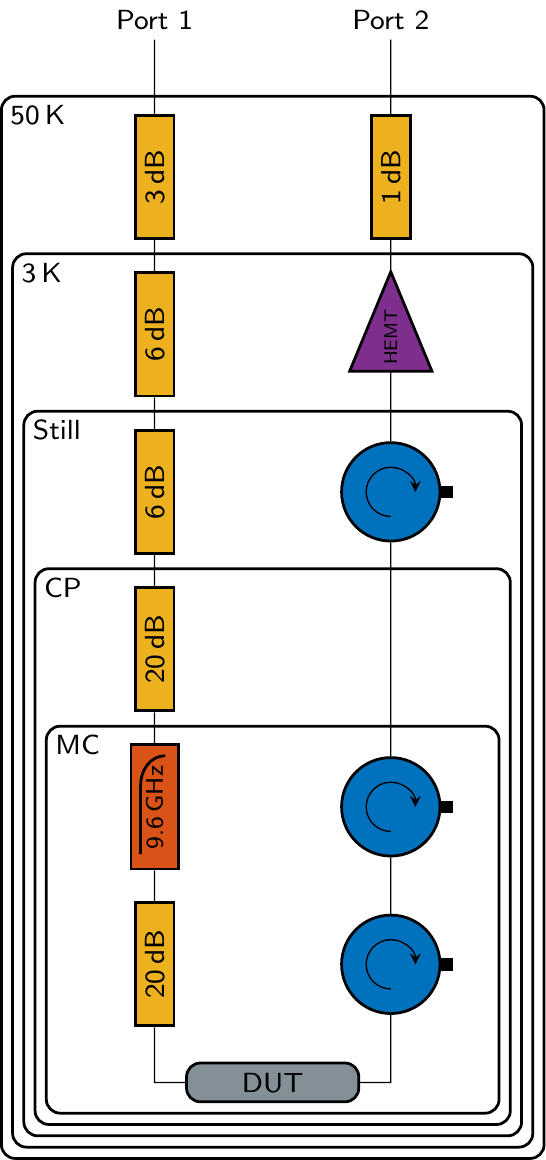}
\caption{DR setup. Infrared radiation shields~\cite{Barends:2011, Corcoles:2011}
and magnetic shields are used, but not shown.}
	\label{Figure19:Bejanin}
\end{figure}

The experimental setup used to measure the superconducting CPW resonators is
shown in Fig.~\ref{Figure19:Bejanin}. The low-temperature system is a
cryogen-free DR from BlueFors Cryogenics Ltd., model~BF-LD250. The DR comprises
five main temperature stages, where microwave components and samples can be
thermally anchored: The RT, \SI{50}{\kelvin}, \SI{3}{\kelvin}, still
(${\sim}\SI{800}{\milli\kelvin}$), cold plate (CP;
${\sim}\SI{50}{\milli\kelvin}$), and MC stage. We will describe the setup
following the input signal through the various temperature stages, from port~$1$
to the input port of the microwave package (where the resonator sample is
mounted) and from the output port of the package to port~$2$. The two ports are
connected to the PNA-X, which serves as both the microwave source and readout
apparatus. Port~$1$ is connected to the RT stage of the DR with SucoFlex
flexible cables followed by a series of two semi-rigid coaxial cables from EZ
Form, model EZ~86-Cu-TP/M17 (each approximately \SI{1.2}{\meter}~long, with
silver-coated copper center conductor, solid PTFE dielectric, and tin-plated
seamless copper outer conductor). Except for the PNA-X ports, which
feature~\SI{3.5}{\milli\meter} connectors, all the connectors and bulkhead
adapters are SMA type. In particular, the RT stage of the DR features a set of
hermetic SMA bulkhead adapters from Huber+Suhner, model 34\_SMA-50-0-3/111\_N,
with a tested leak rate for helium-4 lower
than~\SI{1e-9}{\milli\bar\litre\per\second}.

The DR stages, all the way to the MC stage, are connected by the series of five
semi-rigid coaxial cables from Coax Co., Ltd., model SC-219/50-SS-SS (with
stainless steel~(SUS304) center and outer conductor and solid PTFE dielectric;
the cable lengths from RT to MC are: \SI{39.6}{\centi\meter},
\SI{48.0}{\centi\meter}, \SI{39.9}{\centi\meter}, \SI{27.4}{\centi\meter}, and
\SI{20.0}{\centi\meter}, respectively). The cables are thermalized to the DR
stages by way of cryogenic attenuators from the XMA Corporation-Omni
Spectra\textregistered, model 2082-6418-XX-CRYO, where XX is the attenuation
level in dB; for each stage between RT and MC, we chose~$\textrm{XX} = 03 , 06,
06, 20$, and $20$, respectively. The input signals are filtered by means of a
low-pass filter from Marki Microwave, Inc., model FLP-0960-2S, with bandpass
from DC to~\SI{9.6}{\giga\hertz}. The filter is heat sunk at the MC stage by
anchoring it to a hardware module, which is bolted to the MC stage. The filter
module, and similarly all the other modules used to heat sink microwave
components in the DR, are made from C10100 OFE copper alloy.

A non-magnetic semi-rigid coaxial cable EZ~86-Cu-TP/M17 connects the output port
of the Marki filter to an SMP~086 connector on the mounting plate; the cable
is~\SI{18}{\centi\meter} long and enters the A4K shield through one of the
chimneys on the shield lid. The shield, which is thermalized to the MC stage, is
characterized by a DC relative permeability close to~$80000$ at~\SI{4}{\kelvin}.
The SMP~086 connector is mated to the input port of the DUT shown schematically
in Fig.~\ref{Figure08:Bejanin}. The DUT used in the DR features SMP~047
connectors in lieu of SMA connectors. The DUT when connected to the mounting
plate is shown in Fig.~\ref{Figure01:Bejanin}~(d) and
Fig.~\ref{Figure04:Bejanin}~(c).

The output port of the DUT is then connected to a series of two cryogenic
circulators from Raditek Inc., model RADC-4.0-8.0-Cryo-4-77K-S3-1WR-b (with
\mbox{CryoPerm} magnetic shielding) by means of a semi-rigid superconducting
coaxial cable from Coax Co., model SC-219/50-Nb-Nb, of
length~\SI{17.3}{\centi\meter}. The circulators are thermalized to the MC stage
and are connected to each other by means of a semi-rigid superconducting coaxial
cable from Coax Co., model SC-219/50-Nb-Nb, of length~\SI{20.7}{\centi\meter};
the spare port of each circulator is terminated with an XMA
cryogenic~\SI{50}{\ohm} load, model 2001-7010-02-CRYO, which is thermalized to
the MC stage. The output port of the second circulator is connected by way of a
\SI{51.9}{\centi\meter}~long SC-219/50-Nb-Nb cable to a third circulator at the
still stage (the spare port is terminated with a \SI{50}{\ohm}~load thermalized
to the still). A \SI{42.7}{\centi\meter}~long SC-219/50-Nb-Nb cable connects the
output port of the third circulator to a cryogenic microwave amplifier from Low
Noise Factory AB, model LNF-LNC1\_12A. The amplifier, which is thermalized to
the~\SI{3}{\kelvin} stage, is characterized by a nominal gain of
approximately~\SI{39}{\deci\bel} and a noise temperature of~\SI{5}{\kelvin} at
an operating temperature of~\SI{12}{\kelvin} in the~\SI{4}{\giga\hertz} to
\SI{8}{\giga\hertz} frequency range. Finally, the amplifier output port is
connected to the~\SI{50}{\kelvin} and RT stages by a series of two
SC-219/50-SS-SS cables of length~\SI{38.6}{\centi\meter} and
\SI{31.2}{\centi\meter}, respectively; the cables are thermalized to
the~\SI{50}{\kelvin} stage by means of a \SI{1}{\deci\bel}~XMA attenuator. Two
EZ Form copper cables in series, followed by SucoFlex flexible cables, complete
the network to port~$2$.

The input channel described here is one of three equivalent channels dedicated
to resonator measurements. The three channels share the output line; this is
possible thanks to a microwave switch from Radiall, model~R573.423.605, which is
operated at the MC stage. The switch is located after the DUT but before the two
MC circulators (two of the three input channels and the switch are not shown in
Fig.~\ref{Figure19:Bejanin}). The switch has six inputs and one output, making
it possible to further extend the number of input microwave channels.


%

\clearpage

\pagebreak

\begin{center}
\textbf{\large Supplemental Material for ``The Quantum Socket: Three-Dimensional
Wiring for Extensible Quantum Computing''}
\end{center}

\newcommand{\beginsupplement}{%
	\setcounter{section}{0}
	\renewcommand{\thesection}{S\arabic{section}}%
	\setcounter{table}{0}
	\renewcommand{\thetable}{S\arabic{table}}%
	\setcounter{figure}{0}
	\renewcommand{\thefigure}{S\arabic{figure}}%
	\setcounter{equation}{0}
	\renewcommand{\theequation}{S\arabic{equation}}%
	}
\beginsupplement

\setcounter{page}{1}
\makeatletter

\renewcommand{\bibnumfmt}[1]{[S#1]}
\renewcommand{\citenumfont}[1]{S#1}

\section*{S1: INTRODUCTION}
	\label{INTRODUCTION}

This Supplemental Material is organized as follows. In
Sec.~\ref{ANIMATED:ELECTRIC:FIELD:SIMULATIONS}, we show a set of simulated
electric field animations. In Sec.~\ref{SCREW:IN:MICRO:CONNECTOR:ANOMALY}, we
characterize the microwave anomaly due to the screw-in micro connector. In
Sec.~\ref{SUPERCONDUCTING:RESONATORS:COUPLING:AND:FIT}, we discuss in detail the
geometry of the coupling region between a CPW transmission line and a
superconducting resonator; additionally, we show the fit required to extract the
internal quality factor of the resonators. In Sec.~\ref{SPRING:TESTS}, we
present a movie of the cryogenic spring tests. In
Sec.~\ref{QUANTUM:SOCKET:SURFACE:CODE}, we propose a surface code architecture
that makes use of two readout resonators per qubit. Finally, in
Sec.~\ref{MICROWAVE:PULSE:TRANSMISSION}, we present the quantum socket behavior
when fast microwave pulses are applied.

\section*{S2: ANIMATED ELECTRIC FIELD SIMULATIONS}
	\label{ANIMATED:ELECTRIC:FIELD:SIMULATIONS}

The electric fields discussed in Subsec.~II~D of the main text were generated
for multiple phases, enabling a time-domain analysis. This made it possible to
better determine the behavior of the simulated fields associated with the
various components of the quantum socket. The time-domain animations of the
simulated electric field for the three-dimensional wire, the~\SI{90}{\degree}
transition between the wire and the on-chip pad, and the first box mode are
shown in Movies~1, 2, and 3, respectively.

The simulation of the three-dimensional wire shows a clear impact to the
electric field at the dielectric spacers, giving an indication of where
potential future improvements of the wire should focus.

The simulation of the~\SI{90}{\degree} transition reveals a surprisingly clean
transmission, with no immediately apparent regions of poor performance. The
broad field distribution below the contact pad does suggest the appearance of a
stray capacitance to the cavity below. However, the TDR results in Subsec.~IV~C
indicate that such a capacitance must be small because of a very good impedance
matching at the transition region.

The box mode animation shows the ``pinching'' effect to the electric field
caused by the metallic pillar, leading to a significantly higher relative
permittivity and, in turn, a much lower frequency of the first box mode
($\sim~\SI{6.3}{\giga\hertz}$). Some other regions of interest in Movie~3 are
the outer edges of the chip recess, where the field becomes fully confined
within the substrate. As the magnitude of the field in this region is quite low,
this effect should have only a marginal impact on the performance of the quantum
socket. The animation also reveals some field interaction between the box mode
and the contact pads (which are characterized by a relatively large area),
suggesting these as the main reason for any spurious coupling between on-chip
structures and box modes.

\section*{S3: SCREW-IN MICRO CONNECTOR ANOMALY}
	\label{SCREW:IN:MICRO:CONNECTOR:ANOMALY}

\begin{figure}[t!]
	\centering
	\includegraphics[width=0.99\columnwidth]{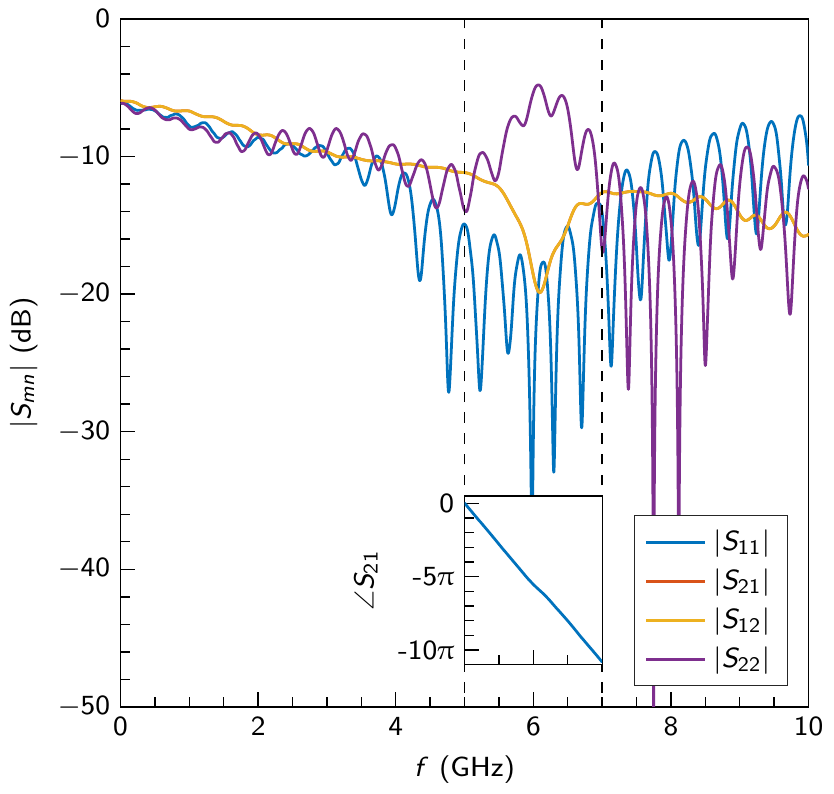}
\caption{S-parameter measurements of the Au sample at room temperature in the
case of a loose screw-in micro connector. The inset shows the unwrapped phase
angle~$\angle S_{21}$; the black dashed lines delimit the frequency region
between~\SI{5}{\giga\hertz} and \SI{7}{\giga\hertz}. The dip at
approximately~\SI{6}{\giga\hertz} corresponds to the loose micro-connector.}
	\label{Figure01S:Bejanin}
\end{figure}

In order to gain further insight into the electrical behavior of the screw-in
micro connector, we report here a set of room temperature measurements showing
the appearance of a microwave anomaly associated with the micro connector. These
measurements complement the discussion associated with Fig.~9~(b) in
Subsec.~IV~B of the main text.

Figure~\ref{Figure01S:Bejanin} shows the magnitude and unwrapped phase angle of
the S-parameters as a function of frequency for an Au sample in the case of a
slightly loose screw-in micro connector; the measurement was performed at room
temperature (cf.~ Subsec.~IV~B of the main text for details on the DUT used for
the measurements). The microwave dip is centered at
approximately~\SI{6}{\giga\hertz} and has a~\SI{3}{\deci\bel} bandwidth of
approximately~\SI{400}{\mega\hertz}. The dip is likely not a Lorentzian-type
feature due to a resonance mode in the package. This is clearly demonstrated by
the phase angle of the coefficients~$S_{21}$ and $S_{12}$, which has the
characteristic frequency dependence of a transmission line (i.e., no phase shift
associated with a resonance). Note that a similar behavior is encountered when
performing a microwave measurement with, e.g., a not well torqued SMA connector.
As expected from energy conservation, the dip in~$S_{21}$ corresponds to a peak
in~$S_{11}$; however, the absence of a peak in~$S_{22}$ indicates that only the
micro connector at the DUT input line was loose in this instance. As explained
in the main text, curing the dip at room temperature is rather straightforward.
However, it is much harder at low temperatures. Hence, we decided to cool down a
DUT without curing the dip that appeared at room temperature. A set of data
at~\SI{77}{\kelvin} was taken immediately after the room temperature data, as
shown in Fig.~9~(b) in Subsec.~IV~B of the main text. We monitored the
S-parameters throughout the cooldown, observing how the dip shifted from
approximately~\SI{6}{\giga\hertz} to \SI{2}{\giga\hertz} while cooling. It is
reasonable to assume that, when cooling down, the micro connector became
slightly more loose due to thermal contractions, thus giving rise to a microwave
transmission dip at lower frequency.

\section*{S4: SUPERCONDUCTING RESONATORS COUPLING AND FIT}
	\label{SUPERCONDUCTING:RESONATORS:COUPLING:AND:FIT}

\begin{figure}[b!]
	\centering
	\includegraphics[width=0.99\columnwidth]{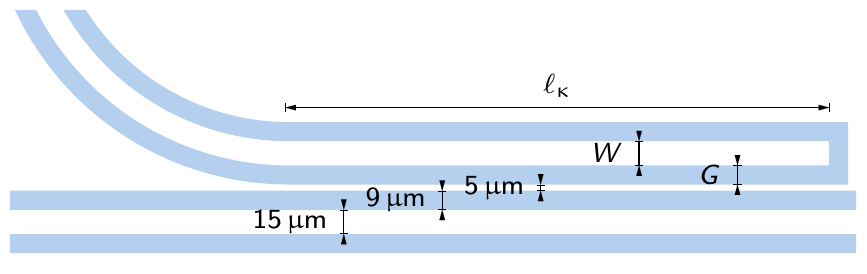}
\caption{Coupling region between a CPW transmission line and a~$\lambda /
4$-wave resonator. White surfaces represent conductive material and blue
surfaces dielectrics.}
	\label{Figure02S:Bejanin}
\end{figure}

Figure~\ref{Figure02S:Bejanin} shows a detail of the coupling region between one
of the CPW transmission lines and a~$\lambda / 4$-wave resonator shown in
Fig.~14 of the main text. The width and gap of the CPW transmission line are
indicated. The resonator center conductor width~$W$ and dielectric gap~$G$ are
also indicated, as well as the coupling length~$\ell_{\kappa}$ and the
line-resonator ground separation (\SI{5}{\micro\meter}). The
length~$\ell_{\kappa}$ determines how strongly the resonator is coupled to the
transmission line and, thus, the resonator external quality
factor~\cite{Frunzio:2005}.

\begin{figure}[ht]
	\centering
	\includegraphics[width=0.99\columnwidth]{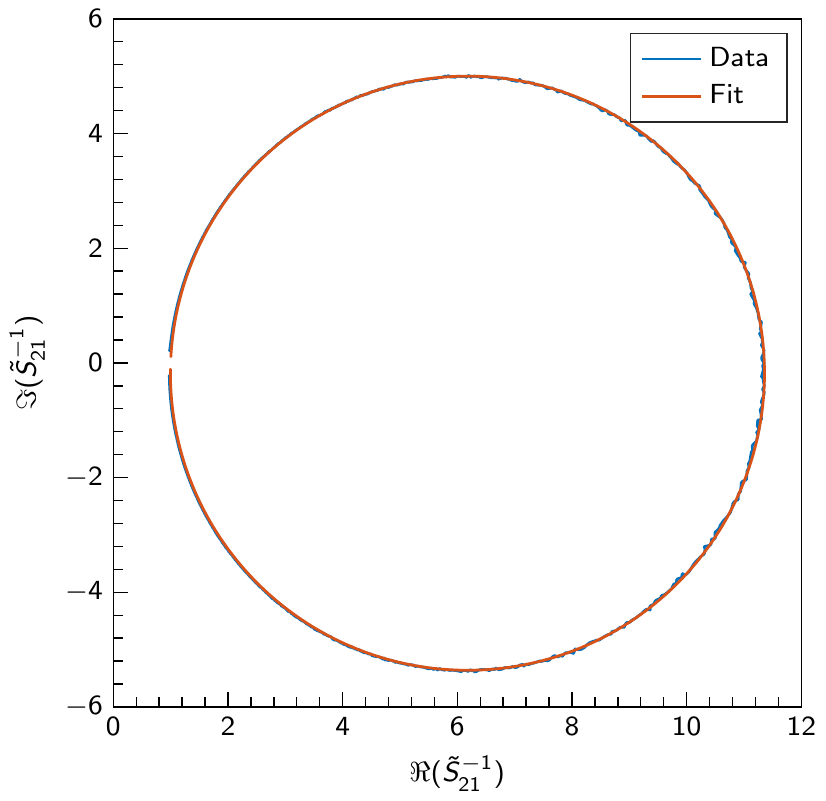}
\caption{Polar plot of the normalized inverse transmission
coefficient~$\tilde{S}_{21}^{-1}$ for superconducting resonator~$3$ discussed in
the main text. Data in blue and fit from Eq.~(\ref{Equation:01S}) in red.}
	\label{Figure03S:Bejanin}
\end{figure}

Figure~\ref{Figure03S:Bejanin} shows a polar plot of the real and imaginary part
of the normalized inverse transmission coefficient~$\tilde{S}_{21}^{-1}$ of the
data shown in Fig.~15~(d) of the main text, along with the fit. The equation
used for the fit is~\cite{Megrant:2012}
\begin{equation}
\tilde{S}_{21}^{-1} = 1 + \frac{Q_{\textrm{i}}}{Q_{\textrm{c}}^*} e^{i \phi} 
\frac{1}{1 + i 2 Q_{\textrm{i}} \delta x} \quad ,
	\label{Equation:01S}
\end{equation}
where~$\phi$ is an offset angle, $\delta x = ( f - f_0) / f_0$, and $i =
\sqrt{-1}$. A normalization is applied before the fit to set the off-resonance
transmission magnitude and phase to~\SI{0}{\deci\bel} and \SI{0}{\radian},
respectively. The fit parameters for the resonator in Fig.~15~(d) of the main
text were found to be~$f_0 = \SI{5064513933(6)}{\hertz}$, $Q_{\textrm{i}} =
165790(60)$, $Q_{\textrm{c}}^* = 16002(4)$, and $\phi = -0.0347(3)$.

\section*{S5: SPRING TESTS}
	\label{SPRING:TESTS}

As discussed in Appendix~D of the main text, Movie~4 shows the setup used to
test the springs at low temperature. In the movie, two springs mounted in series
are compressed, after having been submerged in liquid helium. This process was
repeated multiple times, without any damage occurring to the spring or any
measurable change in the spring constant. The same setup also made it possible
to preform compression measurements while the springs were submerged in a
cryogenic liquid (nitrogen or helium); the compressive force was applied at the
top of the rod, which always remained at room temperature.

\begin{figure}[ht]
	\centering
	\includegraphics[width=0.99\columnwidth]{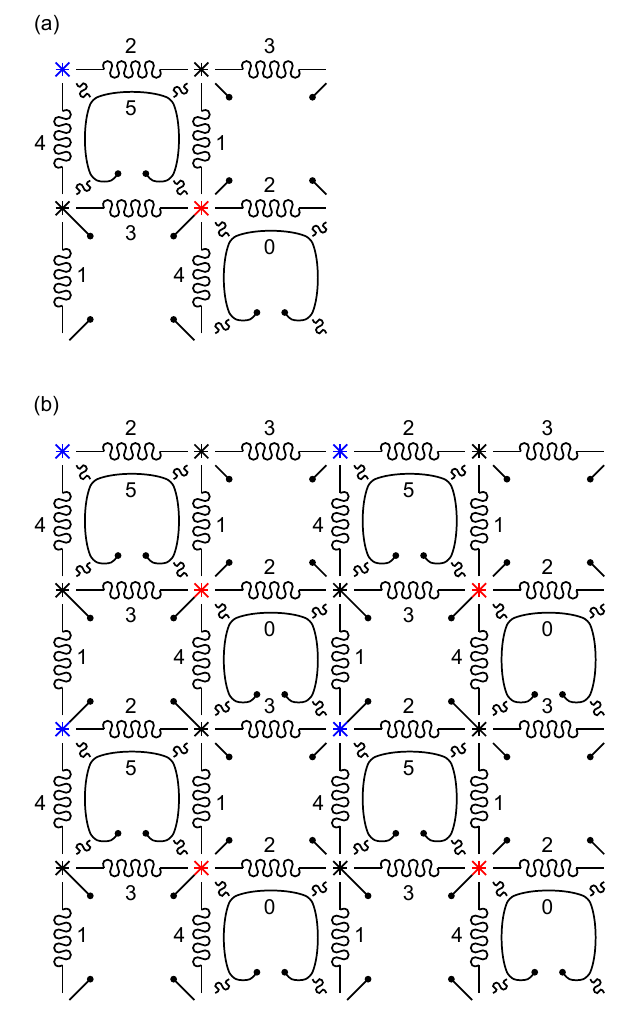}
\caption{Generalized surface code architecture compatible with the quantum
socket. (a) The main difference compared to the architecture presented in the
main text is the addition of an extra measurement resonator for each qubit. In
this case, the qubits are named \textit{octaton} due to their eight arms. The
various frequencies from low to high are indicated by the numbers~$0, 1, 2, 3,
4$, and $5$. The CZ gates follow the same numeration~\cite{Fowler:2012}.
Measurement qubits are indicated in red and blue and data qubits in black. (b)
Extended two-dimensional lattice.}
	\label{Figure04S:Bejanin}
\end{figure}

\section*{S6: QUANTUM SOCKET SURFACE CODE}
	\label{QUANTUM:SOCKET:SURFACE:CODE}

\begin{figure*}[ht]
	\centering
	\includegraphics[width=0.99\textwidth]{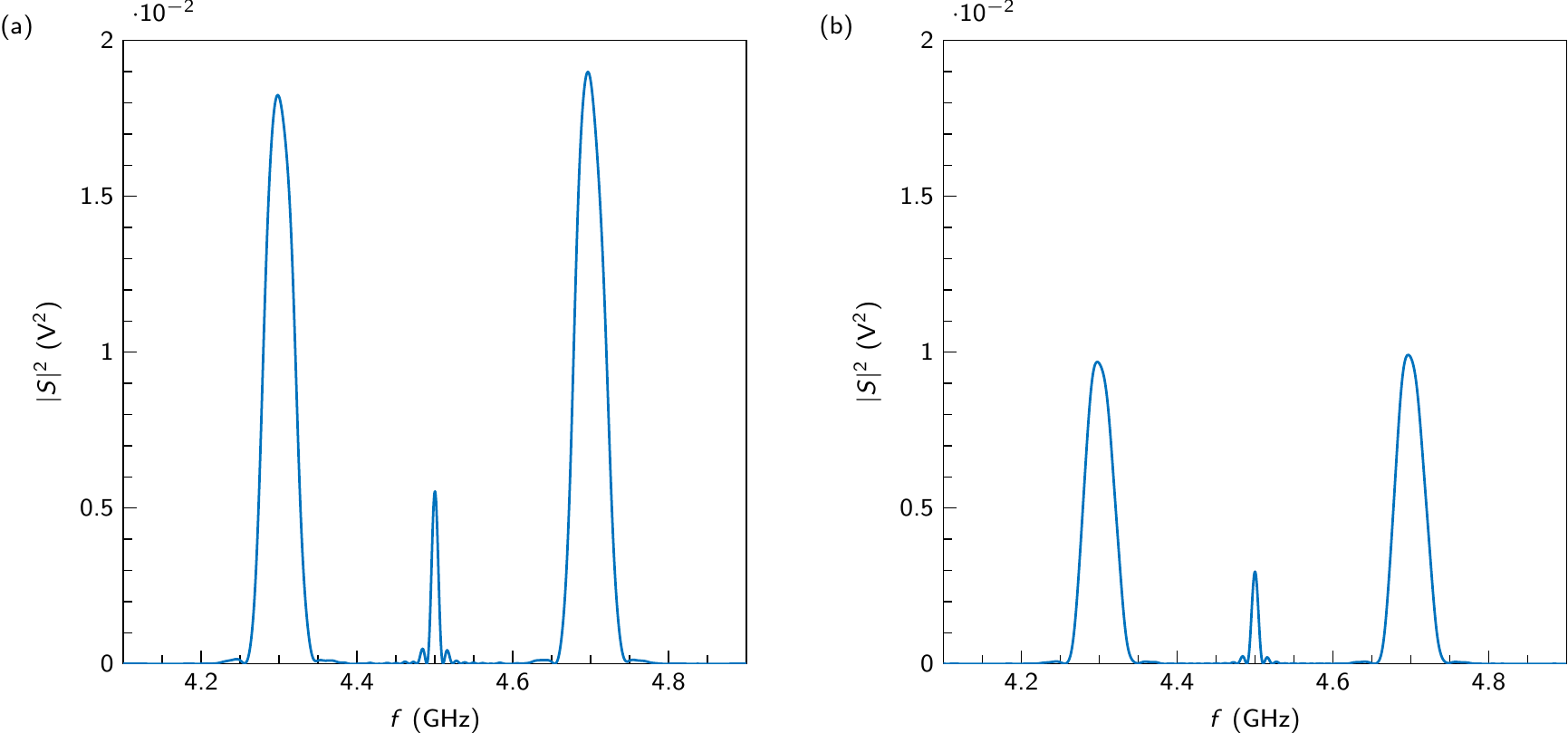}
\caption{Magnitude of a Fourier-transformed signal. (a) Pulse transmitted
through a reference flexible coaxial cable. (b) Pulse transmitted through the
quantum socket.}
	\label{Figure05S:Bejanin}
\end{figure*}

Figure~\ref{Figure04S:Bejanin} shows a more general version of the surface code
architecture in Fig.~16 of the main text. In panel~(a), numbers from~$0$ to~$5$
represent arbitrary frequencies from low to high ($0$ means a lower frequency
than the others). Every qubit is coupled to two measurement resonators; note
that label~$0$ is associated with a CPW transmission line that is coupled to a
set of four measurement resonators, all at slightly different frequencies (to
allow multiplexing) centered around frequency~$0$. Similarly, label~$5$ is
associated with a CPW transmission line that is coupled to a set of four
measurement resonators, all at slightly different frequencies centered around
frequency~$5$. In this architecture, all qubits start at a low frequency then
increase in frequency in unison to perform CZ gates through the coupling
resonators in the correct sequence; in the figure, this sequence is indicated by
the labels (coupling frequencies) $1, 2, 3$, and $4$. Finally, the measurement
qubits (red and blue) are readout at frequency~$5$. This is one surface code
cycle~\cite{Fowler:2012}. The process is then reversed and during the downward
uniform frequency sweep again an appropriate sequence of CZ interactions is
executed, followed by readout at frequency~$0$, implementing the next cycle.
This process avoids sweeping either measurement or data qubits past resonators
during the cycles. Panel~(b) shows an extended portion of the surface code
lattice.

\section*{S7: MICROWAVE PULSE TRANSMISSION}
	\label{MICROWAVE:PULSE:TRANSMISSION}

In Appendix~G of the main text, we have shown that the quantum socket is
characterized by a small frequency dispersion. In this section, we describe a
set of experiments where actual fast microwave pulses were applied to the
quantum socket as well as a reference cable. This makes it possible to assess
the effects of dispersion directly on pulses similar to those that will be used
for the manipulation of superconducting qubits.

Figure~\ref{Figure05S:Bejanin} shows the magnitude of the Fourier transform of a
pulse transmitted through the reference cable (panel~(a)) and the quantum socket
(panel~(b)), both at room temperature. The pulse was generated by mixing a
sinusoidal carrier signal at~\SI{4.5}{\giga\hertz} with a Gaussian-modulated
sinusoidal pulse at~\SI{200}{\mega\hertz}. The time length of the Gaussian
envelope was~\SI{15}{\nano\second} (full width at half maximum (FWHM)). The
carrier signal was generated by means of a microwave source from Keysight,
model~E8257D~PSG; the mixer was an in-phase and quadrature~(IQ) mixer from
Marki, model~IQ-0307LXP; the modulated Gaussian was generated by means of a
customized arbitrary waveform generator~(AWG) from Tabor Electronics Ltd.,
model~WX2184C+; all instruments were synchronized by way of a rubidium atomic
timebase embedded within a digital delay generator from Stanford Research
Systems, Inc., model~DG645/15, which was also used to trigger the AWG through
an~SRD1 module from Stanford Research Systems. The generated signal was split
with a microwave power divider from Krytar, Inc., model~6005180, with one output
sent through a reference SucoFlex flexible coaxial cable from Huber+Suhner of
length~\SI{1.5}{\meter} and the other output through the quantum socket (DUT of
Fig.~8 of the main text). The output of each path was acquired and digitized
using a real-time oscilloscope from Keysight, model~DSO91204A, also synchronized
with the rubidium timebase. The pulse going through the quantum socket is
attenuated due to the resistance of the Ag CPW transmission line, hence, the
magnitude of the Fourier transform is lower than of the reference signal.
However, the overall shape of the two pulses is very similar, with marginal
distortions.

\newpage


\begin{thebibliography}{84}%
\makeatletter
\providecommand \@ifxundefined [1]{%
 \@ifx{#1\undefined}
}%
\providecommand \@ifnum [1]{%
 \ifnum #1\expandafter \@firstoftwo
 \else \expandafter \@secondoftwo
 \fi
}%
\providecommand \@ifx [1]{%
 \ifx #1\expandafter \@firstoftwo
 \else \expandafter \@secondoftwo
 \fi
}%
\providecommand \natexlab [1]{#1}%
\providecommand \enquote  [1]{``#1''}%
\providecommand \bibnamefont  [1]{#1}%
\providecommand \bibfnamefont [1]{#1}%
\providecommand \citenamefont [1]{#1}%
\providecommand \href@noop [0]{\@secondoftwo}%
\providecommand \href [0]{\begingroup \@sanitize@url \@href}%
\providecommand \@href[1]{\@@startlink{#1}\@@href}%
\providecommand \@@href[1]{\endgroup#1\@@endlink}%
\providecommand \@sanitize@url [0]{\catcode `\\12\catcode `\$12\catcode
  `\&12\catcode `\#12\catcode `\^12\catcode `\_12\catcode `\%12\relax}%
\providecommand \@@startlink[1]{}%
\providecommand \@@endlink[0]{}%
\providecommand \url  [0]{\begingroup\@sanitize@url \@url }%
\providecommand \@url [1]{\endgroup\@href {#1}{\urlprefix }}%
\providecommand \urlprefix  [0]{URL }%
\providecommand \Eprint [0]{\href }%
\providecommand \doibase [0]{http://dx.doi.org/}%
\providecommand \selectlanguage [0]{\@gobble}%
\providecommand \bibinfo  [0]{\@secondoftwo}%
\providecommand \bibfield  [0]{\@secondoftwo}%
\providecommand \translation [1]{[#1]}%
\providecommand \BibitemOpen [0]{}%
\providecommand \bibitemStop [0]{}%
\providecommand \bibitemNoStop [0]{.\EOS\space}%
\providecommand \EOS [0]{\spacefactor3000\relax}%
\providecommand \BibitemShut  [1]{\csname bibitem#1\endcsname}%
\let\auto@bib@innerbib\@empty
\bibitem [{\citenamefont {Gambetta}\ \emph {et~al.}(2015)\citenamefont
  {Gambetta}, \citenamefont {Chow},\ and\ \citenamefont
  {Steffen}}]{Gambetta:2015}%
  \BibitemOpen
  \bibfield  {author} {\bibinfo {author} {\bibfnamefont {J.~M.}\ \bibnamefont
  {Gambetta}}, \bibinfo {author} {\bibfnamefont {J.~M.}\ \bibnamefont {Chow}},
  \ and\ \bibinfo {author} {\bibfnamefont {M.}~\bibnamefont {Steffen}},\ }\href
  {https://arxiv.org/abs/1510.04375} {\enquote {\bibinfo {title} {Building
  logical qubits in a superconducting quantum computing system},}\ } (\bibinfo
  {year} {2015}),\ \Eprint {http://arxiv.org/abs/arXiv:1510.04375}
  {arXiv:1510.04375} \BibitemShut {NoStop}%
\bibitem [{\citenamefont {Deutsch}(1985)}]{Deutsch:1985}%
  \BibitemOpen
  \bibfield  {author} {\bibinfo {author} {\bibfnamefont {D.}~\bibnamefont
  {Deutsch}},\ }\href {\doibase 10.1098/rspa.1985.0070} {\bibfield  {journal}
  {\bibinfo  {journal} {Proceedings of the Royal Society A: Mathematical,
  Physical, and Engineering Sciences}\ }\textbf {\bibinfo {volume} {400}},\
  \bibinfo {pages} {97} (\bibinfo {year} {1985})}\BibitemShut {NoStop}%
\bibitem [{\citenamefont {Clarke}\ and\ \citenamefont
  {Wilhelm}(2008)}]{Clarke:2008}%
  \BibitemOpen
  \bibfield  {author} {\bibinfo {author} {\bibfnamefont {J.}~\bibnamefont
  {Clarke}}\ and\ \bibinfo {author} {\bibfnamefont {F.~K.}\ \bibnamefont
  {Wilhelm}},\ }\href {\doibase 10.1038/nature07128} {\bibfield  {journal}
  {\bibinfo  {journal} {Nature}\ }\textbf {\bibinfo {volume} {453}},\ \bibinfo
  {pages} {1031} (\bibinfo {year} {2008})}\BibitemShut {NoStop}%
\bibitem [{\citenamefont {Devoret}\ and\ \citenamefont
  {Schoelkopf}(2013)}]{Devoret:2013}%
  \BibitemOpen
  \bibfield  {author} {\bibinfo {author} {\bibfnamefont {M.~H.}\ \bibnamefont
  {Devoret}}\ and\ \bibinfo {author} {\bibfnamefont {R.~J.}\ \bibnamefont
  {Schoelkopf}},\ }\href {\doibase 10.1126/science.1231930} {\bibfield
  {journal} {\bibinfo  {journal} {Science}\ }\textbf {\bibinfo {volume}
  {339}},\ \bibinfo {pages} {1169} (\bibinfo {year} {2013})}\BibitemShut
  {NoStop}%
\bibitem [{\citenamefont {Martinis}(2015)}]{Martinis:2015}%
  \BibitemOpen
  \bibfield  {author} {\bibinfo {author} {\bibfnamefont {J.~M.}\ \bibnamefont
  {Martinis}},\ }\href {\doibase 10.1038/npjqi.2015.5} {\bibfield  {journal}
  {\bibinfo  {journal} {npj Quantum Information}\ }\textbf {\bibinfo {volume}
  {1}},\ \bibinfo {pages} {15005} (\bibinfo {year} {2015})}\BibitemShut
  {NoStop}%
\bibitem [{\citenamefont {Miller}\ \emph {et~al.}(2012)\citenamefont {Miller},
  \citenamefont {Jhabvala}, \citenamefont {Leong}, \citenamefont {Costen},
  \citenamefont {Sharp}, \citenamefont {Adachi},\ and\ \citenamefont
  {Benford}}]{Miller:2012}%
  \BibitemOpen
  \bibfield  {author} {\bibinfo {author} {\bibfnamefont {T.~M.}\ \bibnamefont
  {Miller}}, \bibinfo {author} {\bibfnamefont {C.~A.}\ \bibnamefont
  {Jhabvala}}, \bibinfo {author} {\bibfnamefont {E.}~\bibnamefont {Leong}},
  \bibinfo {author} {\bibfnamefont {N.~P.}\ \bibnamefont {Costen}}, \bibinfo
  {author} {\bibfnamefont {E.}~\bibnamefont {Sharp}}, \bibinfo {author}
  {\bibfnamefont {T.}~\bibnamefont {Adachi}}, \ and\ \bibinfo {author}
  {\bibfnamefont {D.~J.}\ \bibnamefont {Benford}},\ }in\ \href {\doibase
  10.1117/12.926491} {\emph {\bibinfo {booktitle} {High Energy, Optical, and
  Infrared Detectors for Astronomy V}}},\ Vol.\ \bibinfo {volume} {8453},\
  \bibinfo {editor} {edited by\ \bibinfo {editor} {\bibfnamefont {A.~D.}\
  \bibnamefont {Holland}}\ and\ \bibinfo {editor} {\bibfnamefont {J.~W.}\
  \bibnamefont {Beletic}}}\ (\bibinfo  {publisher} {{SPIE}-Intl. Soc. Optical
  Eng.},\ \bibinfo {year} {2012})\ p.\ \bibinfo {pages} {84532H}\BibitemShut
  {NoStop}%
\bibitem [{\citenamefont {Abraham}\ \emph
  {et~al.}(2014{\natexlab{a}})\citenamefont {Abraham}, \citenamefont {Chow},
  \citenamefont {Gonzalez}, \citenamefont {Keefe}, \citenamefont {Rothwell},
  \citenamefont {Rozen},\ and\ \citenamefont {Steffen}}]{Abraham:2014:a}%
  \BibitemOpen
  \bibfield  {author} {\bibinfo {author} {\bibfnamefont {D.~W.}\ \bibnamefont
  {Abraham}}, \bibinfo {author} {\bibfnamefont {J.~M.}\ \bibnamefont {Chow}},
  \bibinfo {author} {\bibfnamefont {A.~D.~C.}\ \bibnamefont {Gonzalez}},
  \bibinfo {author} {\bibfnamefont {G.~A.}\ \bibnamefont {Keefe}}, \bibinfo
  {author} {\bibfnamefont {M.~E.}\ \bibnamefont {Rothwell}}, \bibinfo {author}
  {\bibfnamefont {J.~R.}\ \bibnamefont {Rozen}}, \ and\ \bibinfo {author}
  {\bibfnamefont {M.}~\bibnamefont {Steffen}},\ }\href
  {https://www.google.com/patents/US20140264287} {\enquote {\bibinfo {title}
  {Removal of spurious microwave modes via flip-chip crossover},}\ } (\bibinfo
  {year} {2014}{\natexlab{a}}),\ \bibinfo {note} {{US} Patent App.
  13/838,324}\BibitemShut {NoStop}%
\bibitem [{\citenamefont {Brecht}\ \emph {et~al.}(2015)\citenamefont {Brecht},
  \citenamefont {Reagor}, \citenamefont {Chu}, \citenamefont {Pfaff},
  \citenamefont {Wang}, \citenamefont {Frunzio}, \citenamefont {Devoret},\ and\
  \citenamefont {Schoelkopf}}]{Brecht:2015}%
  \BibitemOpen
  \bibfield  {author} {\bibinfo {author} {\bibfnamefont {T.}~\bibnamefont
  {Brecht}}, \bibinfo {author} {\bibfnamefont {M.}~\bibnamefont {Reagor}},
  \bibinfo {author} {\bibfnamefont {Y.}~\bibnamefont {Chu}}, \bibinfo {author}
  {\bibfnamefont {W.}~\bibnamefont {Pfaff}}, \bibinfo {author} {\bibfnamefont
  {C.}~\bibnamefont {Wang}}, \bibinfo {author} {\bibfnamefont {L.}~\bibnamefont
  {Frunzio}}, \bibinfo {author} {\bibfnamefont {M.~H.}\ \bibnamefont
  {Devoret}}, \ and\ \bibinfo {author} {\bibfnamefont {R.~J.}\ \bibnamefont
  {Schoelkopf}},\ }\href {\doibase 10.1063/1.4935541} {\bibfield  {journal}
  {\bibinfo  {journal} {Appl. Phys. Lett.}\ }\textbf {\bibinfo {volume}
  {107}},\ \bibinfo {pages} {192603} (\bibinfo {year} {2015})}\BibitemShut
  {NoStop}%
\bibitem [{\citenamefont {Rosenberg}\ \emph {et~al.}(2016)\citenamefont
  {Rosenberg}, \citenamefont {Yost}, \citenamefont {Das}, \citenamefont
  {Hover}, \citenamefont {Racz}, \citenamefont {Weber}, \citenamefont {Yoder},
  \citenamefont {Kerman},\ and\ \citenamefont {Oliver}}]{Rosenberg:2016}%
  \BibitemOpen
  \bibfield  {author} {\bibinfo {author} {\bibfnamefont {D.}~\bibnamefont
  {Rosenberg}}, \bibinfo {author} {\bibfnamefont {D.-R.}\ \bibnamefont {Yost}},
  \bibinfo {author} {\bibfnamefont {R.}~\bibnamefont {Das}}, \bibinfo {author}
  {\bibfnamefont {D.}~\bibnamefont {Hover}}, \bibinfo {author} {\bibfnamefont
  {L.}~\bibnamefont {Racz}}, \bibinfo {author} {\bibfnamefont {S.}~\bibnamefont
  {Weber}}, \bibinfo {author} {\bibfnamefont {J.}~\bibnamefont {Yoder}},
  \bibinfo {author} {\bibfnamefont {A.}~\bibnamefont {Kerman}}, \ and\ \bibinfo
  {author} {\bibfnamefont {W.}~\bibnamefont {Oliver}},\ }\href
  {http://meetings.aps.org/link/BAPS.2016.MAR.X48.3} {\enquote {\bibinfo
  {title} {3{D} integration for superconducting qubits},}\ } (\bibinfo {year}
  {2016}),\ \Eprint {http://arxiv.org/abs/{BAPS}.2016.{MAR}.{X}48.3}
  {{BAPS}.2016.{MAR}.{X}48.3} \BibitemShut {NoStop}%
\bibitem [{\citenamefont {Bruno}\ \emph {et~al.}(2016)\citenamefont {Bruno},
  \citenamefont {Poletto}, \citenamefont {Haider},\ and\ \citenamefont
  {DiCarlo}}]{Bruno:2016}%
  \BibitemOpen
  \bibfield  {author} {\bibinfo {author} {\bibfnamefont {A.}~\bibnamefont
  {Bruno}}, \bibinfo {author} {\bibfnamefont {S.}~\bibnamefont {Poletto}},
  \bibinfo {author} {\bibfnamefont {N.}~\bibnamefont {Haider}}, \ and\ \bibinfo
  {author} {\bibfnamefont {L.}~\bibnamefont {DiCarlo}},\ }\href
  {http://meetings.aps.org/link/BAPS.2016.MAR.X48.4} {\enquote {\bibinfo
  {title} {Extensible circuit {QED} processor architecture with vertical
  {I/O}},}\ } (\bibinfo {year} {2016}),\ \Eprint
  {http://arxiv.org/abs/{BAPS}.2016.{MAR}.{X}48.4} {{BAPS}.2016.{MAR}.{X}48.4}
  \BibitemShut {NoStop}%
\bibitem [{\citenamefont {Brecht}\ \emph {et~al.}(2016)\citenamefont {Brecht},
  \citenamefont {Pfaff}, \citenamefont {Wang}, \citenamefont {Chu},
  \citenamefont {Frunzio}, \citenamefont {Devoret},\ and\ \citenamefont
  {Schoelkopf}}]{Brecht:2016}%
  \BibitemOpen
  \bibfield  {author} {\bibinfo {author} {\bibfnamefont {T.}~\bibnamefont
  {Brecht}}, \bibinfo {author} {\bibfnamefont {W.}~\bibnamefont {Pfaff}},
  \bibinfo {author} {\bibfnamefont {C.}~\bibnamefont {Wang}}, \bibinfo {author}
  {\bibfnamefont {Y.}~\bibnamefont {Chu}}, \bibinfo {author} {\bibfnamefont
  {L.}~\bibnamefont {Frunzio}}, \bibinfo {author} {\bibfnamefont {M.~H.}\
  \bibnamefont {Devoret}}, \ and\ \bibinfo {author} {\bibfnamefont {R.~J.}\
  \bibnamefont {Schoelkopf}},\ }\href {\doibase 10.1038/npjqi.2016.2}
  {\bibfield  {journal} {\bibinfo  {journal} {npj Quantum Information}\
  }\textbf {\bibinfo {volume} {2}},\ \bibinfo {pages} {16002} (\bibinfo {year}
  {2016})}\BibitemShut {NoStop}%
\bibitem [{\citenamefont {Di{V}incenzo}(1995)}]{DiVincenzo:1995}%
  \BibitemOpen
  \bibfield  {author} {\bibinfo {author} {\bibfnamefont {D.~P.}\ \bibnamefont
  {Di{V}incenzo}},\ }\href {\doibase 10.1126/science.270.5234.255} {\bibfield
  {journal} {\bibinfo  {journal} {Science}\ }\textbf {\bibinfo {volume}
  {270}},\ \bibinfo {pages} {255} (\bibinfo {year} {1995})}\BibitemShut
  {NoStop}%
\bibitem [{\citenamefont {Nielsen}\ and\ \citenamefont
  {Chuang}(2000)}]{Nielsen:2000}%
  \BibitemOpen
  \bibfield  {author} {\bibinfo {author} {\bibfnamefont {M.~A.}\ \bibnamefont
  {Nielsen}}\ and\ \bibinfo {author} {\bibfnamefont {I.~L.}\ \bibnamefont
  {Chuang}},\ }\href {\doibase 10.1017/CBO9780511976667} {\emph {\bibinfo
  {title} {Quantum Computation and Quantum Information}}}\ (\bibinfo
  {publisher} {Cambridge University Press ({CUP})},\ \bibinfo {address}
  {Cambridge, {UK}},\ \bibinfo {year} {2000})\BibitemShut {NoStop}%
\bibitem [{\citenamefont {Di{V}incenzo}(2000)}]{DiVincenzo:2000}%
  \BibitemOpen
  \bibfield  {author} {\bibinfo {author} {\bibfnamefont {D.~P.}\ \bibnamefont
  {Di{V}incenzo}},\ }\href {\doibase
  10.1002/1521-3978(200009)48:9/11<771::AID-PROP771>3.0.CO;2-E} {\bibfield
  {journal} {\bibinfo  {journal} {Fortschritte der Physik}\ }\textbf {\bibinfo
  {volume} {48}},\ \bibinfo {pages} {771} (\bibinfo {year} {2000})}\BibitemShut
  {NoStop}%
\bibitem [{\citenamefont {Mermin}(2007)}]{Mermin:2007}%
  \BibitemOpen
  \bibfield  {author} {\bibinfo {author} {\bibfnamefont {N.~D.}\ \bibnamefont
  {Mermin}},\ }\href {\doibase 10.1017/CBO9780511813870} {\emph {\bibinfo
  {title} {Quantum Computer Science - An Introduction}}}\ (\bibinfo
  {publisher} {Cambridge University Press ({CUP})},\ \bibinfo {year}
  {2007})\BibitemShut {NoStop}%
\bibitem [{\citenamefont {Ladd}\ \emph {et~al.}(2010)\citenamefont {Ladd},
  \citenamefont {Jelezko}, \citenamefont {Laflamme}, \citenamefont {Nakamura},
  \citenamefont {Monroe},\ and\ \citenamefont {O'Brien}}]{Ladd:2010}%
  \BibitemOpen
  \bibfield  {author} {\bibinfo {author} {\bibfnamefont {T.~D.}\ \bibnamefont
  {Ladd}}, \bibinfo {author} {\bibfnamefont {F.}~\bibnamefont {Jelezko}},
  \bibinfo {author} {\bibfnamefont {R.}~\bibnamefont {Laflamme}}, \bibinfo
  {author} {\bibfnamefont {Y.}~\bibnamefont {Nakamura}}, \bibinfo {author}
  {\bibfnamefont {C.}~\bibnamefont {Monroe}}, \ and\ \bibinfo {author}
  {\bibfnamefont {J.~L.}\ \bibnamefont {O'Brien}},\ }\href {\doibase
  10.1038/nature08812} {\bibfield  {journal} {\bibinfo  {journal} {Nature}\
  }\textbf {\bibinfo {volume} {464}},\ \bibinfo {pages} {45} (\bibinfo {year}
  {2010})}\BibitemShut {NoStop}%
\bibitem [{\citenamefont {P{\'{e}}rez-Delgado}\ and\ \citenamefont
  {Kok}(2011)}]{PerezDelgado:2011}%
  \BibitemOpen
  \bibfield  {author} {\bibinfo {author} {\bibfnamefont {C.~A.}\ \bibnamefont
  {P{\'{e}}rez-Delgado}}\ and\ \bibinfo {author} {\bibfnamefont
  {P.}~\bibnamefont {Kok}},\ }\href {\doibase 10.1103/PhysRevA.83.012303}
  {\bibfield  {journal} {\bibinfo  {journal} {Phys. Rev. A}\ }\textbf {\bibinfo
  {volume} {83}},\ \bibinfo {pages} {012303} (\bibinfo {year}
  {2011})}\BibitemShut {NoStop}%
\bibitem [{\citenamefont {Montanaro}(2016)}]{Montanaro:2016}%
  \BibitemOpen
  \bibfield  {author} {\bibinfo {author} {\bibfnamefont {A.}~\bibnamefont
  {Montanaro}},\ }\href {\doibase 10.1038/npjqi.2015.23} {\bibfield  {journal}
  {\bibinfo  {journal} {npj Quantum Information}\ }\textbf {\bibinfo {volume}
  {2}},\ \bibinfo {pages} {15023} (\bibinfo {year} {2016})}\BibitemShut
  {NoStop}%
\bibitem [{\citenamefont {Gottesman}(2010)}]{Gottesman:2010}%
  \BibitemOpen
  \bibfield  {author} {\bibinfo {author} {\bibfnamefont {D.}~\bibnamefont
  {Gottesman}},\ }\href {\doibase 10.1090/psapm/068/2762145} {\enquote
  {\bibinfo {title} {An introduction to quantum error correction and
  fault-tolerant quantum computation},}\ } (\bibinfo {year} {2010})\BibitemShut
  {NoStop}%
\bibitem [{\citenamefont {Raussendorf}\ and\ \citenamefont
  {Harrington}(2007)}]{Raussendorf:2007}%
  \BibitemOpen
  \bibfield  {author} {\bibinfo {author} {\bibfnamefont {R.}~\bibnamefont
  {Raussendorf}}\ and\ \bibinfo {author} {\bibfnamefont {J.}~\bibnamefont
  {Harrington}},\ }\href {\doibase 10.1103/PhysRevLett.98.190504} {\bibfield
  {journal} {\bibinfo  {journal} {Phys. Rev. Lett.}\ }\textbf {\bibinfo
  {volume} {98}},\ \bibinfo {pages} {190504} (\bibinfo {year}
  {2007})}\BibitemShut {NoStop}%
\bibitem [{\citenamefont {Fowler}\ \emph {et~al.}(2012)\citenamefont {Fowler},
  \citenamefont {Mariantoni}, \citenamefont {Martinis},\ and\ \citenamefont
  {Cleland}}]{Fowler:2012}%
  \BibitemOpen
  \bibfield  {author} {\bibinfo {author} {\bibfnamefont {A.~G.}\ \bibnamefont
  {Fowler}}, \bibinfo {author} {\bibfnamefont {M.}~\bibnamefont {Mariantoni}},
  \bibinfo {author} {\bibfnamefont {J.~M.}\ \bibnamefont {Martinis}}, \ and\
  \bibinfo {author} {\bibfnamefont {A.~N.}\ \bibnamefont {Cleland}},\ }\href
  {\doibase 10.1103/PhysRevA.86.032324} {\bibfield  {journal} {\bibinfo
  {journal} {Phys. Rev. A}\ }\textbf {\bibinfo {volume} {86}},\ \bibinfo
  {pages} {032324} (\bibinfo {year} {2012})}\BibitemShut {NoStop}%
\bibitem [{\citenamefont {Shor}(1994)}]{Shor:1994}%
  \BibitemOpen
  \bibfield  {author} {\bibinfo {author} {\bibfnamefont {P.~W.}\ \bibnamefont
  {Shor}},\ }in\ \href {\doibase 10.1109/SFCS.1994.365700} {\emph {\bibinfo
  {booktitle} {Proceedings of the 35th Annual Symposium on Foundations of
  Computer Science}}},\ \bibinfo {editor} {edited by\ \bibinfo {editor}
  {\bibfnamefont {S.}~\bibnamefont {Goldwasser}}}\ (\bibinfo  {publisher}
  {{IEEE} Computer Society},\ \bibinfo {address} {Washington, {DC}, {USA}},\
  \bibinfo {year} {1994})\ pp.\ \bibinfo {pages} {124--134}\BibitemShut
  {NoStop}%
\bibitem [{\citenamefont {Kok}\ \emph {et~al.}(2007)\citenamefont {Kok},
  \citenamefont {Munro}, \citenamefont {Nemoto}, \citenamefont {Ralph},
  \citenamefont {Dowling},\ and\ \citenamefont {Milburn}}]{Kok:2007}%
  \BibitemOpen
  \bibfield  {author} {\bibinfo {author} {\bibfnamefont {P.}~\bibnamefont
  {Kok}}, \bibinfo {author} {\bibfnamefont {W.~J.}\ \bibnamefont {Munro}},
  \bibinfo {author} {\bibfnamefont {K.}~\bibnamefont {Nemoto}}, \bibinfo
  {author} {\bibfnamefont {T.~C.}\ \bibnamefont {Ralph}}, \bibinfo {author}
  {\bibfnamefont {J.~P.}\ \bibnamefont {Dowling}}, \ and\ \bibinfo {author}
  {\bibfnamefont {G.~J.}\ \bibnamefont {Milburn}},\ }\href {\doibase
  10.1103/RevModPhys.79.135} {\bibfield  {journal} {\bibinfo  {journal} {Rev.
  Mod. Phys.}\ }\textbf {\bibinfo {volume} {79}},\ \bibinfo {pages} {135}
  (\bibinfo {year} {2007})}\BibitemShut {NoStop}%
\bibitem [{\citenamefont {O'Brien}\ \emph {et~al.}(2009)\citenamefont
  {O'Brien}, \citenamefont {Furusawa},\ and\ \citenamefont
  {Vu{\v{c}}kovi{\'{c}}}}]{OBrien:2009}%
  \BibitemOpen
  \bibfield  {author} {\bibinfo {author} {\bibfnamefont {J.~L.}\ \bibnamefont
  {O'Brien}}, \bibinfo {author} {\bibfnamefont {A.}~\bibnamefont {Furusawa}}, \
  and\ \bibinfo {author} {\bibfnamefont {J.}~\bibnamefont
  {Vu{\v{c}}kovi{\'{c}}}},\ }\href {\doibase 10.1038/NPHOTON.2009.229}
  {\bibfield  {journal} {\bibinfo  {journal} {Nature Photonics}\ }\textbf
  {\bibinfo {volume} {3}},\ \bibinfo {pages} {687} (\bibinfo {year}
  {2009})}\BibitemShut {NoStop}%
\bibitem [{\citenamefont {Monroe}\ and\ \citenamefont
  {Kim}(2013)}]{Monroe:2013}%
  \BibitemOpen
  \bibfield  {author} {\bibinfo {author} {\bibfnamefont {C.}~\bibnamefont
  {Monroe}}\ and\ \bibinfo {author} {\bibfnamefont {J.}~\bibnamefont {Kim}},\
  }\href {\doibase 10.1126/science.1231298} {\bibfield  {journal} {\bibinfo
  {journal} {Science}\ }\textbf {\bibinfo {volume} {339}},\ \bibinfo {pages}
  {1164} (\bibinfo {year} {2013})}\BibitemShut {NoStop}%
\bibitem [{\citenamefont {Cory}(2006)}]{Cory:2006}%
  \BibitemOpen
  \bibfield  {author} {\bibinfo {author} {\bibfnamefont {D.}~\bibnamefont
  {Cory}},\ }in\ \href {\doibase 10.1364/LS.2006.LTuI3} {\emph {\bibinfo
  {booktitle} {Frontiers in Optics, {OSA} Technical Digest ({CD})}}}\ (\bibinfo
   {publisher} {The Optical Society of America},\ \bibinfo {year} {2006})\ p.\
  \bibinfo {pages} {{LT}u{I}3}\BibitemShut {NoStop}%
\bibitem [{\citenamefont {Jones}\ \emph {et~al.}(2012)\citenamefont {Jones},
  \citenamefont {Meter}, \citenamefont {Fowler}, \citenamefont {McMahon},
  \citenamefont {Kim}, \citenamefont {Ladd},\ and\ \citenamefont
  {Yamamoto}}]{Cody:2012}%
  \BibitemOpen
  \bibfield  {author} {\bibinfo {author} {\bibfnamefont {N.~C.}\ \bibnamefont
  {Jones}}, \bibinfo {author} {\bibfnamefont {R.~V.}\ \bibnamefont {Meter}},
  \bibinfo {author} {\bibfnamefont {A.~G.}\ \bibnamefont {Fowler}}, \bibinfo
  {author} {\bibfnamefont {P.~L.}\ \bibnamefont {McMahon}}, \bibinfo {author}
  {\bibfnamefont {J.}~\bibnamefont {Kim}}, \bibinfo {author} {\bibfnamefont
  {T.~D.}\ \bibnamefont {Ladd}}, \ and\ \bibinfo {author} {\bibfnamefont
  {Y.}~\bibnamefont {Yamamoto}},\ }\href {\doibase 10.1103/PhysRevX.2.031007}
  {\bibfield  {journal} {\bibinfo  {journal} {Phys. Rev. {X}}\ }\textbf
  {\bibinfo {volume} {2}},\ \bibinfo {pages} {031007} (\bibinfo {year}
  {2012})}\BibitemShut {NoStop}%
\bibitem [{\citenamefont {Hanson}\ \emph {et~al.}(2007)\citenamefont {Hanson},
  \citenamefont {Kouwenhoven}, \citenamefont {Petta}, \citenamefont {Tarucha},\
  and\ \citenamefont {Vandersypen}}]{Hanson:2007}%
  \BibitemOpen
  \bibfield  {author} {\bibinfo {author} {\bibfnamefont {R.}~\bibnamefont
  {Hanson}}, \bibinfo {author} {\bibfnamefont {L.~P.}\ \bibnamefont
  {Kouwenhoven}}, \bibinfo {author} {\bibfnamefont {J.~R.}\ \bibnamefont
  {Petta}}, \bibinfo {author} {\bibfnamefont {S.}~\bibnamefont {Tarucha}}, \
  and\ \bibinfo {author} {\bibfnamefont {L.~M.}\ \bibnamefont {Vandersypen}},\
  }\href {\doibase 10.1103/RevModPhys.79.1217} {\bibfield  {journal} {\bibinfo
  {journal} {Rev. Mod. Phys.}\ }\textbf {\bibinfo {volume} {79}},\ \bibinfo
  {pages} {1217} (\bibinfo {year} {2007})}\BibitemShut {NoStop}%
\bibitem [{\citenamefont {Maune}\ \emph {et~al.}(2012)\citenamefont {Maune},
  \citenamefont {Borselli}, \citenamefont {Huang}, \citenamefont {Ladd},
  \citenamefont {Deelman}, \citenamefont {Holabird}, \citenamefont {Kiselev},
  \citenamefont {Alvarado-Rodriguez}, \citenamefont {Ross}, \citenamefont
  {Schmitz}, \citenamefont {Sokolich}, \citenamefont {Watson}, \citenamefont
  {Gyure},\ and\ \citenamefont {Hunter}}]{Maune:2012}%
  \BibitemOpen
  \bibfield  {author} {\bibinfo {author} {\bibfnamefont {B.~M.}\ \bibnamefont
  {Maune}}, \bibinfo {author} {\bibfnamefont {M.~G.}\ \bibnamefont {Borselli}},
  \bibinfo {author} {\bibfnamefont {B.}~\bibnamefont {Huang}}, \bibinfo
  {author} {\bibfnamefont {T.~D.}\ \bibnamefont {Ladd}}, \bibinfo {author}
  {\bibfnamefont {P.~W.}\ \bibnamefont {Deelman}}, \bibinfo {author}
  {\bibfnamefont {K.~S.}\ \bibnamefont {Holabird}}, \bibinfo {author}
  {\bibfnamefont {A.~A.}\ \bibnamefont {Kiselev}}, \bibinfo {author}
  {\bibfnamefont {I.}~\bibnamefont {Alvarado-Rodriguez}}, \bibinfo {author}
  {\bibfnamefont {R.~S.}\ \bibnamefont {Ross}}, \bibinfo {author}
  {\bibfnamefont {A.~E.}\ \bibnamefont {Schmitz}}, \bibinfo {author}
  {\bibfnamefont {M.}~\bibnamefont {Sokolich}}, \bibinfo {author}
  {\bibfnamefont {C.~A.}\ \bibnamefont {Watson}}, \bibinfo {author}
  {\bibfnamefont {M.~F.}\ \bibnamefont {Gyure}}, \ and\ \bibinfo {author}
  {\bibfnamefont {A.~T.}\ \bibnamefont {Hunter}},\ }\href {\doibase
  10.1038/nature10707} {\bibfield  {journal} {\bibinfo  {journal} {Nature}\
  }\textbf {\bibinfo {volume} {481}},\ \bibinfo {pages} {344} (\bibinfo {year}
  {2012})}\BibitemShut {NoStop}%
\bibitem [{\citenamefont {Zwanenburg}\ \emph {et~al.}(2013)\citenamefont
  {Zwanenburg}, \citenamefont {Dzurak}, \citenamefont {Morello}, \citenamefont
  {Simmons}, \citenamefont {Hollenberg}, \citenamefont {Klimeck}, \citenamefont
  {Rogge}, \citenamefont {Coppersmith},\ and\ \citenamefont
  {Eriksson}}]{Zwanenburg:2013}%
  \BibitemOpen
  \bibfield  {author} {\bibinfo {author} {\bibfnamefont {F.~A.}\ \bibnamefont
  {Zwanenburg}}, \bibinfo {author} {\bibfnamefont {A.~S.}\ \bibnamefont
  {Dzurak}}, \bibinfo {author} {\bibfnamefont {A.}~\bibnamefont {Morello}},
  \bibinfo {author} {\bibfnamefont {M.~Y.}\ \bibnamefont {Simmons}}, \bibinfo
  {author} {\bibfnamefont {L.~C.}\ \bibnamefont {Hollenberg}}, \bibinfo
  {author} {\bibfnamefont {G.}~\bibnamefont {Klimeck}}, \bibinfo {author}
  {\bibfnamefont {S.}~\bibnamefont {Rogge}}, \bibinfo {author} {\bibfnamefont
  {S.~N.}\ \bibnamefont {Coppersmith}}, \ and\ \bibinfo {author} {\bibfnamefont
  {M.~A.}\ \bibnamefont {Eriksson}},\ }\href {\doibase
  10.1103/RevModPhys.85.961} {\bibfield  {journal} {\bibinfo  {journal} {Rev.
  Mod. Phys.}\ }\textbf {\bibinfo {volume} {85}},\ \bibinfo {pages} {961}
  (\bibinfo {year} {2013})}\BibitemShut {NoStop}%
\bibitem [{\citenamefont {Higginbotham}\ \emph {et~al.}(2014)\citenamefont
  {Higginbotham}, \citenamefont {Kuemmeth}, \citenamefont {Hanson},
  \citenamefont {Gossard},\ and\ \citenamefont {Marcus}}]{Higginbotham:2014}%
  \BibitemOpen
  \bibfield  {author} {\bibinfo {author} {\bibfnamefont {A.~P.}\ \bibnamefont
  {Higginbotham}}, \bibinfo {author} {\bibfnamefont {F.}~\bibnamefont
  {Kuemmeth}}, \bibinfo {author} {\bibfnamefont {M.~P.}\ \bibnamefont
  {Hanson}}, \bibinfo {author} {\bibfnamefont {A.~C.}\ \bibnamefont {Gossard}},
  \ and\ \bibinfo {author} {\bibfnamefont {C.~M.}\ \bibnamefont {Marcus}},\
  }\href {\doibase 10.1103/PhysRevLett.112.026801} {\bibfield  {journal}
  {\bibinfo  {journal} {Phys. Rev. Lett.}\ }\textbf {\bibinfo {volume} {112}},\
  \bibinfo {pages} {026801} (\bibinfo {year} {2014})}\BibitemShut {NoStop}%
\bibitem [{\citenamefont {O'Gorman}\ \emph {et~al.}(2016)\citenamefont
  {O'Gorman}, \citenamefont {Nickerson}, \citenamefont {Ross}, \citenamefont
  {Morton},\ and\ \citenamefont {Benjamin}}]{OGorman:2016}%
  \BibitemOpen
  \bibfield  {author} {\bibinfo {author} {\bibfnamefont {J.}~\bibnamefont
  {O'Gorman}}, \bibinfo {author} {\bibfnamefont {N.~H.}\ \bibnamefont
  {Nickerson}}, \bibinfo {author} {\bibfnamefont {P.}~\bibnamefont {Ross}},
  \bibinfo {author} {\bibfnamefont {J.~J.}\ \bibnamefont {Morton}}, \ and\
  \bibinfo {author} {\bibfnamefont {S.~C.}\ \bibnamefont {Benjamin}},\ }\href
  {\doibase 10.1038/npjqi.2015.19} {\bibfield  {journal} {\bibinfo  {journal}
  {npj Quantum Information}\ }\textbf {\bibinfo {volume} {2}},\ \bibinfo
  {pages} {15019} (\bibinfo {year} {2016})}\BibitemShut {NoStop}%
\bibitem [{\citenamefont {Barends}\ \emph {et~al.}(2014)\citenamefont
  {Barends}, \citenamefont {Kelly}, \citenamefont {Megrant}, \citenamefont
  {Veitia}, \citenamefont {Sank}, \citenamefont {Jeffrey}, \citenamefont
  {White}, \citenamefont {Mutus}, \citenamefont {Fowler}, \citenamefont
  {Campbell}, \citenamefont {Chen}, \citenamefont {Chen}, \citenamefont
  {Chiaro}, \citenamefont {Dunsworth}, \citenamefont {Neill}, \citenamefont
  {O'Malley}, \citenamefont {Roushan}, \citenamefont {Vainsencher},
  \citenamefont {Wenner}, \citenamefont {Korotkov}, \citenamefont {Cleland},\
  and\ \citenamefont {Martinis}}]{Barends:2014}%
  \BibitemOpen
  \bibfield  {author} {\bibinfo {author} {\bibfnamefont {R.}~\bibnamefont
  {Barends}}, \bibinfo {author} {\bibfnamefont {J.}~\bibnamefont {Kelly}},
  \bibinfo {author} {\bibfnamefont {A.}~\bibnamefont {Megrant}}, \bibinfo
  {author} {\bibfnamefont {A.}~\bibnamefont {Veitia}}, \bibinfo {author}
  {\bibfnamefont {D.}~\bibnamefont {Sank}}, \bibinfo {author} {\bibfnamefont
  {E.}~\bibnamefont {Jeffrey}}, \bibinfo {author} {\bibfnamefont {T.~C.}\
  \bibnamefont {White}}, \bibinfo {author} {\bibfnamefont {J.}~\bibnamefont
  {Mutus}}, \bibinfo {author} {\bibfnamefont {A.~G.}\ \bibnamefont {Fowler}},
  \bibinfo {author} {\bibfnamefont {B.}~\bibnamefont {Campbell}}, \bibinfo
  {author} {\bibfnamefont {Y.}~\bibnamefont {Chen}}, \bibinfo {author}
  {\bibfnamefont {Z.}~\bibnamefont {Chen}}, \bibinfo {author} {\bibfnamefont
  {B.}~\bibnamefont {Chiaro}}, \bibinfo {author} {\bibfnamefont
  {A.}~\bibnamefont {Dunsworth}}, \bibinfo {author} {\bibfnamefont
  {C.}~\bibnamefont {Neill}}, \bibinfo {author} {\bibfnamefont
  {P.}~\bibnamefont {O'Malley}}, \bibinfo {author} {\bibfnamefont
  {P.}~\bibnamefont {Roushan}}, \bibinfo {author} {\bibfnamefont
  {A.}~\bibnamefont {Vainsencher}}, \bibinfo {author} {\bibfnamefont
  {J.}~\bibnamefont {Wenner}}, \bibinfo {author} {\bibfnamefont {A.~N.}\
  \bibnamefont {Korotkov}}, \bibinfo {author} {\bibfnamefont {A.~N.}\
  \bibnamefont {Cleland}}, \ and\ \bibinfo {author} {\bibfnamefont {J.~M.}\
  \bibnamefont {Martinis}},\ }\href {\doibase 10.1038/nature13171} {\bibfield
  {journal} {\bibinfo  {journal} {Nature}\ }\textbf {\bibinfo {volume} {508}},\
  \bibinfo {pages} {500} (\bibinfo {year} {2014})}\BibitemShut {NoStop}%
\bibitem [{\citenamefont {C{\'{o}}rcoles}\ \emph {et~al.}(2015)\citenamefont
  {C{\'{o}}rcoles}, \citenamefont {Magesan}, \citenamefont {Srinivasan},
  \citenamefont {Cross}, \citenamefont {Steffen}, \citenamefont {Gambetta},\
  and\ \citenamefont {Chow}}]{Corcoles:2015}%
  \BibitemOpen
  \bibfield  {author} {\bibinfo {author} {\bibfnamefont {A.~D.}\ \bibnamefont
  {C{\'{o}}rcoles}}, \bibinfo {author} {\bibfnamefont {E.}~\bibnamefont
  {Magesan}}, \bibinfo {author} {\bibfnamefont {S.~J.}\ \bibnamefont
  {Srinivasan}}, \bibinfo {author} {\bibfnamefont {A.~W.}\ \bibnamefont
  {Cross}}, \bibinfo {author} {\bibfnamefont {M.}~\bibnamefont {Steffen}},
  \bibinfo {author} {\bibfnamefont {J.~M.}\ \bibnamefont {Gambetta}}, \ and\
  \bibinfo {author} {\bibfnamefont {J.~M.}\ \bibnamefont {Chow}},\ }\href
  {\doibase 10.1038/ncomms7979} {\bibfield  {journal} {\bibinfo  {journal}
  {Nature Communications}\ }\textbf {\bibinfo {volume} {6}},\ \bibinfo {pages}
  {6979} (\bibinfo {year} {2015})}\BibitemShut {NoStop}%
\bibitem [{\citenamefont {Rist{\`{e}}}\ \emph {et~al.}(2015)\citenamefont
  {Rist{\`{e}}}, \citenamefont {Poletto}, \citenamefont {Huang}, \citenamefont
  {Bruno}, \citenamefont {Vesterinen}, \citenamefont {Saira},\ and\
  \citenamefont {DiCarlo}}]{Riste:2015}%
  \BibitemOpen
  \bibfield  {author} {\bibinfo {author} {\bibfnamefont {D.}~\bibnamefont
  {Rist{\`{e}}}}, \bibinfo {author} {\bibfnamefont {S.}~\bibnamefont
  {Poletto}}, \bibinfo {author} {\bibfnamefont {M.-Z.}\ \bibnamefont {Huang}},
  \bibinfo {author} {\bibfnamefont {A.}~\bibnamefont {Bruno}}, \bibinfo
  {author} {\bibfnamefont {V.}~\bibnamefont {Vesterinen}}, \bibinfo {author}
  {\bibfnamefont {O.-P.}\ \bibnamefont {Saira}}, \ and\ \bibinfo {author}
  {\bibfnamefont {L.}~\bibnamefont {DiCarlo}},\ }\href {\doibase
  10.1038/ncomms7983} {\bibfield  {journal} {\bibinfo  {journal} {Nature
  Communications}\ }\textbf {\bibinfo {volume} {6}},\ \bibinfo {pages} {6983}
  (\bibinfo {year} {2015})}\BibitemShut {NoStop}%
\bibitem [{\citenamefont {Kelly}\ \emph {et~al.}(2015)\citenamefont {Kelly},
  \citenamefont {Barends}, \citenamefont {Fowler}, \citenamefont {Megrant},
  \citenamefont {Jeffrey}, \citenamefont {White}, \citenamefont {Sank},
  \citenamefont {Mutus}, \citenamefont {Campbell}, \citenamefont {Chen},
  \citenamefont {Chen}, \citenamefont {Chiaro}, \citenamefont {Dunsworth},
  \citenamefont {Hoi}, \citenamefont {Neill}, \citenamefont {O'Malley},
  \citenamefont {Quintana}, \citenamefont {Roushan}, \citenamefont
  {Vainsencher}, \citenamefont {Wenner}, \citenamefont {Cleland},\ and\
  \citenamefont {Martinis}}]{Kelly:2015}%
  \BibitemOpen
  \bibfield  {author} {\bibinfo {author} {\bibfnamefont {J.}~\bibnamefont
  {Kelly}}, \bibinfo {author} {\bibfnamefont {R.}~\bibnamefont {Barends}},
  \bibinfo {author} {\bibfnamefont {A.~G.}\ \bibnamefont {Fowler}}, \bibinfo
  {author} {\bibfnamefont {A.}~\bibnamefont {Megrant}}, \bibinfo {author}
  {\bibfnamefont {E.}~\bibnamefont {Jeffrey}}, \bibinfo {author} {\bibfnamefont
  {T.~C.}\ \bibnamefont {White}}, \bibinfo {author} {\bibfnamefont
  {D.}~\bibnamefont {Sank}}, \bibinfo {author} {\bibfnamefont {J.~Y.}\
  \bibnamefont {Mutus}}, \bibinfo {author} {\bibfnamefont {B.}~\bibnamefont
  {Campbell}}, \bibinfo {author} {\bibfnamefont {Y.}~\bibnamefont {Chen}},
  \bibinfo {author} {\bibfnamefont {Z.}~\bibnamefont {Chen}}, \bibinfo {author}
  {\bibfnamefont {B.}~\bibnamefont {Chiaro}}, \bibinfo {author} {\bibfnamefont
  {A.}~\bibnamefont {Dunsworth}}, \bibinfo {author} {\bibfnamefont {I.-C.}\
  \bibnamefont {Hoi}}, \bibinfo {author} {\bibfnamefont {C.}~\bibnamefont
  {Neill}}, \bibinfo {author} {\bibfnamefont {P.~J.}\ \bibnamefont {O'Malley}},
  \bibinfo {author} {\bibfnamefont {C.}~\bibnamefont {Quintana}}, \bibinfo
  {author} {\bibfnamefont {P.}~\bibnamefont {Roushan}}, \bibinfo {author}
  {\bibfnamefont {A.}~\bibnamefont {Vainsencher}}, \bibinfo {author}
  {\bibfnamefont {J.}~\bibnamefont {Wenner}}, \bibinfo {author} {\bibfnamefont
  {A.~N.}\ \bibnamefont {Cleland}}, \ and\ \bibinfo {author} {\bibfnamefont
  {J.~M.}\ \bibnamefont {Martinis}},\ }\href {\doibase 10.1038/nature14270}
  {\bibfield  {journal} {\bibinfo  {journal} {Nature}\ }\textbf {\bibinfo
  {volume} {519}},\ \bibinfo {pages} {66} (\bibinfo {year} {2015})}\BibitemShut
  {NoStop}%
\bibitem [{\citenamefont {Mutus}\ \emph {et~al.}(2014)\citenamefont {Mutus},
  \citenamefont {White}, \citenamefont {Barends}, \citenamefont {Chen},
  \citenamefont {Chen}, \citenamefont {Chiaro}, \citenamefont {Dunsworth},
  \citenamefont {Jeffrey}, \citenamefont {Kelly}, \citenamefont {Megrant},
  \citenamefont {Neill}, \citenamefont {O'Malley}, \citenamefont {Roushan},
  \citenamefont {Sank}, \citenamefont {Vainsencher}, \citenamefont {Wenner},
  \citenamefont {Sundqvist}, \citenamefont {Cleland},\ and\ \citenamefont
  {Martinis}}]{Mutus:2014}%
  \BibitemOpen
  \bibfield  {author} {\bibinfo {author} {\bibfnamefont {J.~Y.}\ \bibnamefont
  {Mutus}}, \bibinfo {author} {\bibfnamefont {T.~C.}\ \bibnamefont {White}},
  \bibinfo {author} {\bibfnamefont {R.}~\bibnamefont {Barends}}, \bibinfo
  {author} {\bibfnamefont {Y.}~\bibnamefont {Chen}}, \bibinfo {author}
  {\bibfnamefont {Z.}~\bibnamefont {Chen}}, \bibinfo {author} {\bibfnamefont
  {B.}~\bibnamefont {Chiaro}}, \bibinfo {author} {\bibfnamefont
  {A.}~\bibnamefont {Dunsworth}}, \bibinfo {author} {\bibfnamefont
  {E.}~\bibnamefont {Jeffrey}}, \bibinfo {author} {\bibfnamefont
  {J.}~\bibnamefont {Kelly}}, \bibinfo {author} {\bibfnamefont
  {A.}~\bibnamefont {Megrant}}, \bibinfo {author} {\bibfnamefont
  {C.}~\bibnamefont {Neill}}, \bibinfo {author} {\bibfnamefont {P.~J.}\
  \bibnamefont {O'Malley}}, \bibinfo {author} {\bibfnamefont {P.}~\bibnamefont
  {Roushan}}, \bibinfo {author} {\bibfnamefont {D.}~\bibnamefont {Sank}},
  \bibinfo {author} {\bibfnamefont {A.}~\bibnamefont {Vainsencher}}, \bibinfo
  {author} {\bibfnamefont {J.}~\bibnamefont {Wenner}}, \bibinfo {author}
  {\bibfnamefont {K.~M.}\ \bibnamefont {Sundqvist}}, \bibinfo {author}
  {\bibfnamefont {A.~N.}\ \bibnamefont {Cleland}}, \ and\ \bibinfo {author}
  {\bibfnamefont {J.~M.}\ \bibnamefont {Martinis}},\ }\href {\doibase
  http://dx.doi.org/10.1063/1.4886408} {\bibfield  {journal} {\bibinfo
  {journal} {Applied Physics Letters}\ }\textbf {\bibinfo {volume} {104}},\
  \bibinfo {pages} {263513} (\bibinfo {year} {2014})}\BibitemShut {NoStop}%
\bibitem [{Note1()}]{Note1}%
  \BibitemOpen
  \bibinfo {note} {Even though state-of-the-art bonding machines can
  significantly mitigate these issues.}\BibitemShut {Stop}%
\bibitem [{\citenamefont {Brock}\ \emph {et~al.}(2000)\citenamefont {Brock},
  \citenamefont {Track},\ and\ \citenamefont {Rowell}}]{Brock:2000}%
  \BibitemOpen
  \bibfield  {author} {\bibinfo {author} {\bibfnamefont {D.~K.}\ \bibnamefont
  {Brock}}, \bibinfo {author} {\bibfnamefont {E.~K.}\ \bibnamefont {Track}}, \
  and\ \bibinfo {author} {\bibfnamefont {J.~M.}\ \bibnamefont {Rowell}},\
  }\href {\doibase 10.1109/6.887595} {\bibfield  {journal} {\bibinfo  {journal}
  {{IEEE} Spectrum}\ }\textbf {\bibinfo {volume} {37}},\ \bibinfo {pages} {40}
  (\bibinfo {year} {2000})}\BibitemShut {NoStop}%
\bibitem [{\citenamefont {Mukhanov}(2011)}]{Mukhanov:2011}%
  \BibitemOpen
  \bibfield  {author} {\bibinfo {author} {\bibfnamefont {O.~A.}\ \bibnamefont
  {Mukhanov}},\ }\href {\doibase 10.1109/TASC.2010.2096792} {\bibfield
  {journal} {\bibinfo  {journal} {{IEEE} Transactions on Applied
  Superconductivity}\ }\textbf {\bibinfo {volume} {21}},\ \bibinfo {pages}
  {760} (\bibinfo {year} {2011})}\BibitemShut {NoStop}%
\bibitem [{\citenamefont {Zapatka}\ and\ \citenamefont
  {Ziser}(2009)}]{Zapatka:2009}%
  \BibitemOpen
  \bibfield  {author} {\bibinfo {author} {\bibfnamefont {M.}~\bibnamefont
  {Zapatka}}\ and\ \bibinfo {author} {\bibfnamefont {R.}~\bibnamefont
  {Ziser}},\ }in\ \href {\doibase 10.1109/ARFTG74.2009.5439097} {\emph
  {\bibinfo {booktitle} {2009 74th {ARFTG} Microwave Measurement Conference}}}\
  (\bibinfo  {publisher} {Institute of Electrical {\&} Electronics Engineers
  ({IEEE})},\ \bibinfo {year} {2009})\ pp.\ \bibinfo {pages} {1--6}\BibitemShut
  {NoStop}%
\bibitem [{\citenamefont {Collin}(2001)}]{Collin:2001}%
  \BibitemOpen
  \bibfield  {author} {\bibinfo {author} {\bibfnamefont {R.~E.}\ \bibnamefont
  {Collin}},\ }\href {\doibase 10.1109/9780470544662} {\emph {\bibinfo {title}
  {Foundations for Microwave Engineering - 2nd Edition}}}\ (\bibinfo
  {publisher} {Institute of Electrical {\&} Electronics Engineers ({IEEE}),
  Inc., and John Wiley {\&} Sons, Inc.},\ \bibinfo {address} {New York, {NY},
  and Hoboken, {NJ}, {USA}},\ \bibinfo {year} {2001})\BibitemShut {NoStop}%
\bibitem [{\citenamefont {Pozar}(2011)}]{Pozar:2011}%
  \BibitemOpen
  \bibfield  {author} {\bibinfo {author} {\bibfnamefont {D.~M.}\ \bibnamefont
  {Pozar}},\ }\href
  {http://ca.wiley.com/WileyCDA/WileyTitle/productCd-EHEP002016.html} {\emph
  {\bibinfo {title} {Microwave Engineering - 4th Edition}}}\ (\bibinfo
  {publisher} {John Wiley {\&} Sons, Inc.},\ \bibinfo {address} {Hoboken, {NJ},
  {USA}},\ \bibinfo {year} {2011})\BibitemShut {NoStop}%
\bibitem [{\citenamefont {Wenner}\ \emph {et~al.}(2011)\citenamefont {Wenner},
  \citenamefont {Neeley}, \citenamefont {Bialczak}, \citenamefont {Lenander},
  \citenamefont {Lucero}, \citenamefont {O'Connell}, \citenamefont {Sank},
  \citenamefont {Wang}, \citenamefont {Weides}, \citenamefont {Cleland},\ and\
  \citenamefont {Martinis}}]{Wenner:2011:a}%
  \BibitemOpen
  \bibfield  {author} {\bibinfo {author} {\bibfnamefont {J.}~\bibnamefont
  {Wenner}}, \bibinfo {author} {\bibfnamefont {M.}~\bibnamefont {Neeley}},
  \bibinfo {author} {\bibfnamefont {R.~C.}\ \bibnamefont {Bialczak}}, \bibinfo
  {author} {\bibfnamefont {M.}~\bibnamefont {Lenander}}, \bibinfo {author}
  {\bibfnamefont {E.}~\bibnamefont {Lucero}}, \bibinfo {author} {\bibfnamefont
  {A.~D.}\ \bibnamefont {O'Connell}}, \bibinfo {author} {\bibfnamefont
  {D.}~\bibnamefont {Sank}}, \bibinfo {author} {\bibfnamefont {H.}~\bibnamefont
  {Wang}}, \bibinfo {author} {\bibfnamefont {M.}~\bibnamefont {Weides}},
  \bibinfo {author} {\bibfnamefont {A.~N.}\ \bibnamefont {Cleland}}, \ and\
  \bibinfo {author} {\bibfnamefont {J.~M.}\ \bibnamefont {Martinis}},\ }\href
  {\doibase 10.1088/0953-2048/24/6/065001} {\bibfield  {journal} {\bibinfo
  {journal} {Superconductor Science and Technology}\ }\textbf {\bibinfo
  {volume} {24}},\ \bibinfo {pages} {065001} (\bibinfo {year}
  {2011})}\BibitemShut {NoStop}%
\bibitem [{Note2()}]{Note2}%
  \BibitemOpen
  \bibinfo {note} {The pillar was included in the initial design as there was
  concern over potential damage to the substrate from mechanical strain due to
  the three-dimensional wires pushing on the top of the chip.}\BibitemShut
  {Stop}%
\bibitem [{Note3()}]{Note3}%
  \BibitemOpen
  \bibinfo {note} {The breaking temperature of those components is lower than
  the melting temperature of available eutectic solders.}\BibitemShut {Stop}%
\bibitem [{\citenamefont {Simons}(2001)}]{Simons:2001}%
  \BibitemOpen
  \bibfield  {author} {\bibinfo {author} {\bibfnamefont {R.~N.}\ \bibnamefont
  {Simons}},\ }\href {\doibase 10.1002/0471224758} {\emph {\bibinfo {title}
  {Coplanar Waveguide Circuits, Components, and Systems}}}\ (\bibinfo
  {publisher} {John Wiley {\&} Sons, Inc.},\ \bibinfo {address} {Hoboken, {NJ},
  {USA}},\ \bibinfo {year} {2001})\BibitemShut {NoStop}%
\bibitem [{Note4()}]{Note4}%
  \BibitemOpen
  \bibinfo {note} {Provided the vacuum still constitutes the majority of the
  volume of the cavity.}\BibitemShut {Stop}%
\bibitem [{Note5()}]{Note5}%
  \BibitemOpen
  \bibinfo {note} {Residual-resistivity ratio larger than~$300$.}\BibitemShut
  {Stop}%
\bibitem [{Note6()}]{Note6}%
  \BibitemOpen
  \bibinfo {note} {No Ni adhesion underlayer and an Au bath with minimum
  magnetic impurities were used. The deposited Au was hard Au, type~I, grade C
  (MIL-G-45204C), with a total measured quantity of Ni, iron, and cobalt of~\SI
  {0.204}{\percent }.}\BibitemShut {Stop}%
\bibitem [{\citenamefont {Frunzio}\ \emph {et~al.}(2005)\citenamefont
  {Frunzio}, \citenamefont {Wallraff}, \citenamefont {Schuster}, \citenamefont
  {Majer},\ and\ \citenamefont {Schoelkopf}}]{Frunzio:2005}%
  \BibitemOpen
  \bibfield  {author} {\bibinfo {author} {\bibfnamefont {L.}~\bibnamefont
  {Frunzio}}, \bibinfo {author} {\bibfnamefont {A.}~\bibnamefont {Wallraff}},
  \bibinfo {author} {\bibfnamefont {D.}~\bibnamefont {Schuster}}, \bibinfo
  {author} {\bibfnamefont {J.}~\bibnamefont {Majer}}, \ and\ \bibinfo {author}
  {\bibfnamefont {R.}~\bibnamefont {Schoelkopf}},\ }\href {\doibase
  10.1109/TASC.2005.850084} {\bibfield  {journal} {\bibinfo  {journal} {{IEEE}
  Trans. Appl. Supercond.}\ }\textbf {\bibinfo {volume} {15}},\ \bibinfo
  {pages} {860} (\bibinfo {year} {2005})}\BibitemShut {NoStop}%
\bibitem [{\citenamefont {Song}\ \emph
  {et~al.}(2009{\natexlab{a}})\citenamefont {Song}, \citenamefont {Heitmann},
  \citenamefont {DeFeo}, \citenamefont {Yu}, \citenamefont {McDermott},
  \citenamefont {Neeley}, \citenamefont {Martinis},\ and\ \citenamefont
  {Plourde}}]{Song:2009:a}%
  \BibitemOpen
  \bibfield  {author} {\bibinfo {author} {\bibfnamefont {C.}~\bibnamefont
  {Song}}, \bibinfo {author} {\bibfnamefont {T.~W.}\ \bibnamefont {Heitmann}},
  \bibinfo {author} {\bibfnamefont {M.~P.}\ \bibnamefont {DeFeo}}, \bibinfo
  {author} {\bibfnamefont {K.}~\bibnamefont {Yu}}, \bibinfo {author}
  {\bibfnamefont {R.}~\bibnamefont {McDermott}}, \bibinfo {author}
  {\bibfnamefont {M.}~\bibnamefont {Neeley}}, \bibinfo {author} {\bibfnamefont
  {J.~M.}\ \bibnamefont {Martinis}}, \ and\ \bibinfo {author} {\bibfnamefont
  {B.~L.}\ \bibnamefont {Plourde}},\ }\href {\doibase
  10.1103/PhysRevB.79.174512} {\bibfield  {journal} {\bibinfo  {journal} {Phys.
  Rev. B}\ }\textbf {\bibinfo {volume} {79}},\ \bibinfo {pages} {174512}
  (\bibinfo {year} {2009}{\natexlab{a}})}\BibitemShut {NoStop}%
\bibitem [{\citenamefont {Song}\ \emph
  {et~al.}(2009{\natexlab{b}})\citenamefont {Song}, \citenamefont {DeFeo},
  \citenamefont {Yu},\ and\ \citenamefont {Plourde}}]{Song:2009:b}%
  \BibitemOpen
  \bibfield  {author} {\bibinfo {author} {\bibfnamefont {C.}~\bibnamefont
  {Song}}, \bibinfo {author} {\bibfnamefont {M.~P.}\ \bibnamefont {DeFeo}},
  \bibinfo {author} {\bibfnamefont {K.}~\bibnamefont {Yu}}, \ and\ \bibinfo
  {author} {\bibfnamefont {B.~L.}\ \bibnamefont {Plourde}},\ }\href {\doibase
  10.1063/1.3271523} {\bibfield  {journal} {\bibinfo  {journal} {Appl. Phys.
  Lett.}\ }\textbf {\bibinfo {volume} {95}},\ \bibinfo {pages} {232501}
  (\bibinfo {year} {2009}{\natexlab{b}})}\BibitemShut {NoStop}%
\bibitem [{\citenamefont {Megrant}\ \emph {et~al.}(2012)\citenamefont
  {Megrant}, \citenamefont {Neill}, \citenamefont {Barends}, \citenamefont
  {Chiaro}, \citenamefont {Chen}, \citenamefont {Feigl}, \citenamefont {Kelly},
  \citenamefont {Lucero}, \citenamefont {Mariantoni}, \citenamefont {O'Malley},
  \citenamefont {Sank}, \citenamefont {Vainsencher}, \citenamefont {Wenner},
  \citenamefont {White}, \citenamefont {Yin}, \citenamefont {Zhao},
  \citenamefont {Palmstr{\o}m}, \citenamefont {Martinis},\ and\ \citenamefont
  {Cleland}}]{Megrant:2012}%
  \BibitemOpen
  \bibfield  {author} {\bibinfo {author} {\bibfnamefont {A.}~\bibnamefont
  {Megrant}}, \bibinfo {author} {\bibfnamefont {C.}~\bibnamefont {Neill}},
  \bibinfo {author} {\bibfnamefont {R.}~\bibnamefont {Barends}}, \bibinfo
  {author} {\bibfnamefont {B.}~\bibnamefont {Chiaro}}, \bibinfo {author}
  {\bibfnamefont {Y.}~\bibnamefont {Chen}}, \bibinfo {author} {\bibfnamefont
  {L.}~\bibnamefont {Feigl}}, \bibinfo {author} {\bibfnamefont
  {J.}~\bibnamefont {Kelly}}, \bibinfo {author} {\bibfnamefont
  {E.}~\bibnamefont {Lucero}}, \bibinfo {author} {\bibfnamefont
  {M.}~\bibnamefont {Mariantoni}}, \bibinfo {author} {\bibfnamefont {P.~J.}\
  \bibnamefont {O'Malley}}, \bibinfo {author} {\bibfnamefont {D.}~\bibnamefont
  {Sank}}, \bibinfo {author} {\bibfnamefont {A.}~\bibnamefont {Vainsencher}},
  \bibinfo {author} {\bibfnamefont {J.}~\bibnamefont {Wenner}}, \bibinfo
  {author} {\bibfnamefont {T.~C.}\ \bibnamefont {White}}, \bibinfo {author}
  {\bibfnamefont {Y.}~\bibnamefont {Yin}}, \bibinfo {author} {\bibfnamefont
  {J.}~\bibnamefont {Zhao}}, \bibinfo {author} {\bibfnamefont {C.~J.}\
  \bibnamefont {Palmstr{\o}m}}, \bibinfo {author} {\bibfnamefont {J.~M.}\
  \bibnamefont {Martinis}}, \ and\ \bibinfo {author} {\bibfnamefont {A.~N.}\
  \bibnamefont {Cleland}},\ }\href {\doibase 10.1063/1.3693409} {\bibfield
  {journal} {\bibinfo  {journal} {Appl. Phys. Lett.}\ }\textbf {\bibinfo
  {volume} {100}},\ \bibinfo {pages} {113510} (\bibinfo {year}
  {2012})}\BibitemShut {NoStop}%
\bibitem [{Note7()}]{Note7}%
  \BibitemOpen
  \bibinfo {note} {Alloy~430, UNS C69300.}\BibitemShut {Stop}%
\bibitem [{Note8()}]{Note8}%
  \BibitemOpen
  \bibinfo {note} {UNS C52100.}\BibitemShut {Stop}%
\bibitem [{\citenamefont {C{\'o}rcoles}\ \emph {et~al.}(2011)\citenamefont
  {C{\'o}rcoles}, \citenamefont {Chow}, \citenamefont {Gambetta}, \citenamefont
  {Rigetti}, \citenamefont {Rozen}, \citenamefont {Keefe}, \citenamefont
  {Rothwell}, \citenamefont {Ketchen},\ and\ \citenamefont
  {Steffen}}]{Corcoles:2011}%
  \BibitemOpen
  \bibfield  {author} {\bibinfo {author} {\bibfnamefont {A.~D.}\ \bibnamefont
  {C{\'o}rcoles}}, \bibinfo {author} {\bibfnamefont {J.~M.}\ \bibnamefont
  {Chow}}, \bibinfo {author} {\bibfnamefont {J.~M.}\ \bibnamefont {Gambetta}},
  \bibinfo {author} {\bibfnamefont {C.}~\bibnamefont {Rigetti}}, \bibinfo
  {author} {\bibfnamefont {J.~R.}\ \bibnamefont {Rozen}}, \bibinfo {author}
  {\bibfnamefont {G.~A.}\ \bibnamefont {Keefe}}, \bibinfo {author}
  {\bibfnamefont {M.~B.}\ \bibnamefont {Rothwell}}, \bibinfo {author}
  {\bibfnamefont {M.~B.}\ \bibnamefont {Ketchen}}, \ and\ \bibinfo {author}
  {\bibfnamefont {M.}~\bibnamefont {Steffen}},\ }\href {\doibase
  10.1063/1.3658630} {\bibfield  {journal} {\bibinfo  {journal} {Appl. Phys.
  Lett.}\ }\textbf {\bibinfo {volume} {99}},\ \bibinfo {pages} {181906}
  (\bibinfo {year} {2011})}\BibitemShut {NoStop}%
\bibitem [{\citenamefont {Fowler}\ and\ \citenamefont
  {Martinis}(2014)}]{Fowler:2014}%
  \BibitemOpen
  \bibfield  {author} {\bibinfo {author} {\bibfnamefont {A.~G.}\ \bibnamefont
  {Fowler}}\ and\ \bibinfo {author} {\bibfnamefont {J.~M.}\ \bibnamefont
  {Martinis}},\ }\href {\doibase 10.1103/PhysRevA.89.032316} {\bibfield
  {journal} {\bibinfo  {journal} {Phys. Rev. A}\ }\textbf {\bibinfo {volume}
  {89}},\ \bibinfo {pages} {032316} (\bibinfo {year} {2014})}\BibitemShut
  {NoStop}%
\bibitem [{Note9()}]{Note9}%
  \BibitemOpen
  \bibinfo {note} {In a few instances, the values were confirmed using a
  precision source-measure unit from Keysight Technologies Inc.,
  model~B2911A.}\BibitemShut {Stop}%
\bibitem [{\citenamefont {Barends}\ \emph {et~al.}(2013)\citenamefont
  {Barends}, \citenamefont {Kelly}, \citenamefont {Megrant}, \citenamefont
  {Sank}, \citenamefont {Jeffrey}, \citenamefont {Chen}, \citenamefont {Yin},
  \citenamefont {Chiaro}, \citenamefont {Mutus}, \citenamefont {Neill},
  \citenamefont {O'Malley}, \citenamefont {Roushan}, \citenamefont {Wenner},
  \citenamefont {White}, \citenamefont {Cleland},\ and\ \citenamefont
  {Martinis}}]{Barends:2013}%
  \BibitemOpen
  \bibfield  {author} {\bibinfo {author} {\bibfnamefont {R.}~\bibnamefont
  {Barends}}, \bibinfo {author} {\bibfnamefont {J.}~\bibnamefont {Kelly}},
  \bibinfo {author} {\bibfnamefont {A.}~\bibnamefont {Megrant}}, \bibinfo
  {author} {\bibfnamefont {D.}~\bibnamefont {Sank}}, \bibinfo {author}
  {\bibfnamefont {E.}~\bibnamefont {Jeffrey}}, \bibinfo {author} {\bibfnamefont
  {Y.}~\bibnamefont {Chen}}, \bibinfo {author} {\bibfnamefont {Y.}~\bibnamefont
  {Yin}}, \bibinfo {author} {\bibfnamefont {B.}~\bibnamefont {Chiaro}},
  \bibinfo {author} {\bibfnamefont {J.}~\bibnamefont {Mutus}}, \bibinfo
  {author} {\bibfnamefont {C.}~\bibnamefont {Neill}}, \bibinfo {author}
  {\bibfnamefont {P.}~\bibnamefont {O'Malley}}, \bibinfo {author}
  {\bibfnamefont {P.}~\bibnamefont {Roushan}}, \bibinfo {author} {\bibfnamefont
  {J.}~\bibnamefont {Wenner}}, \bibinfo {author} {\bibfnamefont {T.~C.}\
  \bibnamefont {White}}, \bibinfo {author} {\bibfnamefont {A.~N.}\ \bibnamefont
  {Cleland}}, \ and\ \bibinfo {author} {\bibfnamefont {J.~M.}\ \bibnamefont
  {Martinis}},\ }\href {\doibase 10.1103/PhysRevLett.111.080502} {\bibfield
  {journal} {\bibinfo  {journal} {Phys. Rev. Lett.}\ }\textbf {\bibinfo
  {volume} {111}},\ \bibinfo {pages} {080502} (\bibinfo {year}
  {2013})}\BibitemShut {NoStop}%
\bibitem [{\citenamefont {Teverovsky}(2004)}]{Teverovsky:2004}%
  \BibitemOpen
  \bibfield  {author} {\bibinfo {author} {\bibfnamefont {A.}~\bibnamefont
  {Teverovsky}},\ }in\ \href {\doibase 10.1109/RELPHY.2004.1315388} {\emph
  {\bibinfo {booktitle} {Proceedings of the 42nd Annual {IEEE} International
  Reliability Physics Symposium, Phoenix, 2004}}}\ (\bibinfo  {publisher}
  {Institute of Electrical {\&} Electronics Engineers ({IEEE})},\ \bibinfo
  {year} {2004})\ pp.\ \bibinfo {pages} {547--556}\BibitemShut {NoStop}%
\bibitem [{Note10()}]{Note10}%
  \BibitemOpen
  \bibinfo {note} {Unless specified, all calibrations were performed using SMA
  female to 3.5~mm male adapters. These adapters introduced negligible
  calibration errors from DC to~\SI {10}{\giga \hertz }.}\BibitemShut {Stop}%
\bibitem [{Note11()}]{Note11}%
  \BibitemOpen
  \bibinfo {note} {Note that the Ag measurements were performed without using
  the SMA female to 3.5~mm male adapters.}\BibitemShut {Stop}%
\bibitem [{\citenamefont {Marquardt}\ \emph {et~al.}(2000)\citenamefont
  {Marquardt}, \citenamefont {Le},\ and\ \citenamefont
  {Radebaugh}}]{Marquardt:2000}%
  \BibitemOpen
  \bibfield  {author} {\bibinfo {author} {\bibfnamefont {E.~D.}\ \bibnamefont
  {Marquardt}}, \bibinfo {author} {\bibfnamefont {J.~P.}\ \bibnamefont {Le}}, \
  and\ \bibinfo {author} {\bibfnamefont {R.}~\bibnamefont {Radebaugh}},\ }in\
  \href {\doibase 10.1007/0-306-47112-4_84} {\emph {\bibinfo {booktitle} {11th
  International Cryocooler Conference}}}\ (\bibinfo  {publisher} {Springer
  Science},\ \bibinfo {year} {2000})\ pp.\ \bibinfo {pages}
  {681--687}\BibitemShut {NoStop}%
\bibitem [{Note12()}]{Note12}%
  \BibitemOpen
  \bibinfo {note} {Note that the DUT is a piecewise transmission line
  inhomogeneously filled with dielectric materials. Transforming the time~$t$
  into distance is only possible with detailed knowledge of geometries and
  materials for all regions of the DUT. Since this information is not known to
  a high degree of accuracy, we prefer to express all measured quantities as a
  function of~$t$.}\BibitemShut {Stop}%
\bibitem [{\citenamefont {Bogatin}(2003)}]{Bogatin:2003}%
  \BibitemOpen
  \bibfield  {author} {\bibinfo {author} {\bibfnamefont {E.}~\bibnamefont
  {Bogatin}},\ }\href@noop {} {\emph {\bibinfo {title} {Signal Integrity -
  Simplified - 1st Edition}}}\ (\bibinfo  {publisher} {Prentice Hall
  Professional Technical Reference},\ \bibinfo {address} {Upper Saddle River,
  {NJ}, {USA}},\ \bibinfo {year} {2003})\BibitemShut {NoStop}%
\bibitem [{Note13()}]{Note13}%
  \BibitemOpen
  \bibinfo {note} {Work in progress.}\BibitemShut {Stop}%
\bibitem [{\citenamefont {Abraham}\ \emph
  {et~al.}(2014{\natexlab{b}})\citenamefont {Abraham}, \citenamefont {Keefe},
  \citenamefont {Lavoie},\ and\ \citenamefont {Rothwell}}]{Abraham:2014:b}%
  \BibitemOpen
  \bibfield  {author} {\bibinfo {author} {\bibfnamefont {D.~W.}\ \bibnamefont
  {Abraham}}, \bibinfo {author} {\bibfnamefont {G.~A.}\ \bibnamefont {Keefe}},
  \bibinfo {author} {\bibfnamefont {C.}~\bibnamefont {Lavoie}}, \ and\ \bibinfo
  {author} {\bibfnamefont {M.~E.}\ \bibnamefont {Rothwell}},\ }\href
  {http://patents.justia.com/patent/20140274725} {\enquote {\bibinfo {title}
  {Chip mode isolation and cross-talk reduction through buried metal layers and
  through-vias},}\ } (\bibinfo {year} {2014}{\natexlab{b}})\BibitemShut
  {NoStop}%
\bibitem [{\citenamefont {Abraham}\ \emph {et~al.}(2015)\citenamefont
  {Abraham}, \citenamefont {Chow},\ and\ \citenamefont
  {Gambetta}}]{Abraham:2015}%
  \BibitemOpen
  \bibfield  {author} {\bibinfo {author} {\bibfnamefont {D.~W.}\ \bibnamefont
  {Abraham}}, \bibinfo {author} {\bibfnamefont {J.~M.}\ \bibnamefont {Chow}}, \
  and\ \bibinfo {author} {\bibfnamefont {J.~M.}\ \bibnamefont {Gambetta}},\
  }\href {https://www.google.com/patents/US20140266406} {\enquote {\bibinfo
  {title} {Symmetric placement of components on a chip to reduce crosstalk
  induced by chip modes},}\ } (\bibinfo {year} {2015}),\ \bibinfo {note} {{US}
  Patent 8,972,921}\BibitemShut {NoStop}%
\bibitem [{Note14()}]{Note14}%
  \BibitemOpen
  \bibinfo {note} {Daniel T.~Sank (private communication).}\BibitemShut {Stop}%
\bibitem [{\citenamefont {Haack}\ \emph {et~al.}(2010)\citenamefont {Haack},
  \citenamefont {Helmer}, \citenamefont {Mariantoni}, \citenamefont
  {Marquardt},\ and\ \citenamefont {Solano}}]{Haack:2010}%
  \BibitemOpen
  \bibfield  {author} {\bibinfo {author} {\bibfnamefont {G.}~\bibnamefont
  {Haack}}, \bibinfo {author} {\bibfnamefont {F.}~\bibnamefont {Helmer}},
  \bibinfo {author} {\bibfnamefont {M.}~\bibnamefont {Mariantoni}}, \bibinfo
  {author} {\bibfnamefont {F.}~\bibnamefont {Marquardt}}, \ and\ \bibinfo
  {author} {\bibfnamefont {E.}~\bibnamefont {Solano}},\ }\href {\doibase
  10.1103/PhysRevB.82.024514} {\bibfield  {journal} {\bibinfo  {journal} {Phys.
  Rev. B}\ }\textbf {\bibinfo {volume} {82}},\ \bibinfo {pages} {024514}
  (\bibinfo {year} {2010})}\BibitemShut {NoStop}%
\bibitem [{\citenamefont {Mariantoni}\ \emph {et~al.}(2011)\citenamefont
  {Mariantoni}, \citenamefont {Wang}, \citenamefont {Yamamoto}, \citenamefont
  {Neeley}, \citenamefont {Bialczak}, \citenamefont {Chen}, \citenamefont
  {Lenander}, \citenamefont {Lucero}, \citenamefont {O'Connell}, \citenamefont
  {Sank}, \citenamefont {Weides}, \citenamefont {Wenner}, \citenamefont {Yin},
  \citenamefont {Zhao}, \citenamefont {Korotkov}, \citenamefont {Cleland},\
  and\ \citenamefont {Martinis}}]{Mariantoni:2011}%
  \BibitemOpen
  \bibfield  {author} {\bibinfo {author} {\bibfnamefont {M.}~\bibnamefont
  {Mariantoni}}, \bibinfo {author} {\bibfnamefont {H.}~\bibnamefont {Wang}},
  \bibinfo {author} {\bibfnamefont {T.}~\bibnamefont {Yamamoto}}, \bibinfo
  {author} {\bibfnamefont {M.}~\bibnamefont {Neeley}}, \bibinfo {author}
  {\bibfnamefont {R.~C.}\ \bibnamefont {Bialczak}}, \bibinfo {author}
  {\bibfnamefont {Y.}~\bibnamefont {Chen}}, \bibinfo {author} {\bibfnamefont
  {M.}~\bibnamefont {Lenander}}, \bibinfo {author} {\bibfnamefont
  {E.}~\bibnamefont {Lucero}}, \bibinfo {author} {\bibfnamefont {A.~D.}\
  \bibnamefont {O'Connell}}, \bibinfo {author} {\bibfnamefont {D.}~\bibnamefont
  {Sank}}, \bibinfo {author} {\bibfnamefont {M.}~\bibnamefont {Weides}},
  \bibinfo {author} {\bibfnamefont {J.}~\bibnamefont {Wenner}}, \bibinfo
  {author} {\bibfnamefont {Y.}~\bibnamefont {Yin}}, \bibinfo {author}
  {\bibfnamefont {J.}~\bibnamefont {Zhao}}, \bibinfo {author} {\bibfnamefont
  {A.~N.}\ \bibnamefont {Korotkov}}, \bibinfo {author} {\bibfnamefont {A.~N.}\
  \bibnamefont {Cleland}}, \ and\ \bibinfo {author} {\bibfnamefont {J.~M.}\
  \bibnamefont {Martinis}},\ }\href {\doibase 10.1126/science.1208517}
  {\bibfield  {journal} {\bibinfo  {journal} {Science}\ }\textbf {\bibinfo
  {volume} {334}},\ \bibinfo {pages} {61} (\bibinfo {year} {2011})}\BibitemShut
  {NoStop}%
\bibitem [{\citenamefont {Mariantoni}\ \emph {et~al.}(2008)\citenamefont
  {Mariantoni}, \citenamefont {Deppe}, \citenamefont {Marx}, \citenamefont
  {Gross}, \citenamefont {Wilhelm},\ and\ \citenamefont
  {Solano}}]{Mariantoni:2008}%
  \BibitemOpen
  \bibfield  {author} {\bibinfo {author} {\bibfnamefont {M.}~\bibnamefont
  {Mariantoni}}, \bibinfo {author} {\bibfnamefont {F.}~\bibnamefont {Deppe}},
  \bibinfo {author} {\bibfnamefont {A.}~\bibnamefont {Marx}}, \bibinfo {author}
  {\bibfnamefont {R.}~\bibnamefont {Gross}}, \bibinfo {author} {\bibfnamefont
  {F.~K.}\ \bibnamefont {Wilhelm}}, \ and\ \bibinfo {author} {\bibfnamefont
  {E.}~\bibnamefont {Solano}},\ }\href {\doibase 10.1103/PhysRevB.78.104508}
  {\bibfield  {journal} {\bibinfo  {journal} {Phys. Rev. B}\ }\textbf {\bibinfo
  {volume} {78}},\ \bibinfo {pages} {104508} (\bibinfo {year}
  {2008})}\BibitemShut {NoStop}%
\bibitem [{\citenamefont {Gupta}\ \emph {et~al.}(2011)\citenamefont {Gupta},
  \citenamefont {Kirichenko}, \citenamefont {Dotsenko}, \citenamefont {Miller},
  \citenamefont {Sarwana}, \citenamefont {Talalaevskii}, \citenamefont
  {Delmas}, \citenamefont {Webber}, \citenamefont {Govorkov}, \citenamefont
  {Kirichenko}, \citenamefont {Vernik},\ and\ \citenamefont
  {Tang}}]{Gupta:2011}%
  \BibitemOpen
  \bibfield  {author} {\bibinfo {author} {\bibfnamefont {D.}~\bibnamefont
  {Gupta}}, \bibinfo {author} {\bibfnamefont {D.~E.}\ \bibnamefont
  {Kirichenko}}, \bibinfo {author} {\bibfnamefont {V.~V.}\ \bibnamefont
  {Dotsenko}}, \bibinfo {author} {\bibfnamefont {R.}~\bibnamefont {Miller}},
  \bibinfo {author} {\bibfnamefont {S.}~\bibnamefont {Sarwana}}, \bibinfo
  {author} {\bibfnamefont {A.}~\bibnamefont {Talalaevskii}}, \bibinfo {author}
  {\bibfnamefont {J.}~\bibnamefont {Delmas}}, \bibinfo {author} {\bibfnamefont
  {R.~J.}\ \bibnamefont {Webber}}, \bibinfo {author} {\bibfnamefont
  {S.}~\bibnamefont {Govorkov}}, \bibinfo {author} {\bibfnamefont {A.~F.}\
  \bibnamefont {Kirichenko}}, \bibinfo {author} {\bibfnamefont {I.~V.}\
  \bibnamefont {Vernik}}, \ and\ \bibinfo {author} {\bibfnamefont
  {J.}~\bibnamefont {Tang}},\ }\href {\doibase 10.1109/TASC.2010.2095399}
  {\bibfield  {journal} {\bibinfo  {journal} {{IEEE} Trans. Appl. Supercond.}\
  }\textbf {\bibinfo {volume} {21}},\ \bibinfo {pages} {883} (\bibinfo {year}
  {2011})}\BibitemShut {NoStop}%
\bibitem [{Note15()}]{Note15}%
  \BibitemOpen
  \bibinfo {note} {Confer~\protect \url
  {https://www.iarpa.gov/index.php/research-programs/c3} .}\BibitemShut {Stop}%
\bibitem [{Note16()}]{Note16}%
  \BibitemOpen
  \bibinfo {note} {Note that we also performed magnetic tests by exposing all
  samples to a ultra-high pull neodymium rectangular magnet, with
  dimensions~$\SI {25.4}{\milli \meter } {} \times {} \SI {25.4}{\milli \meter
  } {} \times {} \SI {9.5}{\milli \meter }$ and a pull of~\SI {10.4}{\kilo
  \gram }. We found magnetic fields with the same order of magnitude as in
  Table~\ref {Table06:Bejanin}.}\BibitemShut {Stop}%
\bibitem [{Note17()}]{Note17}%
  \BibitemOpen
  \bibinfo {note} {Confer~\protect \url
  {http://www.lakeshore.com/Documents/LSTC_appendixI_l.pdf} .}\BibitemShut
  {Stop}%
\bibitem [{\citenamefont {Mueller}\ \emph {et~al.}(1978)\citenamefont
  {Mueller}, \citenamefont {Buchal}, \citenamefont {Oversluizen},\ and\
  \citenamefont {Pobell}}]{Mueller:1978}%
  \BibitemOpen
  \bibfield  {author} {\bibinfo {author} {\bibfnamefont {R.~M.}\ \bibnamefont
  {Mueller}}, \bibinfo {author} {\bibfnamefont {C.}~\bibnamefont {Buchal}},
  \bibinfo {author} {\bibfnamefont {T.}~\bibnamefont {Oversluizen}}, \ and\
  \bibinfo {author} {\bibfnamefont {F.}~\bibnamefont {Pobell}},\ }\href
  {\doibase 10.1063/1.1135452} {\bibfield  {journal} {\bibinfo  {journal} {Rev.
  Sci. Instrum.}\ }\textbf {\bibinfo {volume} {49}},\ \bibinfo {pages} {515}
  (\bibinfo {year} {1978})}\BibitemShut {NoStop}%
\bibitem [{Note18()}]{Note18}%
  \BibitemOpen
  \bibinfo {note} {ISO~AlMg1SiCu; UNS~A96061.}\BibitemShut {Stop}%
\bibitem [{Note19()}]{Note19}%
  \BibitemOpen
  \bibinfo {note} {Note that~$\alpha ( 4 )$ can be accurately estimated from
  the data at~\protect \url
  {http://www.cryogenics.nist.gov/MPropsMAY/6061\%20Aluminum/6061_T6Aluminum_rev.htm}
  .}\BibitemShut {Stop}%
\bibitem [{\citenamefont {Swenson}(1983)}]{Swenson:1983}%
  \BibitemOpen
  \bibfield  {author} {\bibinfo {author} {\bibfnamefont {C.~A.}\ \bibnamefont
  {Swenson}},\ }\href {\doibase 10.1063/1.555681} {\bibfield  {journal}
  {\bibinfo  {journal} {Journal of Physical and Chemical Reference Data}\
  }\textbf {\bibinfo {volume} {12}},\ \bibinfo {pages} {179} (\bibinfo {year}
  {1983})}\BibitemShut {NoStop}%
\bibitem [{Note20()}]{Note20}%
  \BibitemOpen
  \bibinfo {note} {Confer~\protect \url
  {http://www.ece.rutgers.edu/~orfanidi/ewa/} .}\BibitemShut {Stop}%
\bibitem [{Note21()}]{Note21}%
  \BibitemOpen
  \bibinfo {note} {Confer~\protect \url
  {http://www.ece.rutgers.edu/~orfanidi/ewa/} .}\BibitemShut {Stop}%
\bibitem [{\citenamefont {Barends}\ \emph {et~al.}(2011)\citenamefont
  {Barends}, \citenamefont {Wenner}, \citenamefont {Lenander}, \citenamefont
  {Chen}, \citenamefont {Bialczak}, \citenamefont {Kelly}, \citenamefont
  {Lucero}, \citenamefont {O'Malley}, \citenamefont {Mariantoni}, \citenamefont
  {Sank}, \citenamefont {Wang}, \citenamefont {White}, \citenamefont {Yin},
  \citenamefont {Zhao}, \citenamefont {Cleland}, \citenamefont {Martinis},\
  and\ \citenamefont {Baselmans}}]{Barends:2011}%
  \BibitemOpen
  \bibfield  {author} {\bibinfo {author} {\bibfnamefont {R.}~\bibnamefont
  {Barends}}, \bibinfo {author} {\bibfnamefont {J.}~\bibnamefont {Wenner}},
  \bibinfo {author} {\bibfnamefont {M.}~\bibnamefont {Lenander}}, \bibinfo
  {author} {\bibfnamefont {Y.}~\bibnamefont {Chen}}, \bibinfo {author}
  {\bibfnamefont {R.~C.}\ \bibnamefont {Bialczak}}, \bibinfo {author}
  {\bibfnamefont {J.}~\bibnamefont {Kelly}}, \bibinfo {author} {\bibfnamefont
  {E.}~\bibnamefont {Lucero}}, \bibinfo {author} {\bibfnamefont
  {P.}~\bibnamefont {O'Malley}}, \bibinfo {author} {\bibfnamefont
  {M.}~\bibnamefont {Mariantoni}}, \bibinfo {author} {\bibfnamefont
  {D.}~\bibnamefont {Sank}}, \bibinfo {author} {\bibfnamefont {H.}~\bibnamefont
  {Wang}}, \bibinfo {author} {\bibfnamefont {T.~C.}\ \bibnamefont {White}},
  \bibinfo {author} {\bibfnamefont {Y.}~\bibnamefont {Yin}}, \bibinfo {author}
  {\bibfnamefont {J.}~\bibnamefont {Zhao}}, \bibinfo {author} {\bibfnamefont
  {A.~N.}\ \bibnamefont {Cleland}}, \bibinfo {author} {\bibfnamefont {J.~M.}\
  \bibnamefont {Martinis}}, \ and\ \bibinfo {author} {\bibfnamefont {J.~J.~A.}\
  \bibnamefont {Baselmans}},\ }\href {\doibase 10.1063/1.3638063} {\bibfield
  {journal} {\bibinfo  {journal} {Appl. Phys. Lett.}\ }\textbf {\bibinfo
  {volume} {99}},\ \bibinfo {pages} {113507} (\bibinfo {year}
  {2011})}\BibitemShut {NoStop}%
\end{thebibliography}

\begin{thebibliography}{3}%
\makeatletter
\providecommand \@ifxundefined [1]{%
 \@ifx{#1\undefined}
}%
\providecommand \@ifnum [1]{%
 \ifnum #1\expandafter \@firstoftwo
 \else \expandafter \@secondoftwo
 \fi
}%
\providecommand \@ifx [1]{%
 \ifx #1\expandafter \@firstoftwo
 \else \expandafter \@secondoftwo
 \fi
}%
\providecommand \natexlab [1]{#1}%
\providecommand \enquote  [1]{``#1''}%
\providecommand \bibnamefont  [1]{#1}%
\providecommand \bibfnamefont [1]{#1}%
\providecommand \citenamefont [1]{#1}%
\providecommand \href@noop [0]{\@secondoftwo}%
\providecommand \href [0]{\begingroup \@sanitize@url \@href}%
\providecommand \@href[1]{\@@startlink{#1}\@@href}%
\providecommand \@@href[1]{\endgroup#1\@@endlink}%
\providecommand \@sanitize@url [0]{\catcode `\\12\catcode `\$12\catcode
  `\&12\catcode `\#12\catcode `\^12\catcode `\_12\catcode `\%12\relax}%
\providecommand \@@startlink[1]{}%
\providecommand \@@endlink[0]{}%
\providecommand \url  [0]{\begingroup\@sanitize@url \@url }%
\providecommand \@url [1]{\endgroup\@href {#1}{\urlprefix }}%
\providecommand \urlprefix  [0]{URL }%
\providecommand \Eprint [0]{\href }%
\providecommand \doibase [0]{http://dx.doi.org/}%
\providecommand \selectlanguage [0]{\@gobble}%
\providecommand \bibinfo  [0]{\@secondoftwo}%
\providecommand \bibfield  [0]{\@secondoftwo}%
\providecommand \translation [1]{[#1]}%
\providecommand \BibitemOpen [0]{}%
\providecommand \bibitemStop [0]{}%
\providecommand \bibitemNoStop [0]{.\EOS\space}%
\providecommand \EOS [0]{\spacefactor3000\relax}%
\providecommand \BibitemShut  [1]{\csname bibitem#1\endcsname}%
\let\auto@bib@innerbib\@empty
\bibitem [{\citenamefont {Frunzio}\ \emph {et~al.}(2005)\citenamefont
  {Frunzio}, \citenamefont {Wallraff}, \citenamefont {Schuster}, \citenamefont
  {Majer},\ and\ \citenamefont {Schoelkopf}}]{Frunzio:2005}%
  \BibitemOpen
  \bibfield  {author} {\bibinfo {author} {\bibfnamefont {L.}~\bibnamefont
  {Frunzio}}, \bibinfo {author} {\bibfnamefont {A.}~\bibnamefont {Wallraff}},
  \bibinfo {author} {\bibfnamefont {D.}~\bibnamefont {Schuster}}, \bibinfo
  {author} {\bibfnamefont {J.}~\bibnamefont {Majer}}, \ and\ \bibinfo {author}
  {\bibfnamefont {R.}~\bibnamefont {Schoelkopf}},\ }\href {\doibase
  10.1109/TASC.2005.850084} {\bibfield  {journal} {\bibinfo  {journal} {{IEEE}
  Trans. Appl. Supercond.}\ }\textbf {\bibinfo {volume} {15}},\ \bibinfo
  {pages} {860} (\bibinfo {year} {2005})}\BibitemShut {NoStop}%
\bibitem [{\citenamefont {Megrant}\ \emph {et~al.}(2012)\citenamefont
  {Megrant}, \citenamefont {Neill}, \citenamefont {Barends}, \citenamefont
  {Chiaro}, \citenamefont {Chen}, \citenamefont {Feigl}, \citenamefont {Kelly},
  \citenamefont {Lucero}, \citenamefont {Mariantoni}, \citenamefont {O'Malley},
  \citenamefont {Sank}, \citenamefont {Vainsencher}, \citenamefont {Wenner},
  \citenamefont {White}, \citenamefont {Yin}, \citenamefont {Zhao},
  \citenamefont {Palmstr{\o}m}, \citenamefont {Martinis},\ and\ \citenamefont
  {Cleland}}]{Megrant:2012}%
  \BibitemOpen
  \bibfield  {author} {\bibinfo {author} {\bibfnamefont {A.}~\bibnamefont
  {Megrant}}, \bibinfo {author} {\bibfnamefont {C.}~\bibnamefont {Neill}},
  \bibinfo {author} {\bibfnamefont {R.}~\bibnamefont {Barends}}, \bibinfo
  {author} {\bibfnamefont {B.}~\bibnamefont {Chiaro}}, \bibinfo {author}
  {\bibfnamefont {Y.}~\bibnamefont {Chen}}, \bibinfo {author} {\bibfnamefont
  {L.}~\bibnamefont {Feigl}}, \bibinfo {author} {\bibfnamefont
  {J.}~\bibnamefont {Kelly}}, \bibinfo {author} {\bibfnamefont
  {E.}~\bibnamefont {Lucero}}, \bibinfo {author} {\bibfnamefont
  {M.}~\bibnamefont {Mariantoni}}, \bibinfo {author} {\bibfnamefont {P.~J.}\
  \bibnamefont {O'Malley}}, \bibinfo {author} {\bibfnamefont {D.}~\bibnamefont
  {Sank}}, \bibinfo {author} {\bibfnamefont {A.}~\bibnamefont {Vainsencher}},
  \bibinfo {author} {\bibfnamefont {J.}~\bibnamefont {Wenner}}, \bibinfo
  {author} {\bibfnamefont {T.~C.}\ \bibnamefont {White}}, \bibinfo {author}
  {\bibfnamefont {Y.}~\bibnamefont {Yin}}, \bibinfo {author} {\bibfnamefont
  {J.}~\bibnamefont {Zhao}}, \bibinfo {author} {\bibfnamefont {C.~J.}\
  \bibnamefont {Palmstr{\o}m}}, \bibinfo {author} {\bibfnamefont {J.~M.}\
  \bibnamefont {Martinis}}, \ and\ \bibinfo {author} {\bibfnamefont {A.~N.}\
  \bibnamefont {Cleland}},\ }\href {\doibase 10.1063/1.3693409} {\bibfield
  {journal} {\bibinfo  {journal} {Appl. Phys. Lett.}\ }\textbf {\bibinfo
  {volume} {100}},\ \bibinfo {pages} {113510} (\bibinfo {year}
  {2012})}\BibitemShut {NoStop}%
\bibitem [{\citenamefont {Fowler}\ \emph {et~al.}(2012)\citenamefont {Fowler},
  \citenamefont {Mariantoni}, \citenamefont {Martinis},\ and\ \citenamefont
  {Cleland}}]{Fowler:2012}%
  \BibitemOpen
  \bibfield  {author} {\bibinfo {author} {\bibfnamefont {A.~G.}\ \bibnamefont
  {Fowler}}, \bibinfo {author} {\bibfnamefont {M.}~\bibnamefont {Mariantoni}},
  \bibinfo {author} {\bibfnamefont {J.~M.}\ \bibnamefont {Martinis}}, \ and\
  \bibinfo {author} {\bibfnamefont {A.~N.}\ \bibnamefont {Cleland}},\ }\href
  {\doibase 10.1103/PhysRevA.86.032324} {\bibfield  {journal} {\bibinfo
  {journal} {Phys. Rev. A}\ }\textbf {\bibinfo {volume} {86}},\ \bibinfo
  {pages} {032324} (\bibinfo {year} {2012})}\BibitemShut {NoStop}%
\end{thebibliography}

%

\end{document}